%% file: B2G-20-007_temp.tex
\pdfoutput=1
\documentclass[11pt,twoside,a4paper,cmspaper,final,collab]{cms-tdr}
\def\svnVersion{15ee2f2}\def\svnDate{2022/04/27}\def\cmsCernNoTag{CERN-EP-2021-226}\def\cmsCernDate{\today}\def\cmsMessage{Published in the Journal of High Energy Physics as \href{http://dx.doi.org/10.1007/JHEP05(2022)005}{\doi{10.1007/JHEP05(2022)005}.}}
\begin{document}\cmsNoteHeader{B2G-20-007}

\newlength\cmsTabSkip\setlength{\cmsTabSkip}{1ex}
\newcommand{\mx}{\ensuremath{m_{\PX}}\xspace}
\newcommand{\mbb}{\ensuremath{m_{\bbbar}}\xspace}
\newcommand{\mhh}{\ensuremath{m_{\PH\PH}}\xspace}
\newcommand{\qqpr}{\ensuremath{\qqbar^\prime}\xspace}

\newcommand{\hh}{\ensuremath{\PH\PH}\xspace}
\newcommand{\hbb}{\ensuremath{\PH\to\bbbar}\xspace}
\newcommand{\hww}{\ensuremath{\PH\to\PW\PW^*}\xspace}
\newcommand{\wqq}{\ensuremath{\PW\to\qqpr}\xspace}
\newcommand{\hhbbww}{\ensuremath{\PH\PH\to\bbbar\PW\PW^*}\xspace}
\newcommand{\hhbbtautau}{\ensuremath{\PH\PH\to\bbbar\PGt\PGt}\xspace}
\newcommand{\wwlnuqq}{\ensuremath{\PW\PW^*\to\Pell\Pgn\qqpr}\xspace}
\newcommand{\wwlnulnu}{\ensuremath{\PW\PW^*\to\Pell\PGn\Pell\PGn}\xspace}
\newcommand{\tautaulnunulnunu}{\ensuremath{\PGt\PGt \to\Pell\PGn\PGn\Pell\PGn\PGn}\xspace}

\newcommand{\hbbjet}{\ensuremath{\bbbar~\text{jet}}\xspace}
\newcommand{\wqqjet}{\ensuremath{\qqpr~\text{jet}}\xspace}
\newcommand{\wqqjets}{\ensuremath{\qqpr~\text{jets}}\xspace}

\newcommand{\wjets}{\ensuremath{\PW \!\! + \! \text{jets}}\xspace}
\newcommand{\zjets}{\ensuremath{\PZ/\gamma^* \!\! + \! \text{jets}}\xspace}
\newcommand{\vjets}{\ensuremath{\PV\! +\! \text{jets}}\xspace}
\newcommand{\zll}{\ensuremath{\PZ \to\Pell\Pell}\xspace}
\newcommand{\zgll}{\ensuremath{\PZ/\gamma^* \to\Pell\Pell}\xspace}

\newcommand{\tauTO}{\ensuremath{\tau_{2}/\tau_{1}}\xspace}
\newcommand{\ptom}{\ensuremath{\pt/m}\xspace}
\newcommand{\dlnuqq}{\ensuremath{D_{\Pell \PGn \qqpr}}\xspace}

\newcommand{\mtn}{\ensuremath{m_{\cPqt}}\xspace}
\newcommand{\mwn}{\ensuremath{m_{\PW}}\xspace}
\newcommand{\losttwn}{\ensuremath{\text{lost-}\cPqt/\PW}\xspace}
\newcommand{\Losttwn}{\ensuremath{\text{Lost-}\cPqt/\PW}\xspace}
\newcommand{\qgn}{\ensuremath{\Pq/\Pg}\xspace}

\newcommand{\bL}{\ensuremath{\cPqb\text{L}}\xspace}
\newcommand{\bT}{\ensuremath{\cPqb\text{T}}\xspace}

\newcommand{\mtbkg}{\ensuremath{\mtn~\text{background}}\xspace}
\newcommand{\mwbkg}{\ensuremath{\mwn~\text{background}}\xspace}
\newcommand{\losttwbkg}{\ensuremath{\losttwn~\text{background}}\xspace}
\newcommand{\qgbkg}{\ensuremath{\qgn~\text{background}}\xspace}

\newcommand{\pbkg}{\ensuremath{P_{\text{bkg}}}\xspace}
\newcommand{\pbb}{\ensuremath{P_{\bbbar}}\xspace}
\newcommand{\phh}{\ensuremath{P_{\PH\PH}}\xspace}
\newcommand{\psig}{\ensuremath{P_{\text{signal}}}\xspace}

\newcommand{\pnu}{\ensuremath{\vec{p}_\PGn}\xspace}
\newcommand{\vqq}{\ensuremath{V_{\qqpr}}\xspace}
\newcommand{\Rqq}{\ensuremath{R_{\qqpr}}\xspace}
\newcommand{\ptjet}{\ensuremath{p_{\mathrm{T}}^\text{jet}}\xspace}

\newcommand{\mlnuqq}{\ensuremath{m_{\Pell\PGn\qqpr}}\xspace}
\newcommand{\mlnu}{\ensuremath{m_{\Pell\PGn}}\xspace}
\newcommand{\mjet}{\ensuremath{m_{\text{jet}}}\xspace}

\newcommand{\vptreco} {\ensuremath{\vec{p}_{\mathrm{T}}^{\text{reco}}}\xspace}
\newcommand{\ptreco}{\ensuremath{p_{\text{T}}^{\text{reco}}}\xspace}

\newcommand{\llprob}{\ensuremath{P}\xspace}

\newcommand{\ptmissPar}{\ensuremath{p_{\mathrm{T}, \parallel}^{\text{miss}}}\xspace}
\newcommand{\ptmissPrp}{\ensuremath{p_{\mathrm{T}, \perp}^{\text{miss}}}\xspace}

\newcommand{\pinv}{\ensuremath{p_{\text{inv}}}\xspace}
\newcommand{\minv}{\ensuremath{m_{\text{inv}}}\xspace}

\newcommand{\mll}{\ensuremath{m_{\Pell\Pell}}\xspace}
\newcommand{\drll}{\ensuremath{\DR_{\Pell\Pell}}\xspace}
\newcommand{\dphimetll}{\ensuremath{\Delta\phi(\ptvecmiss,\vec{p}_{\Pell\Pell})}\xspace}

\newcommand{\zhbb}{\ensuremath{\PZ/\PH \to \bbbar}\xspace}
\newcommand{\dbb}{\ensuremath{D_{\zhbb}}\xspace}

\cmsNoteHeader{B2G-20-007}
\title{Search for heavy resonances decaying to a pair of Lorentz-boosted Higgs bosons in final states with leptons and a bottom quark pair at \texorpdfstring{$\sqrt{s} = 13\TeV$}{sqrt(s) = 13 TeV}}

\author*[cern]{Brent Stone, David Saltzberg}

\date{\today}

\abstract{
A search for new heavy resonances decaying to a pair of Higgs bosons (\hh) in proton-proton collisions at a center-of-mass energy of 13\TeV is presented. Data were collected with the CMS detector at the LHC in 2016--2018, corresponding to an integrated luminosity of 138\fbinv. Resonances with a mass between 0.8 and 4.5\TeV are considered using events in which one Higgs boson decays into a bottom quark pair and the other into final states with either one or two charged leptons. Specifically, the single-lepton decay channel $\hhbbww \to \bbbar \Pell \PGn \qqpr$ and the dilepton decay channels $\hhbbww \to \bbbar \Pell \PGn \Pell \PGn$ and $\hhbbtautau \to \bbbar \Pell \PGn \PGn \Pell \PGn \PGn$ are examined, where \Pell in the final state corresponds to an electron or muon. The signal is extracted using a two-dimensional maximum likelihood fit of the \hbb jet mass and \hh invariant mass distributions. No significant excess above the standard model expectation is observed in data. Model-independent exclusion limits are placed on the product of the cross section and branching fraction for narrow spin-0 and spin-2 massive bosons decaying to \hh. The results are also interpreted in the context of radion and bulk graviton production in models with a warped extra spatial dimension. The results provide the most stringent limits to date for $\PX\to\hh$ signatures with final-state leptons and at some masses provide the most sensitive limits of all $\PX\to\hh$ searches.
}

\hypersetup{pdfauthor={CMS Collaboration},pdftitle={Search for heavy resonances decaying to a pair of Lorentz-boosted Higgs bosons in final states with leptons and a bottom quark pair at sqrt(s) = 13 TeV},pdfsubject={CMS},pdfkeywords={CMS, BSM, heavy resonances, HH pairs}}

\maketitle 

\section{Introduction\label{sec:intro}}
The discovery of a Higgs boson (\PH) at the CERN LHC~\cite{HiggsDiscoveryAtlas,HiggsDiscoveryCMS,Chatrchyan:2013lba} validated the proposed mass generation mechanism within the standard model (SM)~\cite{Englert:1964et,Higgs}, the so-called ``Brout-Englert-Higgs mechanism.'' A number of theoretical difficulties found in the simple model are ameliorated by an extended Higgs sector~\cite{Branco:2011iw}. Supersymmetry~\cite{Ramond:1971gb,Golfand:1971iw,Neveu:1971rx,Volkov:1972jx,Wess:1973kz,Wess:1974tw,Fayet:1974pd,Nilles:1983ge} requires such an extended Higgs sector that includes additional spin-0 particles. Models with warped extra dimensions, proposed by Randall and Sundrum~\cite{Randall:1999ee}, postulate the existence of a compact fourth spatial dimension with a warped metric. Such compactification creates heavy resonances arising as a tower of Kaluza--Klein excitations, leading to possible spin-0 radions~\cite{Goldberger:1999uk,DeWolfe:1999cp,Csaki:1999mp,Csaki:2000zn} or spin-2 bulk gravitons~\cite{Davoudiasl:1999jd,Agashe:2007zd,Fitzpatrick:2007qr}. The ATLAS~\cite{Aaboud:2018ohp,Aaboud:2017gsl,Aaboud:2017rel,Aaboud:2017cxo,Aaboud:2018zhh,ATLAS:2019nat,ATLAS:2020jgy,ATLAS:2020fry,ATLAS:2020qiz,ATLAS:2020azv,ATLAS:2022hwc,ATLAS:2018sbw} and CMS~\cite{Sirunyan:2017nrt,Sirunyan:2018fuh,Sirunyan:2018hsl,Sirunyan:2018qca,Sirunyan:2017acf,Sirunyan:2017wto,Sirunyan:2018qob,Sirunyan:2017isc,Sirunyan:2016cao,Sirunyan:2017jtu,Khachatryan:2014xja,Khachatryan:2015ywa,CMS:2019kaf,CMS:2021fyk,CMS:2021klu,CMS:2021itu,CMS:2021xor} Collaborations have conducted a number of searches for these particles, where the new bosons decay into vector bosons and/or SM Higgs bosons ($\PW\PW$, $\PZ\PZ$, $\PW\PZ$, \hh, $\PZ\PH$, or $\PW\PH$).

In this paper, we present an expansion of a previous search~\cite{B2G-18-008} for heavy resonances (\PX) decaying to \hh. The previous study considered a smaller data set of proton-proton ($\Pp\Pp$) collisions and searched for a signal in which one Higgs boson decayed to a bottom quark pair ($\bbbar$) and the second decayed to a pair of \PW bosons, with one decaying leptonically and the other hadronically ($\PW\PW^* \to \Pell \Pgn \qqpr$). The data set analyzed in Ref.~\cite{B2G-18-008} corresponded to collisions at $\sqrt{s}=13\TeV$ recorded in 2016 with an integrated luminosity of 36\fbinv. In this new search, in addition to the $\hh \to \bbbar\Pell\PGn\qqpr$ decay channel from Ref.~\cite{B2G-18-008}, two other signal decay channels are included by considering dilepton decays of the Higgs boson that does not decay to $\bbbar$: the $\PH\to\wwlnulnu$ and the $\PH\to\tautaulnunulnunu$ decays. In all three cases, the \Pell denotes an electron or a muon; the analysis is also sensitive to leptonically decaying \PGt leptons in the $\bbbar \PW\PW^*$ decays. Events from $\bbbar\PGt\PGt$ comprise 30--35\% of the total expected dilepton signal yield. The analysis is optimized for the three $\PX\to\hh$ channels just mentioned, but signal events from $\hh\to\bbbar\PZ\PZ^*$ are also included in our acceptance and constitute 1--3\% of the total expected signal yield.

This search is performed on a data set of $\Pp\Pp$ collisions at a center-of-mass energy of 13\TeV, collected in 2016--2018 at the CERN LHC, corresponding to an integrated luminosity of 138\fbinv, and considers narrow resonances in the mass range $0.8<\mx<4.5\TeV$.
The Higgs bosons have a high Lorentz boost because of the large values of $\mx$ considered, so the decay products of each one are contained in a collimated cone. The degree of collimation is enough such that the hadronically decaying bosons (\PH and \PW) are each reconstructed as a single jet that has substructure consistent with a decay to two energetic quarks. The distinguishing characteristic of the signal is a peak in the two-dimensional (2D) plane of the \hbb jet mass \mbb and the reconstructed \hh invariant mass \mhh.

In the single-lepton (SL) channel, the quarks in the $\PH\to\wwlnuqq$ decay are reconstructed as a single large jet (the \wqqjet) with a nearby lepton (\Pe or \PGm). This jet is required to have substructure consistent with a decay to two energetic quarks. This Higgs boson decay chain is reconstructed as the \wqqjet, the lepton, and the missing transverse momentum \ptvecmiss. In the dilepton (DL) channel, two leptons are reconstructed in close proximity to each other, with \ptvecmiss nearby, consistent with the expected neutrinos. In all channels considered, the \hbb decay is reconstructed as a single large jet (the \hbbjet) with substructure and high transverse momentum \pt.

The main SM background in this search arises from top quark pair~(\ttbar) production. This analysis is most sensitive to top quarks that have collimated decay products because of large Lorentz boosts. In the SL channel, the largest background comes from \ttbar decays in which one top quark decays with a charged lepton and a neutrino ($\cPqt\to \PW\cPqb \to \Pell\Pgn\cPqb$), and the other decays exclusively to quarks ($\cPqt \to \PW\cPqb \to \qqpr \cPqb$), which can be mistakenly reconstructed as the \hbbjet candidate. Other significant backgrounds in this channel are the production of \PW bosons in association with jets with $\PW\to \Pell \PGn$ (hereafter referred to as \wjets), and multijet events from quantum chromodynamic processes (QCD multijets), with either a lepton originating from heavy flavor decay or a hadron misidentified as a lepton. In the DL channel, the background yield is smaller than that of the SL channel by a factor of ${\approx}60$. Top quark pair production is the dominant background here too, with approximately equal contributions from \ttbar events with a single lepton in the final state and events in which both top quarks decay leptonically. Single-lepton \ttbar events can fall into the DL channel when some part of the hadronic top quark decay is misidentified as a lepton. The other significant background in this channel is production of $\PZ/\PGg^*$ bosons in association with jets (\zjets), with $\zgll$. These backgrounds are distinguished in data using the \mbb spectrum. Contributions from backgrounds with an SM Higgs boson (\eg, $\ttbar\PH$) are considered but found to be negligible in both channels.

The CMS detector and the simulated samples used to build the analysis are described in Sections~\ref{sec:cmsDetector} and~\ref{sec:samples}, respectively.
Relative to Ref.~\cite{B2G-18-008}, this analysis incorporates new DL signal modes and employs new particle reconstruction and identification techniques in the SL channel. These include more efficient algorithms for identifying electrons and jets with \PQb hadrons (\PQb tagging) as well as an improved reconstruction procedure for the $\hww\to\Pell\PGn\qqpr$ decay. The developments are discussed in Section~\ref{sec:reco}, which details the event reconstruction and identification, including the final-state particles and the intermediate-state bosons.
Section~\ref{sec:selection} discusses the selection criteria used to discriminate signal from background and the division of all events into 12 exclusive categories by the number of leptons, the lepton flavor, the quality of jet flavor tagging, and the \hww decay kinematics.
Section~\ref{sec:modeling} details the model-building process for the signal and the background.
The signal and SM background yields are estimated using a simultaneous maximum likelihood fit to the 2D \mbb and \mhh mass distributions in all 12 categories. All systematic uncertainties are discussed in Section~\ref{sec:syst}, and the post-fit results are presented in Section~\ref{sec:results}.
The analysis is summarized in Section~\ref{sec:summary}.

Tabulated results are provided in the HEPData record for this analysis~\cite{hepdata}.

\section{The CMS detector and global event reconstruction} \label{sec:cmsDetector}
The central feature of the CMS apparatus is a superconducting solenoid of 6\unit{m} internal diameter, providing a magnetic field of 3.8\unit{T}. Within the solenoid volume are a silicon pixel and strip tracker, a lead tungstate crystal electromagnetic calorimeter (ECAL), and a brass and scintillator hadron calorimeter (HCAL), each composed of a barrel and two endcap sections. Forward calorimeters extend the pseudorapidity ($\eta$) coverage provided by the barrel and endcap detectors. Muons are measured in gaseous detectors embedded in the steel flux-return yoke outside the solenoid. A more detailed description of the CMS detector, together with a definition of the coordinate system used and the relevant kinematic variables, can be found in Ref.~\cite{Chatrchyan:2008zzk}.

Events of interest are selected using a two-tiered trigger system. The first level, composed of custom hardware processors, uses information from the calorimeters and muon detectors to select events at a rate of around 100\unit{kHz} within a fixed latency of about 4\mus~\cite{Sirunyan:2020zal}. The second level, known as the high-level trigger, consists of a farm of processors running a version of the full event reconstruction software optimized for fast processing and reduces the event rate to around 1\unit{kHz} before data storage~\cite{Khachatryan:2016bia}.

Event reconstruction relies on a particle-flow (PF) algorithm~\cite{CMS-PRF-14-001}, which aims to identify each individual particle in an event with an optimized combination of information from the various elements of the CMS detector. The vector \ptvecmiss is computed as the negative vector \pt sum of all the PF candidates in an event, and its magnitude is denoted as \ptmiss~\cite{Sirunyan:2019kia}. The \ptvecmiss is modified to account for corrections to the energy scale of the reconstructed jets in the event. In each event, jets are clustered from these PF candidates using the anti-\kt algorithm~\cite{Cacciari:2008gp,Cacciari:2011ma} with a distance parameter of 0.4 (AK4 jets) and of 0.8 (AK8 jets). 

The jet momentum is determined as the vectorial sum of all particle momenta in the jet, and is found from simulation to be, on average, within 5--10\% of the true momentum over the entire \pt spectrum and detector acceptance. Additional $\Pp\Pp$ interactions within the same or nearby bunch crossings (pileup) can contribute extra tracks and calorimetric energy depositions, increasing the apparent jet momentum. To mitigate this effect for AK4 jets, tracks identified as originating from pileup vertices are discarded, and an offset correction is applied to correct for residual contributions~\cite{CMS-PRF-14-001,Cacciari:2011ma}. For AK8 jets, a different pileup per particle identification algorithm~\cite{Sirunyan:2020foa,Bertolini:2014bba} reduces the effect of pileup by considering local shape variables~\cite{Bertolini:2014bba} to rescale the momentum of each jet constituent according to its probability to originate from the primary vertex. Jet energy corrections are derived from simulation studies so that the average measured energy of jets becomes identical to that of particle-level jets. In situ measurements of the momentum balance in dijet, photon+jet, $\PZ$+jet, and multijet events are used to determine any residual differences between the jet energy scale in data and in simulation, and appropriate corrections are made~\cite{Khachatryan:2016kdb}.

\section{Simulated samples\label{sec:samples}}
Signal and background yields are extracted from a fit to the data in the 2D \mbb and \mhh mass distribution using templates obtained from samples generated by Monte Carlo (MC) simulation.

The signal processes $\Pp\Pp\to\PX\to\hh\to\bbbar\PV\PV^{*}$ (where $\PV = \PW$ or \PZ) and $\Pp\Pp\to\PX\to\hh\to\bbbar\Pgt\Pgt$ are simulated for spin-0 radions and spin-2 gravitons in the bulk scenario of Randall--Sundrum models with warped extra dimensions. Only the $\bbbar\PW\PW^{*}$ and $\bbbar\Pgt\Pgt$ events are used to optimize the analysis, but any $\bbbar\PZ\PZ^{*}$ events that pass the full selection are included in the signal acceptance. The simulated \PX bosons are produced via gluon fusion and with a narrow width (1\MeV) that is small compared to the experimental resolution of roughly 5\%. The branching fractions used to normalize the signal correspond to those expected for SM Higgs boson decays. The signal is generated at leading order (LO) using the \MGvATNLO V5 2.4.2 generator~\cite{Alwall:2014hca} with the MLM merging scheme~\cite{Alwall:2007fs} for \mx of 0.8--4.5\TeV.

The \MGvATNLO generator is also used to produce the \wjets, \zgll, and QCD multijet background samples at LO. The \wjets and \zgll samples are normalized using next-to-next-to-LO (NNLO) cross sections, calculated with \FEWZ v3.1~\cite{Li:2012wna}. Samples of $\PW\PZ$ diboson production and of the associated production of \ttbar with either a \PW or \PZ boson are also generated with \MGvATNLO but at next-to-LO (NLO) with the FxFx jet merging scheme~\cite{Frederix2012}. The \POWHEG v2 generator is used to produce samples for \ttbar, $\PW\PW$, $\PZ\PZ$, $\ttbar\PH$, and single top quark production at NLO~\cite{Nason:2004rx,Frixione:2007vw,Alioli:2010xd,Re:2010bp,Melia:2011tj,Nason:2013ydw,Frederix:2012dh,Hartanto:2015uka}. Furthermore, the \ttbar process is normalized to the NNLO cross section, computed with \textsc{Top++} v2.0~\cite{Czakon:2011xx}.

Parton showering and hadronization are simulated in the 2016 samples with \PYTHIA v8.226~\cite{Sjostrand:2014zea} using the CUETP8M1~\cite{Khachatryan:2015pea} tune, except for the \ttbar, $\ttbar\PH$, and $\PX\to\PH\PH\to\bbbar \PV\PV^{*}$ signal samples, which are simulated using the CP5 tune. For 2017--2018, \PYTHIA v8.230 and the CP5 tune~\cite{Sirunyan:2020pea} are used to produce the samples. The parton distribution functions (PDFs) used to produce the samples are the NNPDF~3.0~\cite{Ball:2015mu} set for the 2016 data set and the NNPDF~3.1~\cite{Ball:2017mu} set for the 2017--2018 data sets.
The simulation of the CMS detector is performed with the \GEANTfour~\cite{Agostinelli:2002hh} toolkit. The simulated samples are weighted to have the same multiplicity distribution of pileup interactions as observed in data.

\section{Decay chain reconstruction\label{sec:reco}}

All signal events, regardless of lepton multiplicity, feature a high-\pt jet that has substructure consistent with two \PQb quark decays. This jet is generally opposite in the transverse plane to a collection of other particles from a boosted Higgs boson decay. In the SL channel, signal events feature a lepton originating from a boosted \PW boson decay and a nearby jet that has substructure consistent with a \wqq decay. Even at the lowest considered \mx of 0.8\TeV, the median angular distance $\DR = \sqrt{\smash[b]{(\Delta\eta)^2+(\Delta\phi)^2}}$ (where $\phi$ is the azimuthal angle) between the \wqq decay and the lepton is approximately $\DR=0.5$.
In the DL channel, there are two high-\pt leptons originating from the decay of either a boosted \PW boson pair or a boosted \PGt lepton pair, but there is no jet in the vicinity of the leptons, resulting in a cleaner experimental signature.

Events are first selected by the trigger system with small year-to-year differences in the criteria. Events are triggered if they contain one of the following: an isolated muon with $\pt > 24\GeV$ (27\GeV in 2017), an isolated electron with $\pt > 32\GeV$ (27\GeV in 2016), or $\HT>1050\GeV$ (900\GeV in 2016), where \HT is the scalar sum of jet \pt for all trigger-level AK4 jets with $\pt>30\GeV$.
An inclusive-OR combination of lepton and \HT triggers is used because the high-\mx SL signal does not have leptons that are sufficiently isolated to pass the online lepton isolation selection, as the decay products \wwlnuqq are highly collimated. Additional multiobject triggers that select events with at least one lepton and considerable jet energy supplement these triggers, helping to maintain high trigger efficiency for signal over the entire range of \mx. In particular, these multiobject triggers fire for events with $\HT>450\GeV$ (400\GeV in 2016) and a lepton that has $\pt > 15\GeV$ and looser isolation requirements than for the previously mentioned isolated single-lepton triggers. These multiobject triggers are particularly helpful for the SL signal topology with the lepton close to the jet.
The trigger efficiency is measured for $\Pe\Pgm$ \ttbar events in data for events passing offline selection criteria for \HT in the SL channel and both \HT and lepton \pt in the DL channel. We use \ttbar events because the lepton and jet multiplicities resemble those in signal events. Simulation is corrected such that the trigger efficiency matches that in the data.
In the SL channel, the trigger efficiency for signal events is over 96\% at $\mx=0.8\TeV$ and increases to $>$99\% above $\mx=1.0\TeV$. In the DL channel, the trigger efficiency is $>$99\% over the full range of \mx.

\subsection{Electron and muon identification \label{sec:leptons}}

Different selection criteria are required for the SL and DL channels to identify signal-like leptons because of the different decay topologies. First, however, an event in either channel must contain either a muon with $\pt > 27\GeV$ or an electron with $\pt > 30\GeV$. In the DL channel, the other lepton must have $\pt > 10\GeV$. All muons are required to have $\abs{\eta} < 2.4$. Electrons in the DL channel are required to have $\abs{\eta}<2.5$, but those in the SL channel are restricted to the ECAL barrel region ($\abs{\eta}<1.479$) to suppress a significant contribution from the QCD multijet background with a small loss in signal acceptance.
Leptons must satisfy reconstruction quality and identification requirements that are optimized to maintain high efficiency and low probability for misidentifying hadrons as leptons~\cite{CMS:2020uim,Sirunyan:2018}. Additionally, the impact parameters of lepton tracks with respect to the primary vertex are required to be consistent with those originating from this vertex. Looser constraints on the impact parameter are used in the DL channel because some of the leptons originate from $\PH\to\Pgt\Pgt$ decays and thus have significant displacements from the primary vertex.
Leptons are required to be isolated with an isolation cone size designed for leptons from boosted decays, in which the cone size becomes smaller with larger \pt~\cite{Rehermann:2010vq}. Because less hadronic energy is expected near the leptons in the DL channel than in the SL channel, the allowed extra transverse energy in the isolation cone is smaller.

In the SL channel, as measured with signal simulation, the electron selection efficiency has a maximum of 70\% at $\mx = 0.8\TeV$ and then degrades to 7.5\% at $\mx = 4.5\TeV$. This is caused by a selection imposed at a low-level reconstruction step on the ratio of the energy deposited in the HCAL to that deposited in the ECAL. Electrons in the $\hww \to \Pe\Pgn\qqpr$ decay often fail this selection because of nearby energy deposits from the \wqqjet, which grow with larger boosts. However, the reconstruction of muons does not rely on such HCAL measurements and so is much more efficient than for electrons, but the isolation of muons is still sensitive to the \wqqjet. As a result, the overall selection efficiency for signal muons is better than for electrons but still degrades for larger \mx; the muon efficiency ranges from approximately 90\% at $\mx = 0.8\TeV$ down to 60\% at $\mx = 4.5\TeV$.

In the DL channel, where there is no \wqqjet, the lepton selection efficiency is larger for all \mx than in the SL channel. Because of the increased boost of the system, the efficiency still drops toward high \mx. For electrons, the reconstruction efficiency is much larger than in the SL channel, ranging from approximately 82\% at $\mx = 0.8\TeV$ down to 71\% at $\mx = 4.5\TeV$. The muon efficiency is also larger, ranging from approximately 96\% at $\mx = 0.8\TeV$ down to 91\% at $\mx = 4.5\TeV$. The lepton efficiencies are also measured in simulation and data in a \zll sample, and the simulation is corrected to match the efficiency in data. The systematic uncertainties in these measurements are applied to the normalization of the signal.

\subsection{Reconstruction and flavor identification of jets \label{sec:jetreco}}

Because of the boost imparted to the Higgs bosons by the decay of the much more massive \PX boson, the \hbb and \wqq decays are each reconstructed as a single, merged AK8 jet with two-prong substructure. In order to prevent the \wqqjet from containing the lepton's momentum in the SL channel, the PF candidates associated with the lepton are not included in the clustering of the set of jets from which the \wqqjet is selected. Only the PF candidates associated with a lepton that fulfills the analysis requirements are removed, and the same jet energy corrections described in Section~\ref{sec:cmsDetector} for AK8 jets are applied to these lepton-subtracted AK8 jets. We ensure the validity of applying these corrections to the lepton-subtracted AK8 jets by comparing the jet energy response in simulation between jets that require lepton subtraction and jets that do not. Jets of both types are required to have $\abs{\eta} < 2.4$ so that most of the jet particles are within the acceptance of the tracker.

The Higgs bosons have collimated decays and typically are produced back-to-back in the transverse plane, \ie, $\Delta \phi (\PH,\PH) \approx \pi$. The \hbbjet candidate is required to have $\pt>200\GeV$. In the SL channel, it is required to have a $\Delta \phi > 2.0$ separation from the lepton and a $\DR > 1.6$ separation from the \wqqjet, while in the DL channel, it is required to have a $\Delta \phi > 2.0$ separation from the dilepton momentum and to not contain either lepton within the jet cone. The \wqqjet in the SL channel is chosen as the closest AK8 jet in $\DR$ to the lepton, provided that it is found within $\DR < 1.2$ of the lepton and has $\pt > 50\GeV$. Within both the $\bbbar$~and \wqqjets, two subjets are reconstructed that must each have $\pt > 20\GeV$. Constituents of the AK8 jets are first reclustered using the Cambridge--Aachen algorithm~\cite{Dokshitzer:1997in,Wobisch:1998wt}. The ``modified mass drop tagger'' algorithm~\cite{Dasgupta:2013ihk,Butterworth:2008iy}, also known as the ``soft drop'' (SD) algorithm, with angular exponent $\beta = 0$, soft cutoff threshold $z_{\mathrm{cut}} < 0.1$, and characteristic radius $R_{0} = 0.8$~\cite{Larkoski:2014wba}, is applied to remove soft, wide-angle radiation from the jet. The subjets used are those remaining after the algorithm has removed all recognized soft radiation. The jets in this analysis are required to have exactly two subjets. The SD jet mass is the invariant mass of these two subjets. The SD jet mass of the \hbbjet is used to obtain the search variable \mbb, after applying \pt-dependent corrections, so that \mbb in simulation is on average equal to the Higgs boson mass of 125\GeV.

Identifying the \hbb decay in signal events and discriminating against background events relies on tagging jets as likely to have originated from \PQb~hadron decays. The AK8 jets are identified as consistent with a $\bbbar$ decay using the \textsc{DeepAK8} mass-decorrelated $\PZ/\PH \to \bbbar$ tagger~\cite{Sirunyan:2018jin}, with a discriminator denoted as \dbb, at a working point that has an efficiency of ${\approx} 85\%$ for selecting $\bbbar$ jets and a misidentification probability of $<$1\% for pure light-flavor quark and gluon jets. This is a deep neural network based tagger, designed to discriminate high-\pt jets consistent with a $\bbbar$~substructure against light-flavor quark (\PQu, \PQd, \PQs) or gluon jets. Furthermore, by design the tagger does not sculpt the SD jet mass distributions, thereby enabling the use of the SD jet mass in the background estimation. The \PQb tagging efficiencies are measured in data, and the simulation is corrected for any discrepancies. The uncertainty in this $\bbbar$ tagging efficiency is the dominant systematic uncertainty in the analysis, denoted as ``\hbbjet tagging" in Table~\ref{tab:sig_uncertainties} and discussed later in Section~\ref{sec:signalUncs}.

In \ttbar events, the most common events misreconstructed as signal, the \hbbjet candidate is typically reconstructed around the decay of one of the \PQb quarks, while the other \PQb quark decays into the opposite direction in the transverse plane. Identifying a \PQb-tagged AK4 jet that is separated from the \hbbjet is an effective method of discriminating between such \ttbar events and signal events, in which the \hbbjet is reconstructed from the two \hbb quarks. To be considered a candidate \PQb jet, an AK4 jet must have $\pt>30\GeV$ and be identified using the \textsc{DeepJet} tagger~\cite{BTV-16-002,Bols:2020bkb,CMS-DP-2018-058} at a working point that has an efficiency of ${\approx} 80\%$ for selecting \PQb jets and a misidentification probability of ${\approx} 1\%$ for light-flavor quark and gluon jets.

\subsection{Reconstructing the \texorpdfstring{$\hh$}{HH} system mass}

Depending on whether the final state has one or two leptons, different strategies are employed to reconstruct the four-momentum of the Higgs boson that does not decay to $\bbbar$. The mass \mhh is then the invariant mass of this four-momentum and the \hbbjet four-momentum. The mass of the \hbbjet used in this calculation is not the SD jet mass \mbb but is rather the ungroomed jet mass. In Sections~\ref{sec:hwwReco_1l} and \ref{sec:hReco_2l}, respectively, the reconstruction strategies are described for the SL and DL channels.

\subsubsection{Single-lepton channel} \label{sec:hwwReco_1l}

To reconstruct the Higgs boson four-momentum in the $\PH\to\wwlnuqq$ decay chain from the visible and invisible decay products, a likelihood-based technique that takes the reconstructed lepton, the \ptvecmiss, and the \wqqjet as input is employed. For each event, values for the following five parameters are extracted by maximizing a likelihood function:
\begin{itemize}
\item \pnu: the three components of the neutrino momentum.
\item \Rqq: the jet response correction, a multiplicative scale factor applied to the \pt of the \wqqjet. The jet \pt is allowed to vary because the uncertainty associated with the estimated \pt of this jet is large.
\item \vqq: a boolean indicator of whether the \wqqjet favors a larger or smaller mass than the leptonic \PW boson decay. This is largely a bookkeeping device for the \PW and $\PW^*$ hypotheses.
\end{itemize}
With these parameters, the \hww four-momentum can be fully determined. This four-momentum is then the sum of the neutrino four-momentum $p_{\PGn}$, the \wqqjet four-momentum (with \pt modified by \Rqq), and the four-momentum of the lepton.

The likelihood function is constructed with six probability density functions (pdfs) $P(x|\vec{y})$ estimated from signal simulation, where $x$ is the corresponding observable in the pdf. The symbol $\vec{y}$ represents the set of free parameters associated with that pdf, such as \vqq. These pdfs are represented as one-dimensional (1D) histograms. The full likelihood function is:
\begin{linenomath}
\begin{equation} \label{eq:likeli}
\lumi = \llprob(\mjet|\vqq)\llprob(\ptjet|\Rqq,\vqq)\llprob(\mlnuqq|\pnu,\Rqq,\vqq)\llprob(\mlnu|\pnu,\vqq)\llprob(\ptvecmiss|\pnu,\vqq).
\end{equation}
\end{linenomath}
The observable \mjet is the SD jet mass of the \wqqjet, and its corresponding pdf is coarsely binned to remain insensitive to the precise modeling of the SD algorithm.  
The observable \ptjet is the unmodified \wqqjet \pt, and the pdf $\llprob(\ptjet|\Rqq,\vqq)$ is the jet \pt response.
Two other observables, \mlnu and \mlnuqq, are masses of the lepton-neutrino pair and the lepton-neutrino-\wqqjet system.

The last factor in Eq.~(\ref{eq:likeli}) represents the product of two pdfs, each corresponding to a single component of \ptvecmiss:
\begin{linenomath}
\begin{equation}
\llprob(\ptvecmiss|\pnu,\vqq) = \llprob(\ptmissPar|\pnu,\vqq) \llprob(\ptmissPrp|\pnu,\vqq).
\end{equation}
\end{linenomath}
The two observables \ptmissPar and \ptmissPrp are defined with respect to the reference frame of the \hww decay, along the direction of \vptreco:
\begin{linenomath}
\begin{equation}
\vptreco = \ptvecmiss + (\vec{p}_{\Pell} + \vec{p}_{\wqqjet})_{\text{T}}.
\end{equation}
\end{linenomath}
The two \ptvecmiss pdf factors are parameterized as the components of the extra \ptmiss (relative to the neutrino momentum) that are parallel and perpendicular to this vector \vptreco. The extra \ptmiss along this direction arises primarily from mismeasurement of the \hbbjet, while the orthogonal component arises mostly from pileup and the underlying event.

The pdfs \llprob of the observables are generally independent of \mx, but there is still some residual dependence. We account for this by producing two sets of pdfs, one at low \ptreco ($<$600\GeV) and one at high \ptreco ($>$1400\GeV). Then, event-by-event, the histogram of the pdf is obtained by interpolating between the two histograms at the two regimes of \ptreco. This interpolation is performed linearly as a function of the event \ptreco.
The \llprob are all dependent on whether the hadronically decaying \PW boson is heavier than the leptonically decaying \PW, so each factor is dependent on the free parameter \vqq. Correlations among the observables in the likelihood were studied and found not to affect the sensitivity significantly.

This method gives an \mhh resolution for signal events that is very similar to that from a direct calculation using the Higgs boson mass as a constraint (as in~\cite{B2G-18-008}), but for background events it typically returns lower values of \mhh than in the direct calculation. We take advantage of this fact using an alternative likelihood $\lumi_{\text{alt}}$, which is less constrained by the intermediate masses. Instead of fitting for the neutrino $p_z$, the masses \mlnu and \mlnuqq are included as free parameters. Both likelihoods are used to construct a discriminating variable between signal and background:
\begin{linenomath}
\begin{equation} \label{eq:dlnuqq}
\dlnuqq = -2 \log{ \lumi / \lumi_{\text{alt}} },
\end{equation}
\end{linenomath}
where \lumi is the likelihood described in Eq.~(\ref{eq:likeli}). We discuss how \dlnuqq is used in Section~\ref{sec:selection}.

\subsubsection{Dilepton channel} \label{sec:hReco_2l}

Because of the absence of a \wqqjet, the presence of larger \ptmiss, and much smaller backgrounds, there is no need for a likelihood-based technique to separate signal and background in the DL channel. Instead, we make simple assumptions regarding the decay kinematic distributions in order to reconstruct the full invisible four-momentum \pinv due to neutrinos. First, the transverse components of \pinv are taken directly from the \ptvecmiss. Second, because the decay products of the boosted Higgs boson are collimated, we assume the polar angle $\theta$ of \pinv is equal to that of the dilepton momentum: $\theta_{\text{inv}} = \theta_{\Pell\Pell}$. With this constraint, the $z$-component of \pinv is obtained. Lastly, the invisible invariant mass \minv due to neutrinos is assumed to be 55\GeV, the mean of the distribution from signal simulation. The corresponding Higgs boson four-momentum is the summed four-momentum of \pinv and the dilepton four-momentum $p_{\Pell\Pell}$.

\section{Event selection and categorization\label{sec:selection}}
Events are selected in this search if they pass the following criteria indicating that they could include the production and decay of an \PX boson. They are then divided into 12 distinct categories (eight SL and four DL). A separate set of criteria is applied to define control regions that are used to validate the modeling of background processes.

Offline, all events are required to have $\HT>400\GeV$, either one electron with $\pt>30\GeV$ or one muon with $\pt>27\GeV$, and a selected \hbbjet. Background from \ttbar production is reduced by vetoing all events with an AK4 jet that is $\DR > 1.2$ from the \hbbjet and is identified as a \PQb jet, as described in Section~\ref{sec:jetreco}.

We ensure that the sets of events belonging to the SL and DL channels are disjoint. To accomplish this, we first impose that any event with exactly two oppositely charged lepton candidates passing the DL channel lepton selection be assigned as a DL event. Otherwise, if the event has at least one lepton candidate passing the SL channel lepton selection and also has fewer than two lepton candidates passing the DL channel lepton selection, it is classified as an SL event. In this case, the highest-\pt lepton candidate that passes the SL channel lepton selection is selected for Higgs boson reconstruction. If these two criteria cannot be fulfilled by the set of lepton candidates, the event is not used in the analysis.

The following sections review the event selection and categorization of events into the 12 exclusive search regions. Selections that are used only to discriminate signal from background and not to categorize events are detailed in Sections~\ref{sec:cuts_1l} and~\ref{sec:cuts_2l} for the SL and DL channels, respectively. Section~\ref{sec:categorization} discusses the discriminating selections that are also used to categorize events.

\subsection{Single-lepton channel event selection} \label{sec:cuts_1l}

\begin{figure}[htbp!]
\centering
\includegraphics[width=0.45\textwidth]{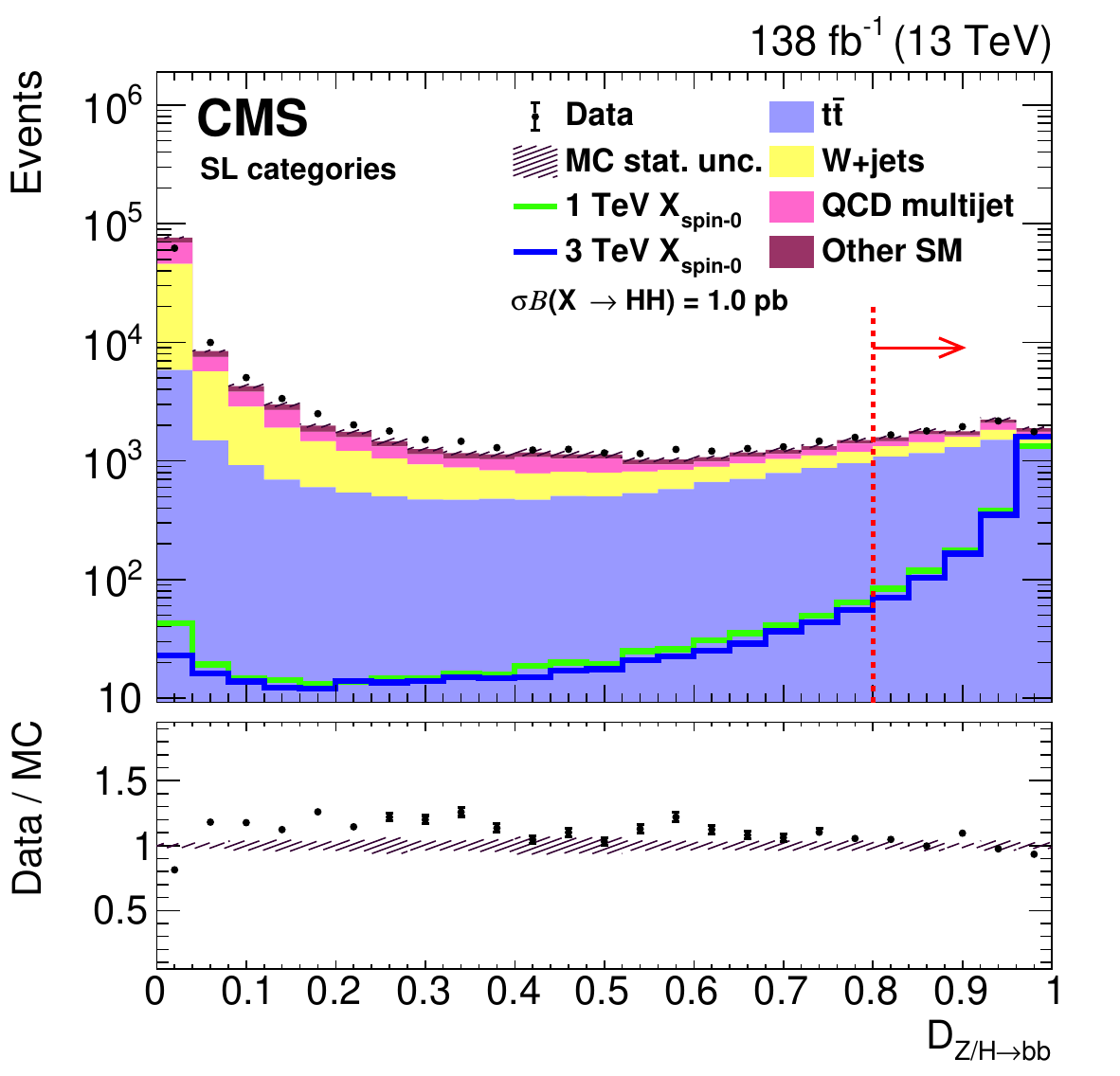}
\includegraphics[width=0.45\textwidth]{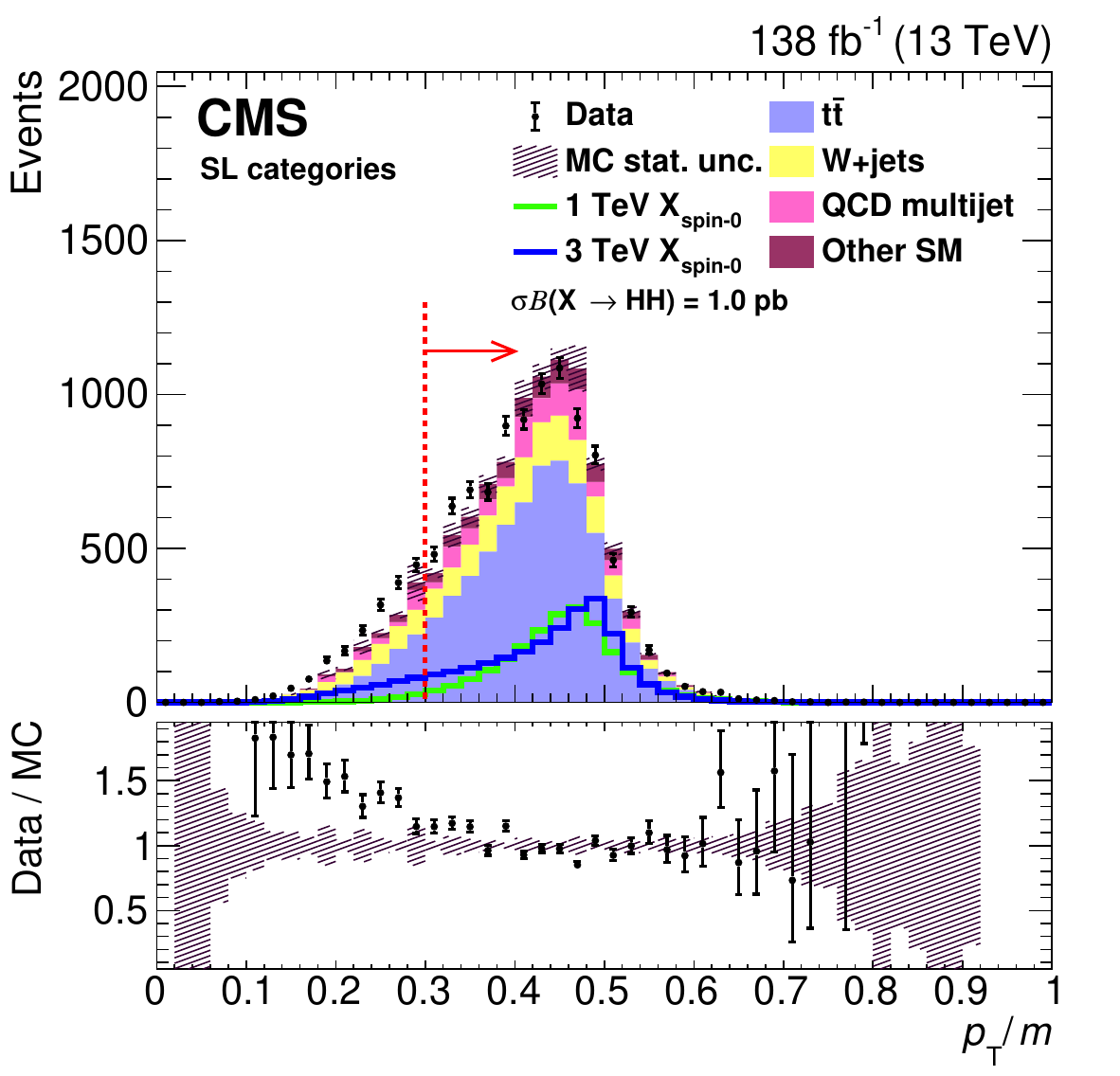}
\includegraphics[width=0.45\textwidth]{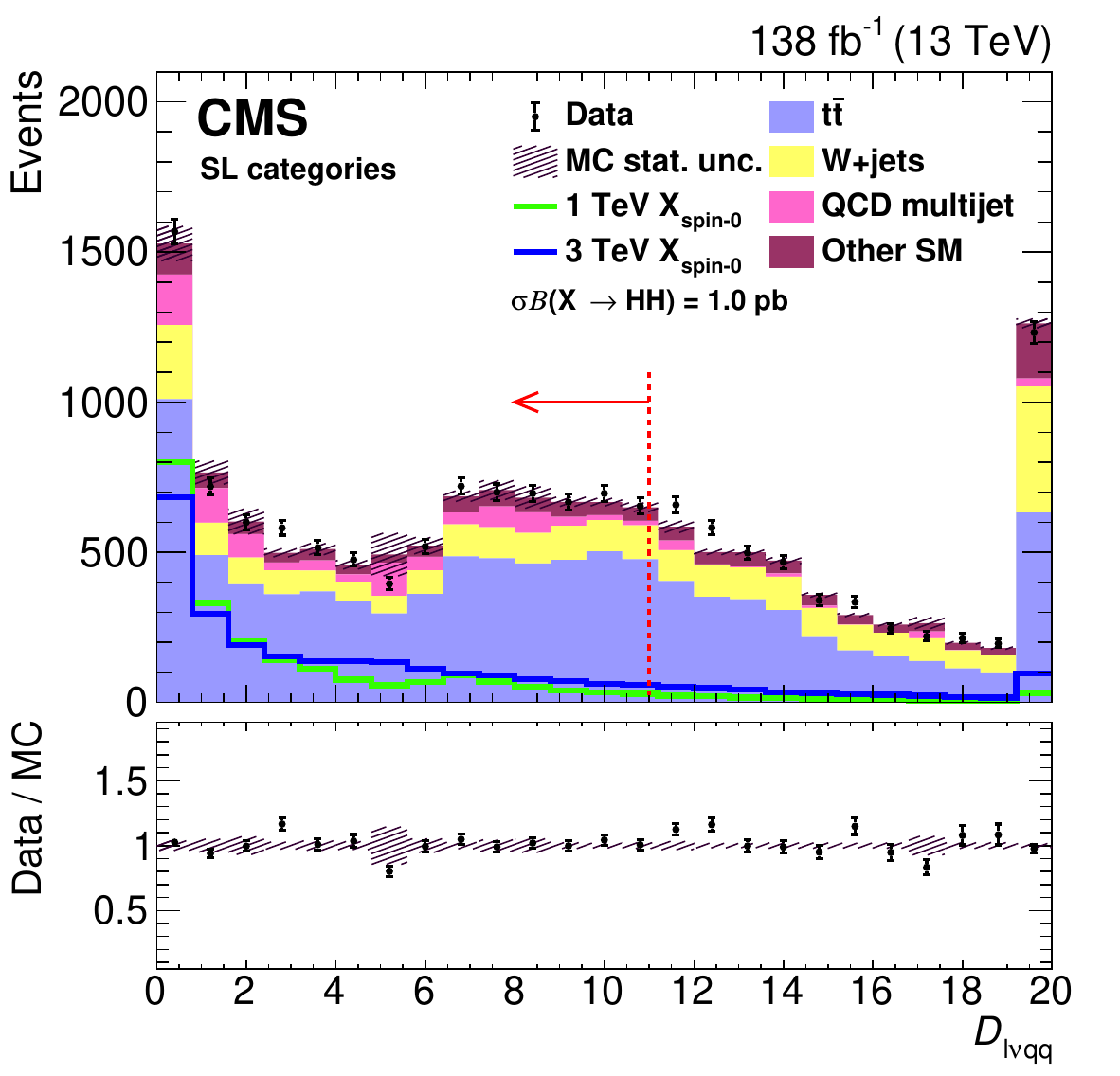}
\includegraphics[width=0.45\textwidth]{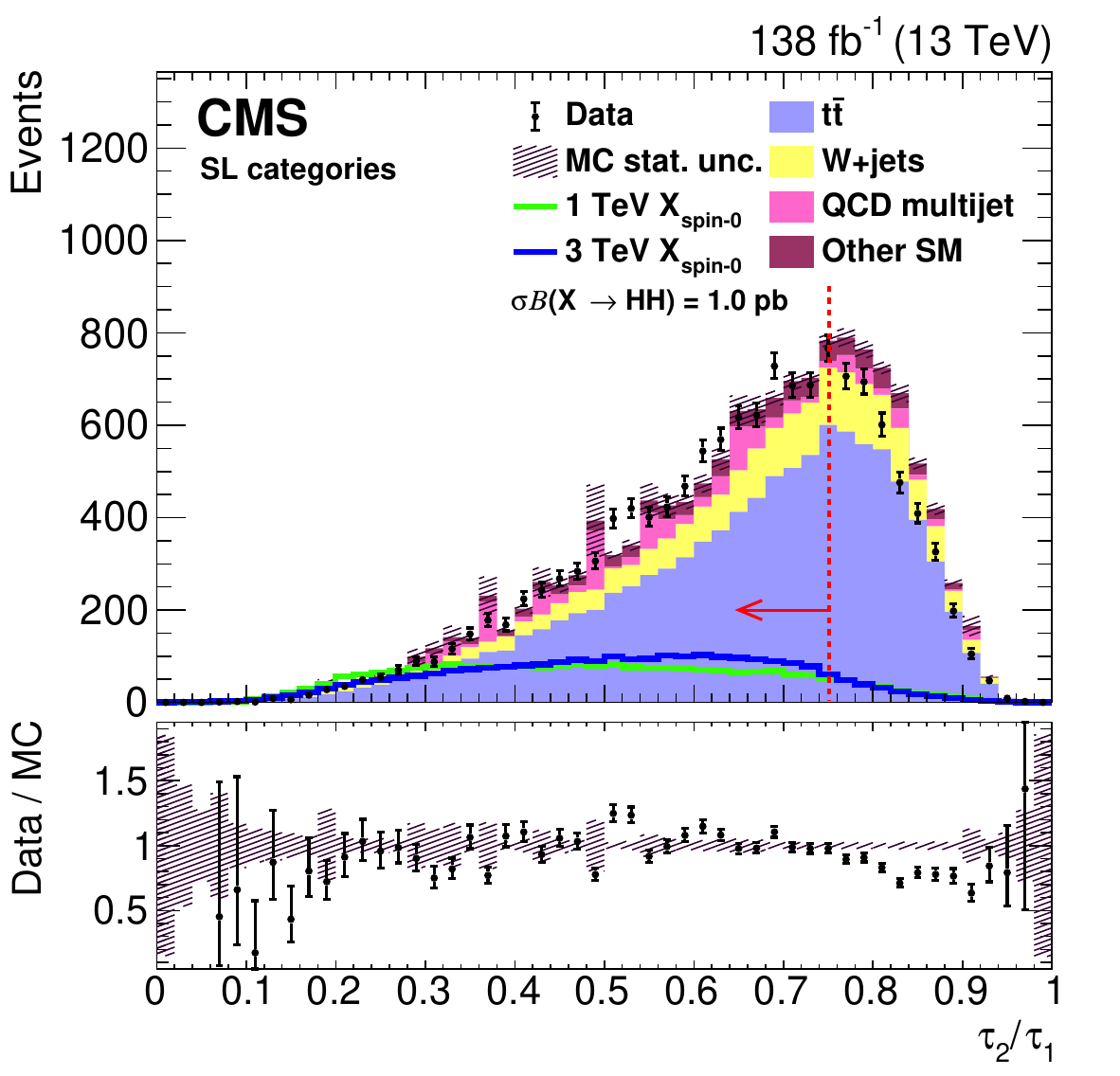}
\caption{Single-lepton channel observables: distributions are shown for data (points), pre-fit simulated SM processes (filled histograms), and simulated signal (solid lines). The statistical uncertainty in the simulated sample is shown as the hatched band. Spin-0 signals for \mx of 1.0 and 3.0\TeV are displayed. The rightmost bin in the \dlnuqq plot contains the overflow events. For both signal models, $\sigma \mathcal{B} (\PX\to\hh)$ is set to 1.0\unit{pb}. The lower panels of each plot show the ratio of the data to the sum of all background processes. The red dashed line and arrow indicate the selected region of the variable of interest.}
\label{fig:event_vars_1l}
\end{figure}

In the SL channel, the \wqqjet is chosen as the closest AK8 jet to the lepton, and it is required to have $\pt > 50\GeV$ and be located within $\DR < 1.2$ of the lepton, where the former requirement is optimized for signal acceptance and the latter for background rejection. Jets in background events tend to be produced at higher $\abs{\eta}$ than those produced in signal events, which contain jets from the decay of a heavy particle. To exploit this property, the ratio of the \pt of \hww divided by \mhh, denoted as \ptom, is required to be $>$0.3. The distribution of \ptom is shown in Fig.~\ref{fig:event_vars_1l} (upper right) for the data, expected pre-fit background, and two signal mass hypotheses with a normalization corresponding to a product of the cross section and branching fraction ($\sigma \mathcal{B}$) of 1.0\unit{pb}.

\subsection{Dilepton channel event selection} \label{sec:cuts_2l}
 
\begin{figure}[htp]
\centering
\includegraphics[width=0.45\textwidth]{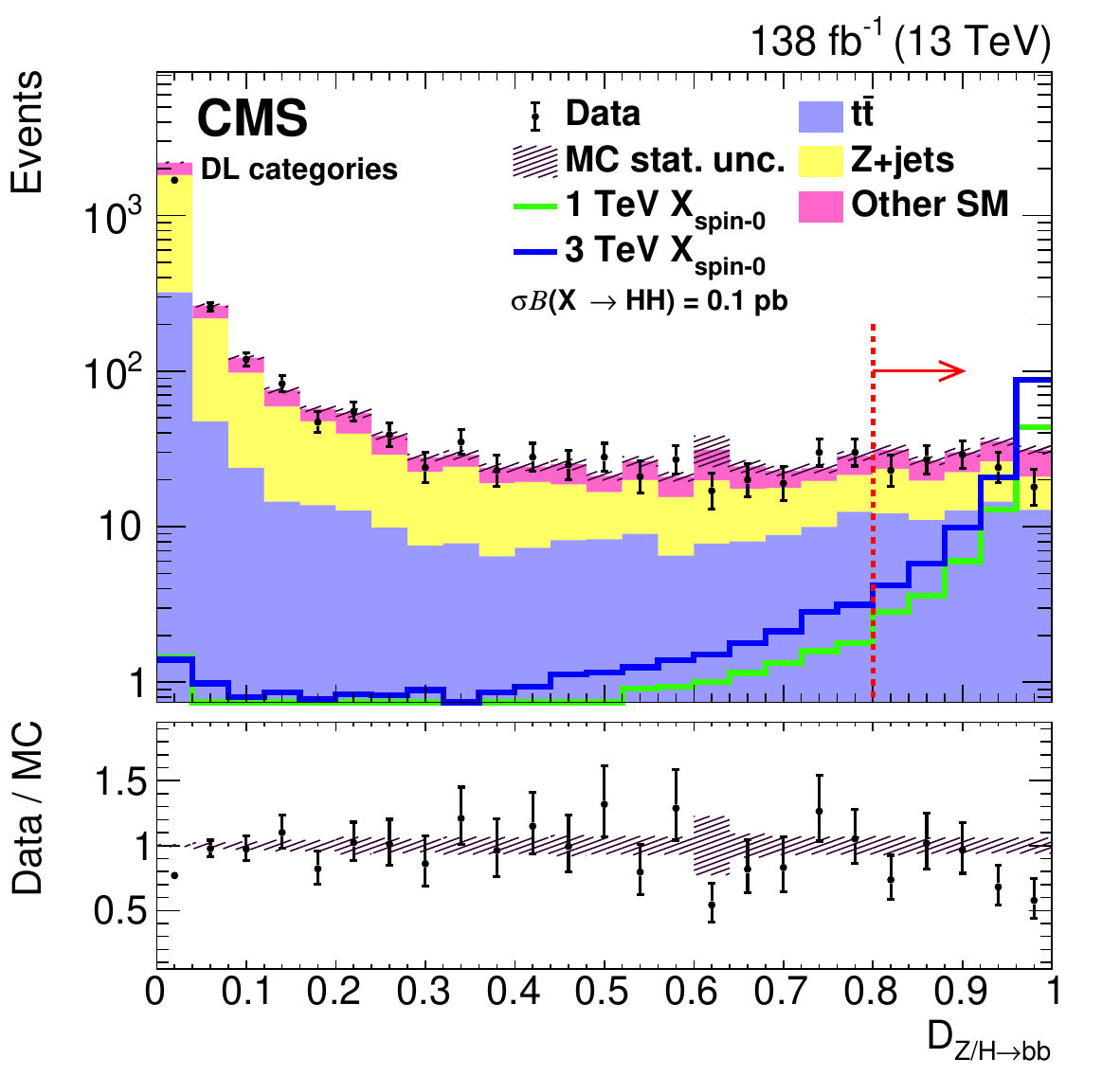}
\includegraphics[width=0.45\textwidth]{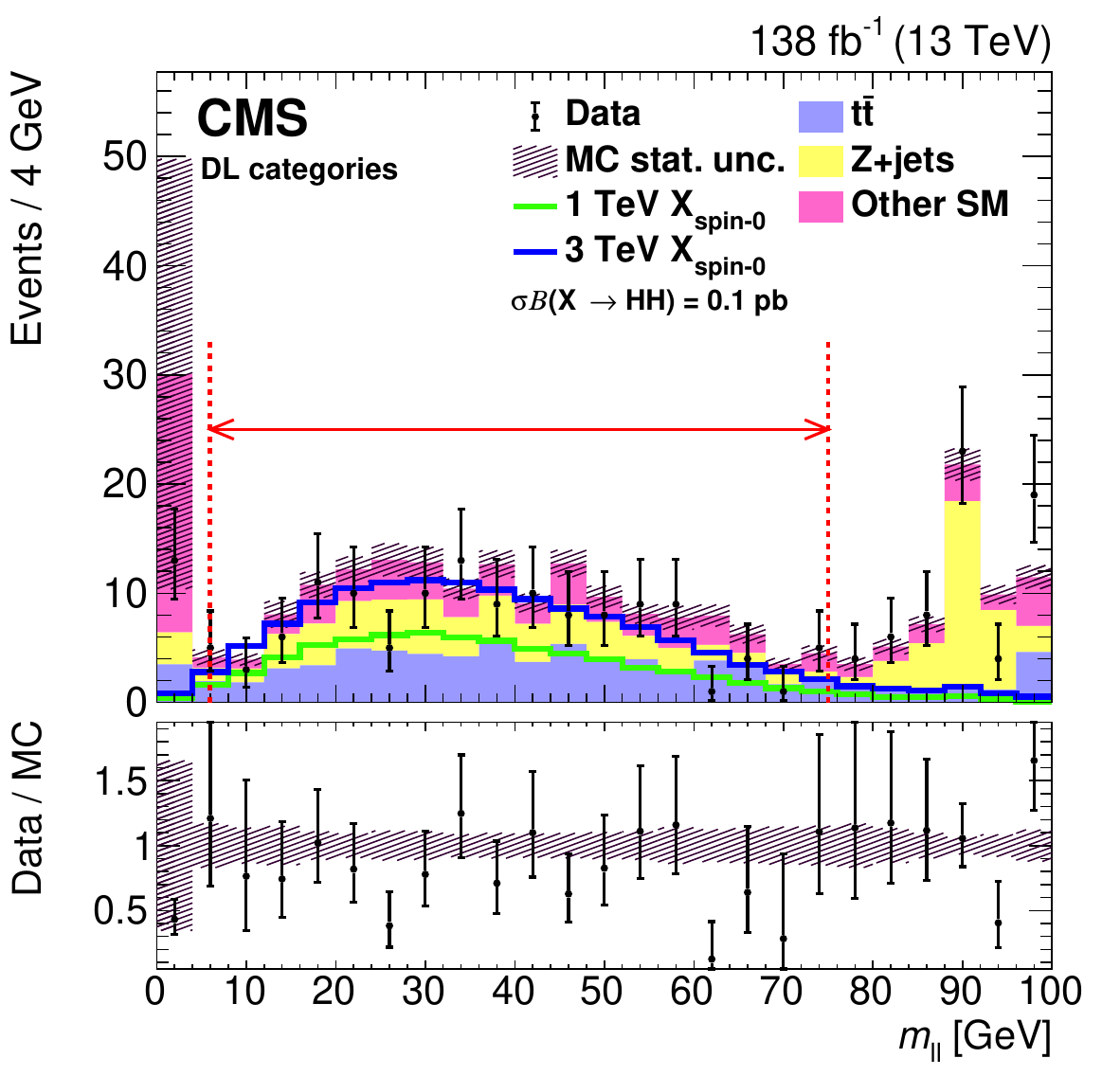}
\includegraphics[width=0.45\textwidth]{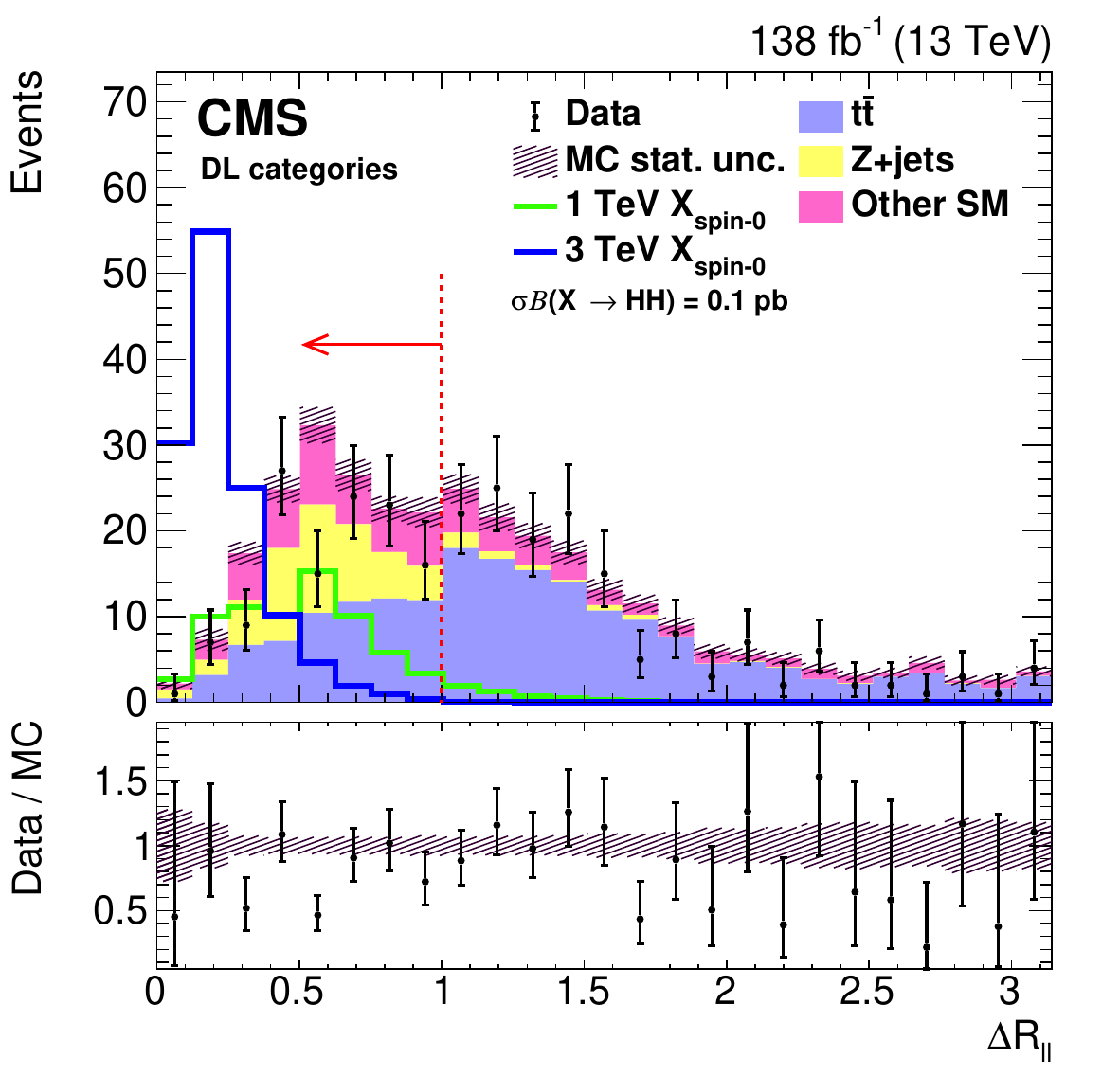}
\includegraphics[width=0.45\textwidth]{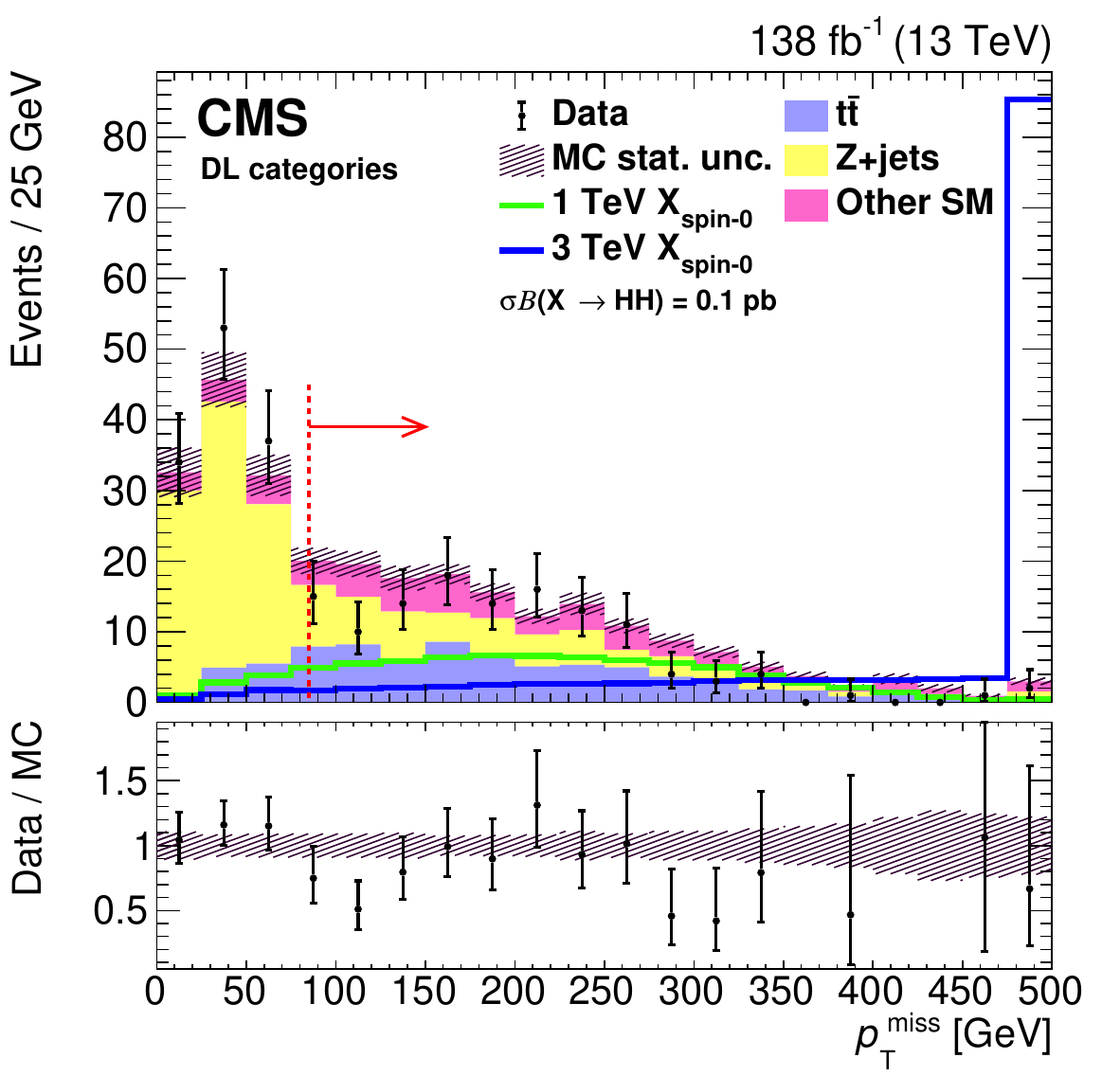}
\includegraphics[width=0.45\textwidth]{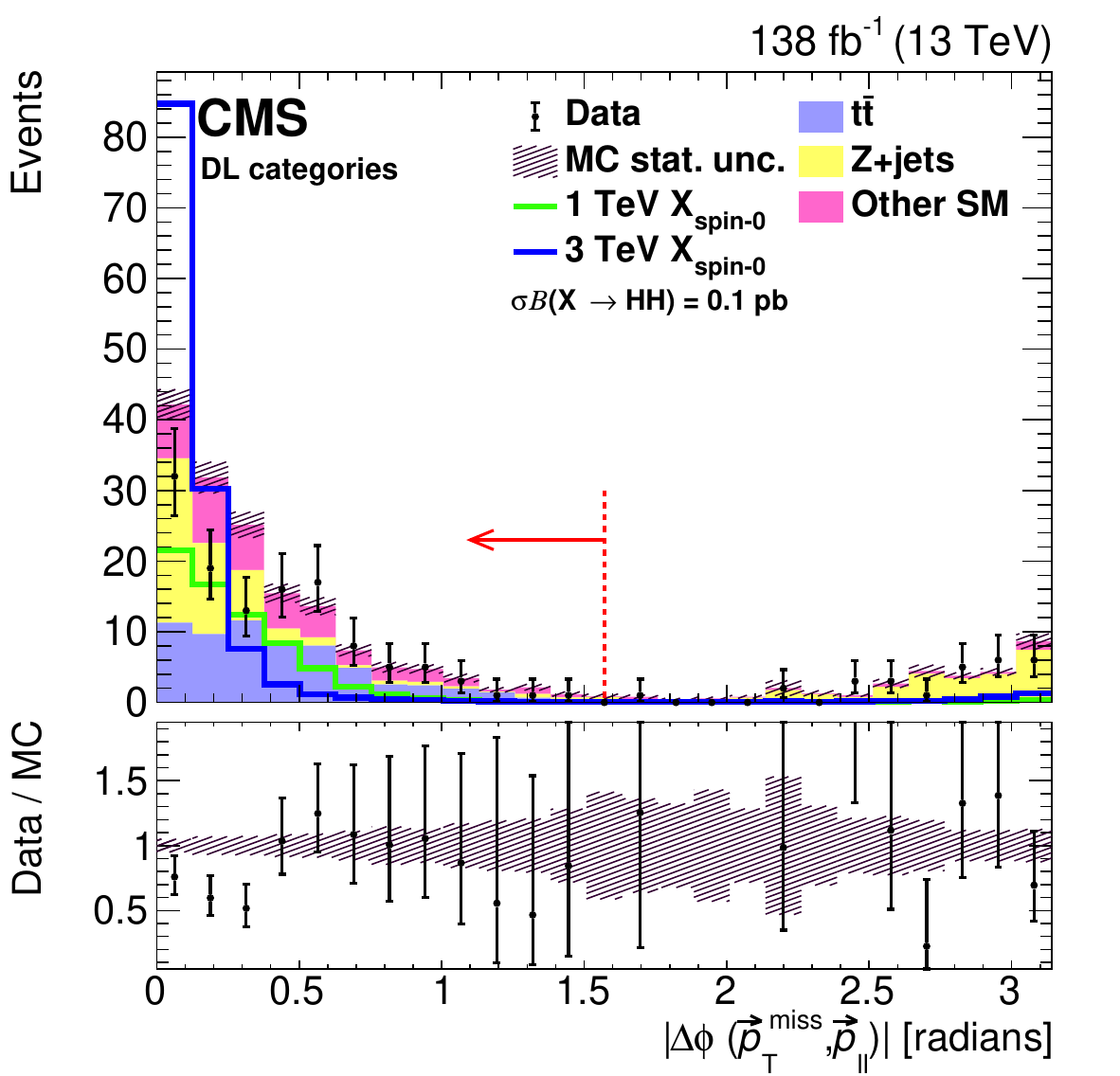}
\caption{Dilepton channel observables: distributions are shown for data (points), pre-fit simulated SM processes (filled histograms), and simulated signal (solid lines). The statistical uncertainty in the simulated sample is shown as the hatched band. Spin-0 signals for \mx of 1.0 and 3.0\TeV are displayed. The rightmost bin in the \mll, \drll, and \ptmiss plots contains the overflow events. For both signal models, $\sigma \mathcal{B} (\PX\to\hh)$ is set to 0.1\unit{pb}. The lower panels of each plot show the ratio of the data to the sum of all background processes. The red dashed line and arrow indicate the selected region of the variable of interest.}
\label{fig:event_vars_2l}
\end{figure}

Events in the DL channel must pass additional criteria. In signal events, the invariant mass of the two leptons is kinematically constrained by the mass of the boosted Higgs boson from which they originate, peaking near 30\GeV. Background in the \mll spectrum from \zjets populates predominantly lower masses from the continuum and higher masses from the \PZ boson. Background from \ttbar also populates higher masses since the leptons are typically opposite each other in the transverse plane. Requiring the dilepton invariant mass to satisfy $6 < \mll < 75\GeV$ reduces these backgrounds while preserving the signal. Requiring that the leptons be close together in $\eta$-$\phi$ space with $\drll < 1.0$ further helps to suppress the \ttbar background. In \zjets the \ptvecmiss can be in the direction of the \hbbjet, away from the leptons, due to jet mismeasurements, while in signal the \ptvecmiss is close to the leptons because of the boosted Higgs boson decay. Thus, we also require that $\abs{\dphimetll} < \pi/2$ to discriminate against \zjets. Background is further separated from signal by requiring $\ptmiss > 85\GeV$. Figure~\ref{fig:event_vars_2l} shows the distributions of the discriminating variables \mll (upper right), \drll (middle left), \ptmiss (middle right), and $\abs{\dphimetll}$ (lower).

\subsection{Event categorization \label{sec:categorization}}

Events are categorized by event properties that reflect the signal purity, and the categorization is the same over the full range of \mx.
In the SL channel, electron and muon events are separated because their reconstruction efficiencies for background and signal are different, resulting in different signal purities. The electron and muon categories are labeled ``\Pe'' and ``\PGm,'' respectively, in the figures. Likewise, in the DL channel, events with leptons of the same flavor and of the opposite (different) flavor are separated because the background composition is different between these two cases. These are labeled ``SF'' and ``OF,'' respectively, in the figures. We do not separate $\Pe\Pe$ from $\PGm\PGm$ in the DL channel because these events have similar ratios of signal to background.
For all events, there are two categories for \hbbjet tagging, constructed from different subsets of the distribution of the \textsc{DeepAK8} mass-decorrelated $\PZ/\PH \to \bbbar$ discriminator \dbb, introduced in Section~\ref{sec:jetreco}. The distribution of \dbb is shown in the upper left plot of Figs.~\ref{fig:event_vars_1l} and~\ref{fig:event_vars_2l} for the SL and DL channels, respectively. The discriminator value ranges from 0 to 1, with larger values indicating that the jet is more consistent with $\bbbar$ substructure. We use two working points that yield a loose category defined by $0.8 < \dbb < 0.97$ (labeled  ``\bL'') and a tight category defined by $\dbb \ge 0.97$ (labeled ``\bT'').
One more criterion for categorization, related to the $\PH\to\wwlnuqq$ decay, is implemented for the SL channel but not the DL channel. This categorization relies on both the \tauTO $N$-subjettiness ratio~\cite{Thaler:2010tr} of the \wqqjet (denoted now as \tauTO) and the \hww likelihood discriminator \dlnuqq that was first introduced in Eq.~(\ref{eq:dlnuqq}). The ratio \tauTO measures how consistent the jet substructure is with a two-prong decay versus a single-prong decay, with lower values more strongly indicating a two-prong decay. Figure~\ref{fig:event_vars_1l} shows the distributions of \dlnuqq (lower left) and \tauTO (lower right). All events in the SL but not in the DL search region are required to satisfy both $\tauTO < 0.75$ and $\dlnuqq < 11.0$. We construct a low-purity category (labeled ``LP'') with events that satisfy either $0.45 < \tauTO < 0.75$ or $2.5 < \dlnuqq < 11.0$ and a high-purity category (labeled ``HP'') with events that satisfy both $\tauTO < 0.45$ and $\dlnuqq < 2.5$. In 2016 data, the lower working point for \tauTO is 0.55 instead of 0.45.

The selections just described are combined to produce 12 distinct search categories (eight SL and four DL). When describing a single category, the label is a combination of those listed above. For example, in the SL channel the tightest \hbbjet tagging category with a low-purity selection on the \hww decay in the electron channel is: ``\Pe,~\bT,~LP.'' The categories and their corresponding labels are summarized in Tables~\ref{tab:cuts_cat} and \ref{tab:cuts_cat_2l}.

\begin{table}[!ht]
\centering
\topcaption{\label{tab:cuts_cat} The SL channel event categorization and corresponding category labels. All combinations of the two lepton flavors, two \hbbjet tagging, and two \hww decay purity selections are used to form eight independent event categories. The lower \tauTO working point is 0.55 (0.45) in 2016 (2017--2018).}
\begin{tabular}{lll}

Categorization type & Selection & Label\\
\hline
\multirow{2}{*}{Lepton flavor}	& Electron & \Pe \\
 	 			& Muon & \PGm \\ [\cmsTabSkip]
\multirow{2}{*}{\hbbjet tagging}	& $0.8 < \dbb < 0.97$ & \bL \\
			& $\dbb > 0.97$ & \bT \\ [\cmsTabSkip]
\multirow{2}{*}{\hww purity}	& $0.45 (0.55) < \tauTO < 0.75$ or $2.5 < \dlnuqq < 11.0$ & LP \\
				& $\tauTO < 0.45 (0.55)$ and $\dlnuqq < 2.5$ & HP \\

\end{tabular}
\end{table}

\begin{table}[!ht]
\centering
\topcaption{\label{tab:cuts_cat_2l} The DL channel event categorization and corresponding category labels. All combinations of the two lepton flavors and two \hbbjet tagging selections are used to form four independent event categories.}
\begin{tabular}{lll}

Categorization type & Selection & Label\\
\hline
\multirow{2}{*}{Lepton flavor}	& Two electrons or two muons & SF \\
 	 		& One electron and one muon & OF \\ [\cmsTabSkip]
\multirow{2}{*}{\hbbjet tagging}	& $0.8 < \dbb < 0.97$ & \bL \\
			& $\dbb > 0.97$ & \bT \\

\end{tabular}
\end{table}

The search is performed for $30 < \mbb < 210\GeV$ and $700 < \mhh < 5050\GeV$. Extending the \mbb mass window down to 30\GeV helps to capture the background in the fit, but events below 30\GeV would be relatively difficult to model since these are events for which the SD algorithm removes nearly all of the jet energy. The \mhh lower bound is chosen such that the \mhh distribution is monotonically decreasing for the full background. The upper bound is several hundred \GeVns above the highest mass event observed in data.

For spin-0 scenarios in the considered \hh modes, the total selection efficiency for an SL channel event to pass the criteria of any event category is 9\% at $\mx=0.8\TeV$. This efficiency includes the branching fraction for \hbb. The efficiency increases with \mx up to 23\% at $\mx=1.5\TeV$ because the Higgs boson decays become more collimated.
Above 1.5\TeV, the selection efficiency decreases to a minimum of 14\% at $\mx=4.5\TeV$ for two main reasons: the \PQb tagging efficiency degrades for high-\pt jets and the lepton isolation worsens for extremely collimated Higgs boson decays.
For DL channel events, the combined selection efficiency to pass the criteria of any event category is 9\% at $\mx=0.8\TeV$, increases sharply with \mx to 30\% at $\mx=1.5\TeV$, and then increases more slowly to 36\% at $\mx=4.5\TeV$. The efficiency grows over the full range of \mx because in the absence of a nearby jet, the leptons become easier to select at high \pt. Tables~\ref{tab:nm1_effs_1l} and~\ref{tab:nm1_effs_2l} show the efficiencies for each individual selection requirement with the full selection otherwise applied.

\begin{table}[!ht]
\centering
\topcaption{\label{tab:nm1_effs_1l} Efficiencies of each selection criterion in the SL channel with the rest of the full selection applied. The efficiencies for the total expected SM background and signals at 1.0 and 3.0\TeV are shown.}
\begin{tabular}{lccc}
\multirow{2}{*}{SL channel selection} & \multirow{2}{*}{Background} & \multicolumn{2}{c}{Signal} \\
 &  & 1\TeV & 3\TeV \\
\hline
\PQb jet veto & 0.31 & 0.87 & 0.82 \\
$\dbb > 0.8$ & 0.07 & 0.81 & 0.84 \\
$\tauTO < 0.75$ & 0.69 & 0.91 & 0.92 \\
$\dlnuqq < 11.0$ & 0.63 & 0.87 & 0.83 \\
$\ptom > 0.3$ & 0.87 & 0.97 & 0.86 \\

\end{tabular}
\end{table}

\begin{table}[!ht]
\centering
\topcaption{\label{tab:nm1_effs_2l} Efficiencies of each selection criterion in the DL channel with the rest of the full selection applied. The efficiencies for the total expected SM background and signals at 1.0 and 3.0\TeV are shown.}
\begin{tabular}{lccc}
\multirow{2}{*}{DL channel selection} & \multirow{2}{*}{Background} & \multicolumn{2}{c}{Signal} \\
 &  & 1\TeV & 3\TeV \\
\hline
\PQb jet veto & 0.45 & 0.86 & 0.84 \\
$\dbb > 0.8$ & 0.05 & 0.81 & 0.83 \\
$\ptmiss > 85\GeV$ & 0.55 & 0.88 & 0.97 \\
$6 < \mll < 75 \GeV$ & 0.62 & 0.95 & 0.94 \\
$\drll < 1.0$ & 0.51 & 0.93 & 0.998 \\
$\abs{\dphimetll} < \pi/2$ & 0.83 & 0.98 & 0.97 \\

\end{tabular}
\end{table}

The Higgs bosons in signal events from a spin-2 \PX boson are produced at lower values of $\abs{\eta}$ than those from a spin-0 \PX, resulting in larger selection efficiencies for spin-2 events. The relative increase in efficiency for spin-2 signal is larger at low mass (${\approx}40\%$) than at high mass (${\approx}15\%$).

\subsection{Control regions}

Two control regions (CRs) are used to validate the SM background estimation and systematic uncertainties. These regions are depleted of signal by construction, and the events within them are not used to search for signal.
The first, labeled ``top CR,'' targets background events with top quarks, particularly \ttbar. Such events are selected by inverting the AK4 jet \PQb tag veto. To increase the statistical power of the sample, the \ptom selection is removed for SL channel events, and the \drll selection is altered to $\drll > 0.4$ for DL channel events. Events in this CR are then divided into the 12 categories previously described in Section~\ref{sec:categorization}. The \mbb and \mhh distributions in this CR are similar to the distributions in the signal region for backgrounds with top quarks.
The top quark \pt spectrum in \ttbar events has been shown to be mismodeled in simulation~\cite{Sirunyan:2017mzl,Khachatryan:2016mnb}. A small \pt-dependent correction, on the order of a few percent, is measured in an expanded version of this CR and applied to the \ttbar simulation.

While the top CR is an adequate probe of processes that involve top quarks, it is not sensitive to background from \zjets, \wjets, or QCD multijet processes. Instead, a second CR, labeled ``non-top CR,'' is used to study the modeling of these processes. The selection of events in this CR is the same as for the signal region, except that the \hbbjet is required to be inconsistent with having $\bbbar$ substructure, \ie, $0.01 < \dbb < 0.04$. We exclude events with $\dbb < 0.01$ because of substantial mismodeling in that region. As a result, events in this CR are not categorized by \hbbjet tagging, yielding half as many (six) categories here as in the top CR. Because it has fewer categories, the non-top CR cannot in principle test the modeling of the \PQb tagging of \qgbkg jets that contain \PQb quarks or are misidentified as containing \PQb quarks. Instead, we rely on the top CR to verify that this modeling is well behaved.

Ultimately, the final values of the normalization and shape of each background component and their corresponding uncertainties are determined in the 2D fit to the data in the search region.

\section{Background and signal modeling\label{sec:modeling}}

The search is performed by simultaneously estimating the signal and background yields with a 2D maximum likelihood fit of the data in the 12 event categories. The data are binned in two dimensions, \mbb and \mhh, within the ranges $30 < \mbb < 210\GeV$ and $700 < \mhh < 5050\GeV$. The \mbb bin width is 6\GeV, and the \mhh bin width is variable: 25\GeV width at the low end of the mass range, 50\GeV width in the middle of the mass range, and 75\GeV at high mass. These bin widths are smaller than the mass resolutions of the signal in the relevant parts of \mhh space.
Signal and background mass distributions are modeled using a number of 2D templates that are created using only simulation, which is smoothed using different strategies described in the rest of this section before the templates are fit to data. Independent templates are used for each event category. Shape and normalization uncertainties that account for possible differences between data and simulation, detailed in Section~\ref{sec:syst}, are included while executing the fit. This fitting method was previously presented in Ref.~\cite{Sirunyan:2018iff}.

\subsection{Background component classification} \label{sec:classification}

To perform the fit to data, we split the background into components and then generate 2D templates in the \mbb and \mhh mass plane for each component independently. The normalization and shape of each component is then allowed to vary in the fit to the data in each search category.

\begin{figure}[ht!]
\centering
\includegraphics[width=0.45\textwidth]{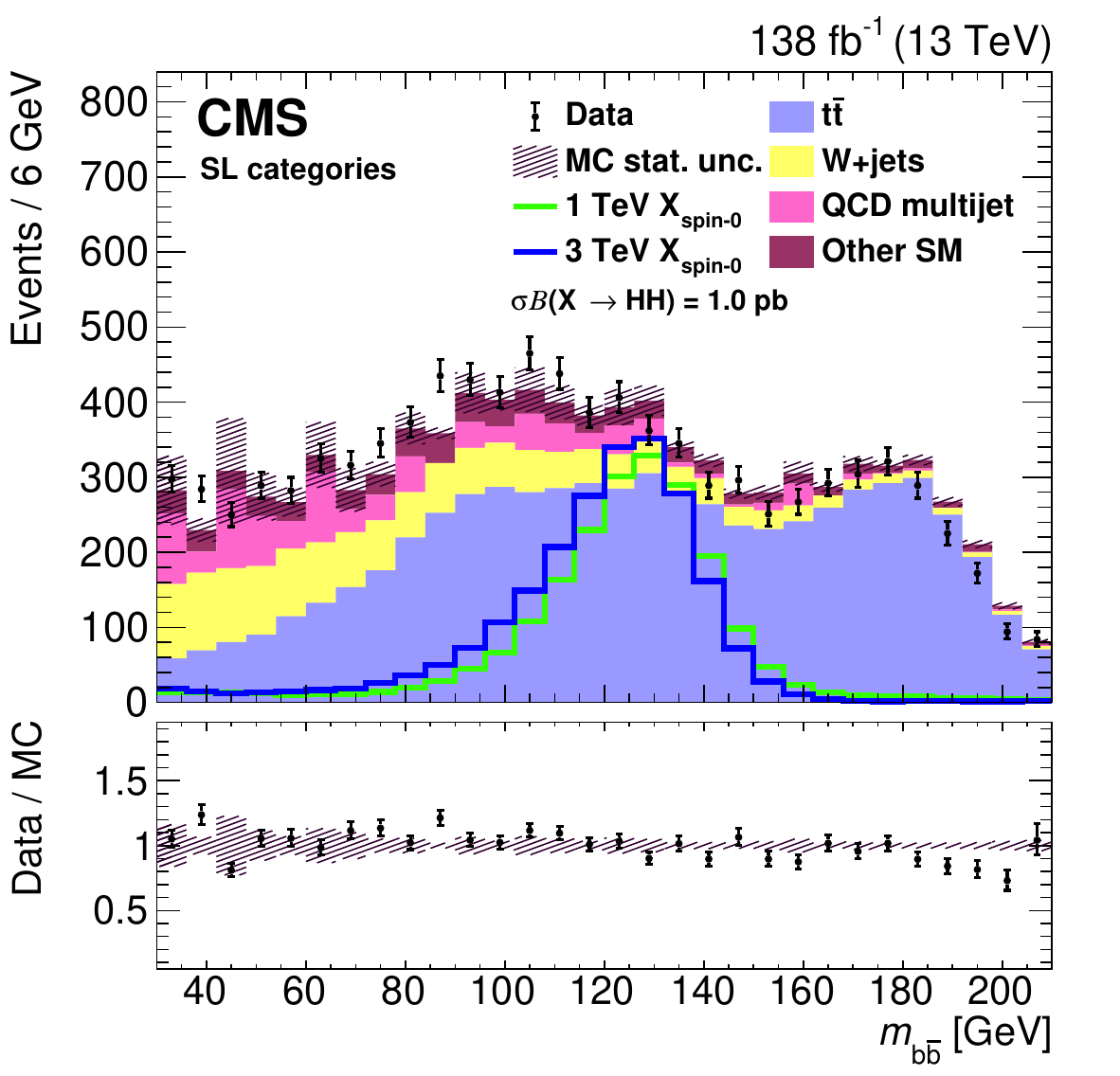}
\includegraphics[width=0.45\textwidth]{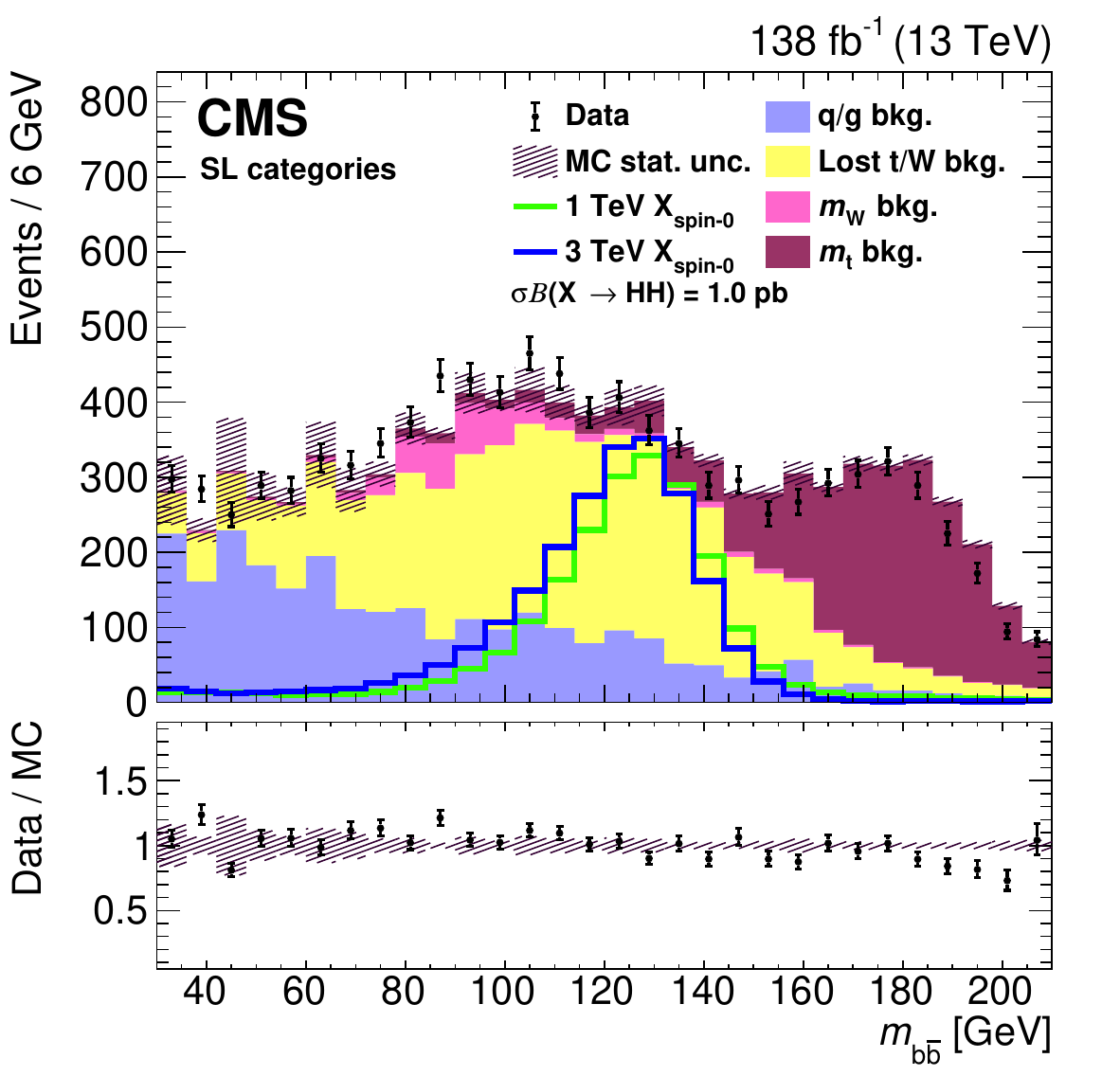}
\includegraphics[width=0.45\textwidth]{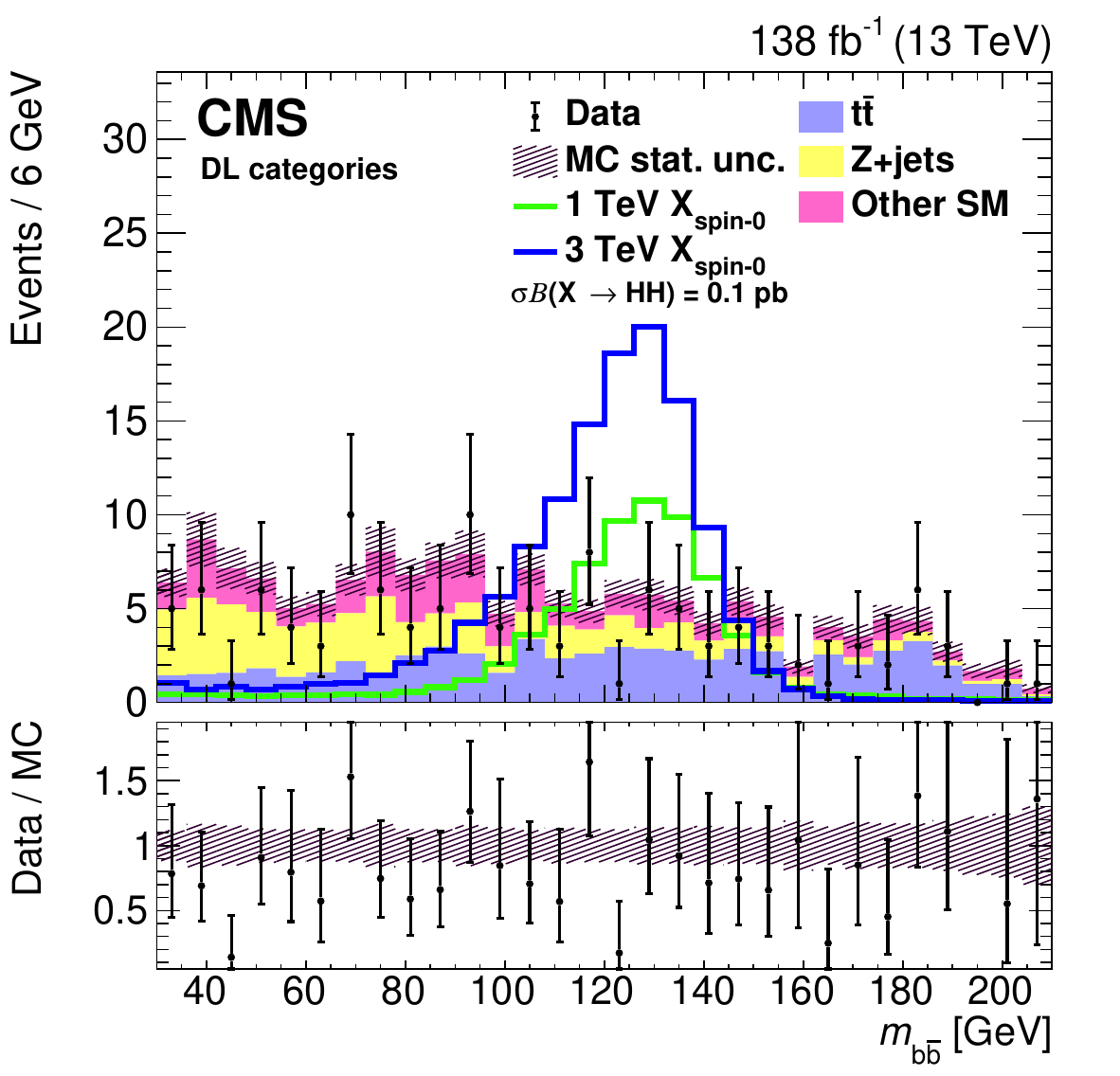}
\includegraphics[width=0.45\textwidth]{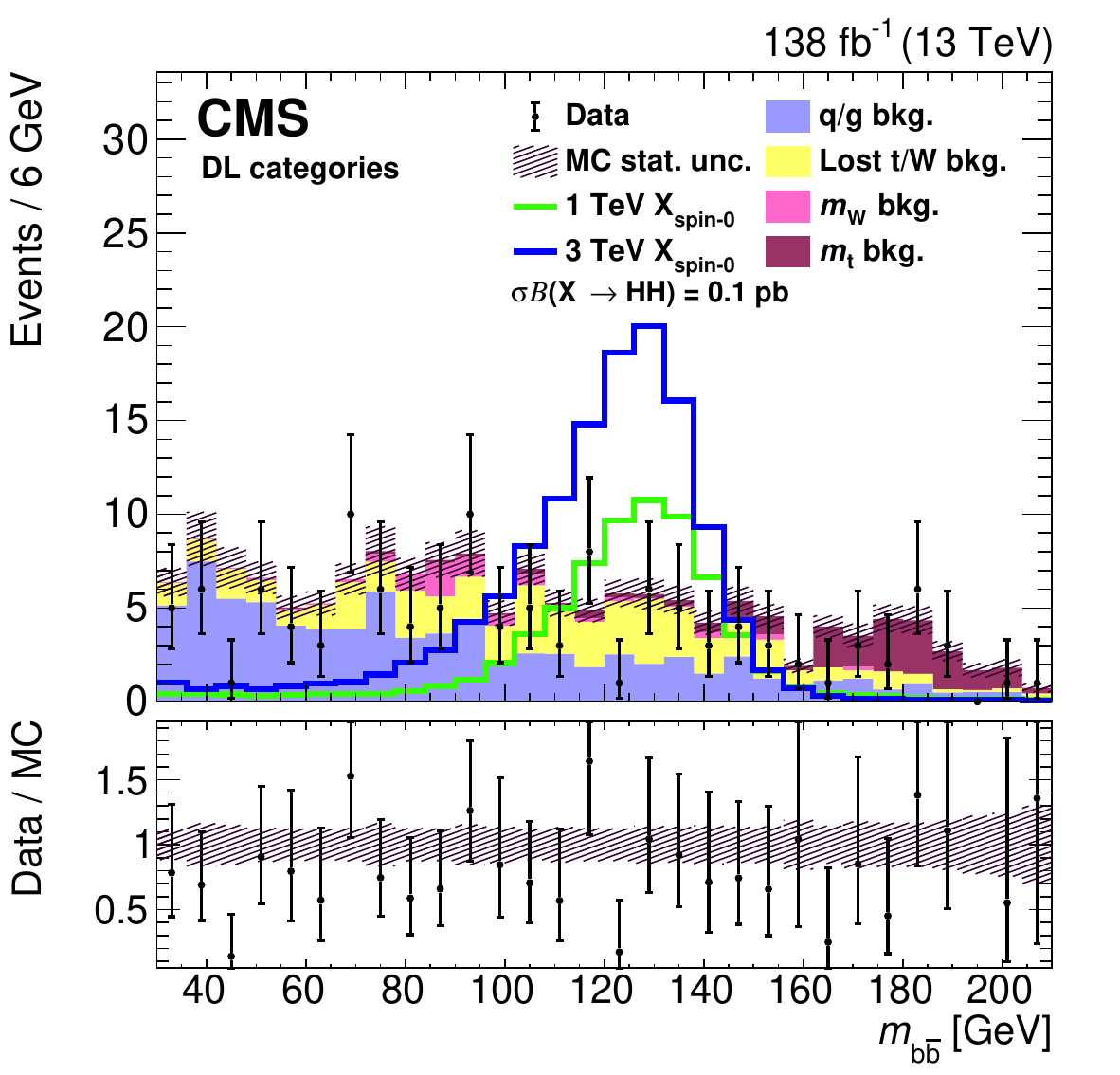}
\caption{ The pre-fit \mbb distributions for the SL (upper row) and DL (lower row) channels. The data are shown as the points with error bars. In each plot, the pre-fit background (filled histograms) is shown broken down either according to the SM process (left) or according to the background classification of Section~\ref{sec:classification} (right). The total simulated background is the same in each case. The statistical uncertainty in the simulated sample is shown as the hatched band. Spin-0 signals for \mx of 1.0 and 3.0\TeV are also shown (solid lines). The product $\sigma \mathcal{B} (\PX\to\hh)$ is set to 1.0\unit{pb} for the SL channel and 0.1\unit{pb} for the DL channel. The lower panels of each plot show the ratio of the data to the sum of all background processes. }
\label{fig:mbbComp}
\end{figure}

Instead of splitting by SM process, we distinguish four components by particle-level information, such that they each have distinct \mbb distribution shapes.
The background is divided by counting in simulation the number of generator-level quarks from the immediate decay of a top quark or vector boson within $\DR<0.8$ of the \hbbjet axis. The first component is the ``\mtbkg,'' in which all three quarks from a single top quark decay fulfill this criterion. The second component is the ``\mwbkg,'' identified as the events that do not fulfill the \mtbkg criterion but in which both quarks from either a \PZ or \PW boson fall within the jet cone. Both of these backgrounds contain resonant peaks in the \mbb shape corresponding to either the top quark or \PW boson mass. The ``\losttwbkg'' contains events in which at least one quark is contained within the \hbbjet cone, but not the full set needed to satisfy one of the previous two requirements. Finally, all other events are designated by the ``\qgbkg''. The first three categories are primarily composed of \ttbar events, while the \qgbkg is composed mostly of \wjets and QCD multijet processes in the SL channel and of \zjets in the DL channel. The background classification is summarized in Table~\ref{tab:bkg-type}. Figure~\ref{fig:mbbComp} shows the pre-fit \mbb spectrum separately for the SL and DL channels. The background components are shown either as SM processes or with the background classification just described.

\begin{table}[ht!]
\centering
\topcaption{\label{tab:bkg-type} The four background components with their kinematical properties and defining number of generator-level quarks within $\DR<0.8$ of the \hbbjet axis.}
\begin{tabular}{llll}
Bkg. category & Dominant SM processes & Resonant in \mbb & Num. of particle-level quarks\\
\hline\\ [-10pt]
\mtn & \ttbar & top quark mass & 3 from top quark\\
\mwn & \ttbar & \PW boson mass & 2 from \PW boson\\
\losttwn & \ttbar & No & 1 or 2 \\
\qgn & \vjets and QCD multijet & No & 0\\
\end{tabular}
\end{table}

\subsection{Template construction strategy}

For each of the four background components, a unique template in the \mbb and \mhh mass plane is produced for each of the 12 event categories. First, we produce a small set of inclusive templates that have more statistical power than the set of events in each individual search category. These inclusive templates are made by combining events in multiple categories and by relaxing selections, provided that the inclusive shape remains consistent with the shape for the full selection. Then, for each of the 12 event categories, the inclusive templates are fit to the simulated mass distributions to produce templates with their own individual shapes.
This fit is performed in a similar manner and with a similar parameterization of the template shape as is done for the fit to data. The background templates and associated systematic uncertainties are ultimately validated by fitting to data in dedicated CRs, a procedure described in Section~\ref{sec:validation}.

In the SL channel, a modified approach is used when building templates that reduces fluctuations due to the limited size of the QCD multijet simulated event sample. The \hbbjet reconstruction in the QCD multijet simulation is similar to that in \wjets, and the \wjets simulation has much more statistical power. Both processes contribute significantly to the \qgn background, with light-flavored quark or gluon AK8 jets that are misidentified as \PQb jets, yielding very similar falling shapes in the \mbb spectrum and similar \hbbjet tagging distributions. Instead of using the QCD simulation directly in the \qgn background modeling, a combined distribution is created by measuring the ratio of QCD to \wjets event yields as a function of \mhh and then using these corrections to scale up the \wjets simulation. Corrections and distributions are obtained for each lepton flavor and \hww purity category, since the \hbbjet tagging between \wjets and QCD is equivalent. This distribution is then used as input to the \qgn background modeling to account for both processes.

\subsection{Background modeling}

The background templates are modeled using conditional probabilities of \mbb as a function of \mhh so that the templates include the correlation of these two variables, fully described in Ref.~\cite{B2G-18-008}. The full 2D template is defined as:
\begin{linenomath}
\begin{equation}\label{eq:bkg_model}
\pbkg(\mbb,\mhh) = \pbb(\mbb|\mhh,\theta_1) \phh(\mhh|\theta_2),
\end{equation}
\end{linenomath}
where \pbb is a 2D conditional probability distribution, \phh is a 1D probability distribution, and $\theta_1$ and $\theta_2$ are sets of nuisance parameters used to account for background shape uncertainties. The sets $\theta_1$ and $\theta_2$ do not share any common nuisance parameters.

The \phh templates are produced by smoothing 1D \mhh histograms with kernel density estimation (KDE)~\cite{Rosenblatt,Silverman,Scott}. To produce these templates, we use Gaussian kernels with adaptive bandwidths, which are parameters of the KDE that control the smoothing and are dependent on the local event density. We do this to apply less smoothing to regions of the distribution with many events and more smoothing to regions with few events. For $\mhh\gtrsim 2\TeV$, where there are very few events in simulation or in data, the \mhh tail is further smoothed by fitting with an exponential function.

The 2D templates \pbb are obtained with different methods for the resonant and nonresonant background components. For each of the resonant backgrounds (\mtn and \mwn), we fit the \mbb distributions with a double Crystal Ball function~\cite{Oreglia:1980cs,Gaiser:1982yw} centered around \mtn and \mwn, respectively. This function has a Gaussian core, which is used to model the bulk of the \mbb resonance, and power-law tails, which account for the effects of jet misreconstruction. The fits are performed for events binned in \mhh to capture the dependence of the \mbb shape on \mhh. For the nonresonant backgrounds (\losttwn and \qgn), the \pbb are estimated from 2D histograms using 2D KDE. Independent KDE parameters are used for each dimension and each background when building the \pbb templates. As done for the \phh tail modeling, the high-mass \mhh distribution tail here is exponentially smoothed. The normalizations from simulation are used as the initial values for the background normalizations in the fit to data.

\subsection{Signal modeling}

The signal templates are also defined following Ref.~\cite{B2G-18-008} using conditional probabilities:
\begin{linenomath}
\begin{equation}
\psig(\mbb,\mhh|\mx) = \phh(\mhh|\mbb,\mx,\theta_1 ^\prime) \pbb(\mbb|\mx,\theta_2 ^\prime).
\end{equation}
\end{linenomath}
The sets $\theta_1 ^\prime$ and $\theta_2 ^\prime$ do not share any common nuisance parameters. However, $\theta_2 ^\prime$ and $\theta_1$ from Eq.~(\ref{eq:bkg_model}) do share two nuisance parameters corresponding to the mass scale and resolution uncertainties of SD jets in the \mbb dimension. This is discussed in more detail in Section~\ref{sec:shapeUncs_bkg}.

The \psig distributions are first obtained for discrete \mx values by fitting histograms of the signal mass distributions. The mass shapes for spin-0 and spin-2 signals are very similar, and so the modeling is performed on the combined set of events and applied to both spin hypotheses. Models continuous in \mx are then produced by interpolating the fit parameters. The 1D \pbb templates are created by fitting the \mbb spectra with a double Crystal Ball function, and the mass resolution is slightly larger than 10\%, with the largest resolution at low mass. The modeling of events in the \bL category also contains an exponential component to model the small fraction of signal events with no resonant peak in the distribution.

The 2D \phh templates are designed to account for correlations between \mhh and \mbb. These \mhh distributions are also modeled with a double Crystal Ball function, but with an additional linear dependence on \mbb, parameterized by $\Delta_{\bbbar} = (\mbb - \mu_{\bbbar})/\sigma_{\bbbar}$. Here, $\mu_{\bbbar}$ and $\sigma_{\bbbar}$ are the mean and width parameters, respectively, in the fit to the \mbb spectra. To accomplish this, the mean parameter $\mu_{\hh}$ in the Crystal Ball function fit is then taken to be
\begin{linenomath}
\begin{equation}
\mu_{\hh} = \mu_0 (1 + \mu_1 \Delta_{\bbbar}),
\end{equation}
\end{linenomath}
where $\mu_{0}$ and $\mu_{1}$ are fit parameters. With this approach, we can account for mismeasurements of the \hbbjet that result in mismeasurements of \mhh. The resolution of the \mhh resonance, denoted as $\sigma_{\hh}$, is also dependent on \mbb such that
\begin{linenomath}
\begin{align}
\sigma_{\hh} &=
\begin{cases}
\sigma_{0} (1 + \sigma_1  \abs{\Delta_{\text{\bbbar}}}), & \Delta_{\text{\bbbar}} < 0 \\
\sigma_{0}, & \Delta_{\text{\bbbar}} > 0
\end{cases}
\end{align}
\end{linenomath}
where $\sigma_0$ and $\sigma_1$ are fit parameters. In the case that the SD algorithm produces an undermeasurement of \mbb by removing too much energy from the Higgs boson decay, the correlation increases, and the \mhh resolution grows wider. For $\abs{\Delta_{\text{\bbbar}}} > 2.5$, we use the value at the boundary, since the correlation does not hold for severe mismeasurements. The \mhh resolution is ${\approx} 5\%$.

The product of the acceptance and efficiency for $\PX\to\hh$ events to fall into any of the individual search categories is taken from simulation. As done for the signal shape parameters, the efficiency is interpolated along \mx. Uncertainties in the relative acceptances and in the integrated luminosity of the sample are included in the 2D maximum likelihood fit that is used to obtain confidence intervals for the $\PX\to\hh$ process.
The signal modeling is tested using pseudo-experiments in which we fit the templates to pseudodata that contain a fixed amount of signal; no significant bias in the fitted signal yield is found.

\subsection{Validation of background models with control region data \label{sec:validation}}

\begin{figure}[ht!]
\centering
\includegraphics[width=0.45\textwidth]{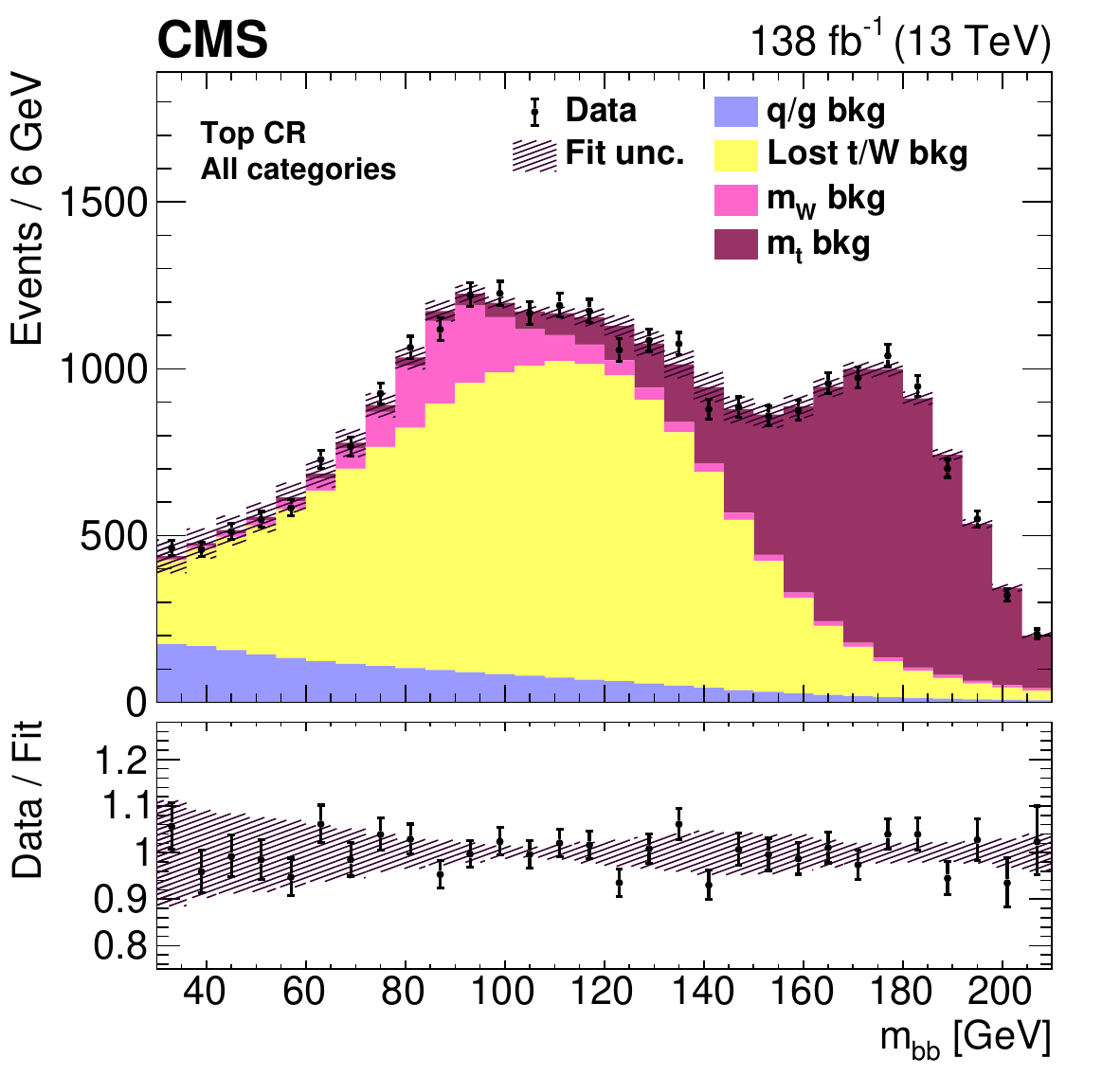}
\includegraphics[width=0.45\textwidth]{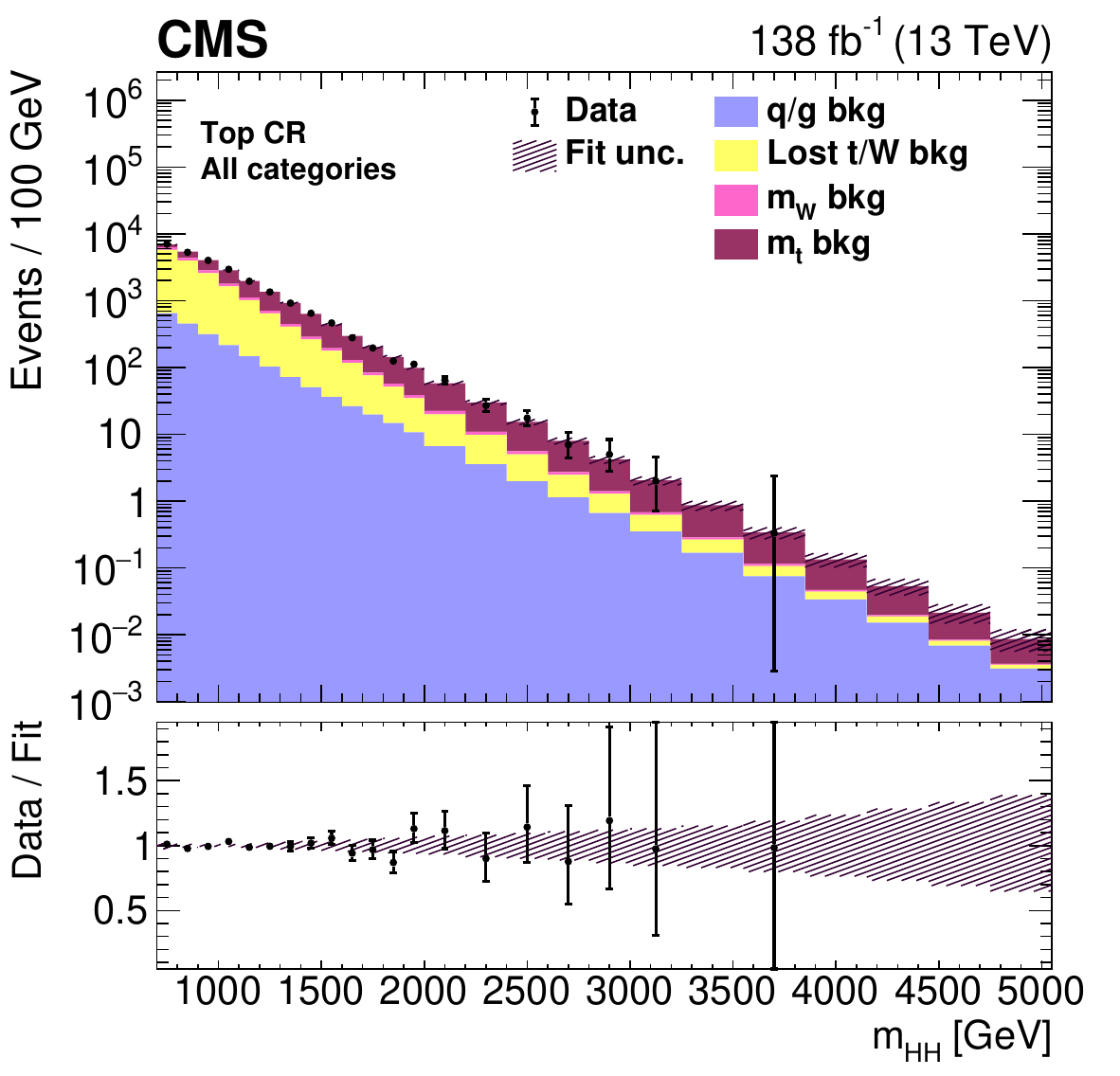}
\includegraphics[width=0.45\textwidth]{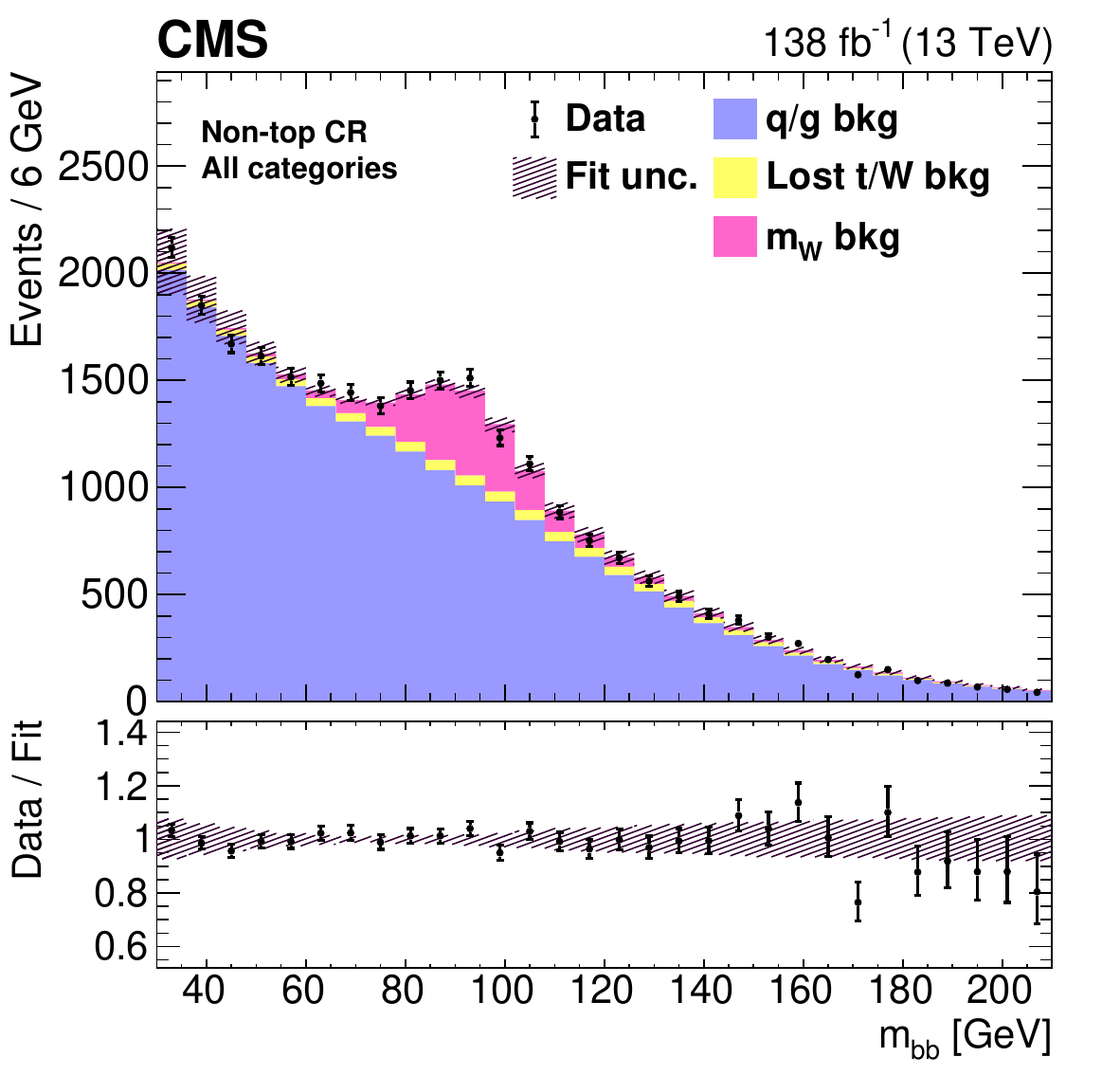}
\includegraphics[width=0.45\textwidth]{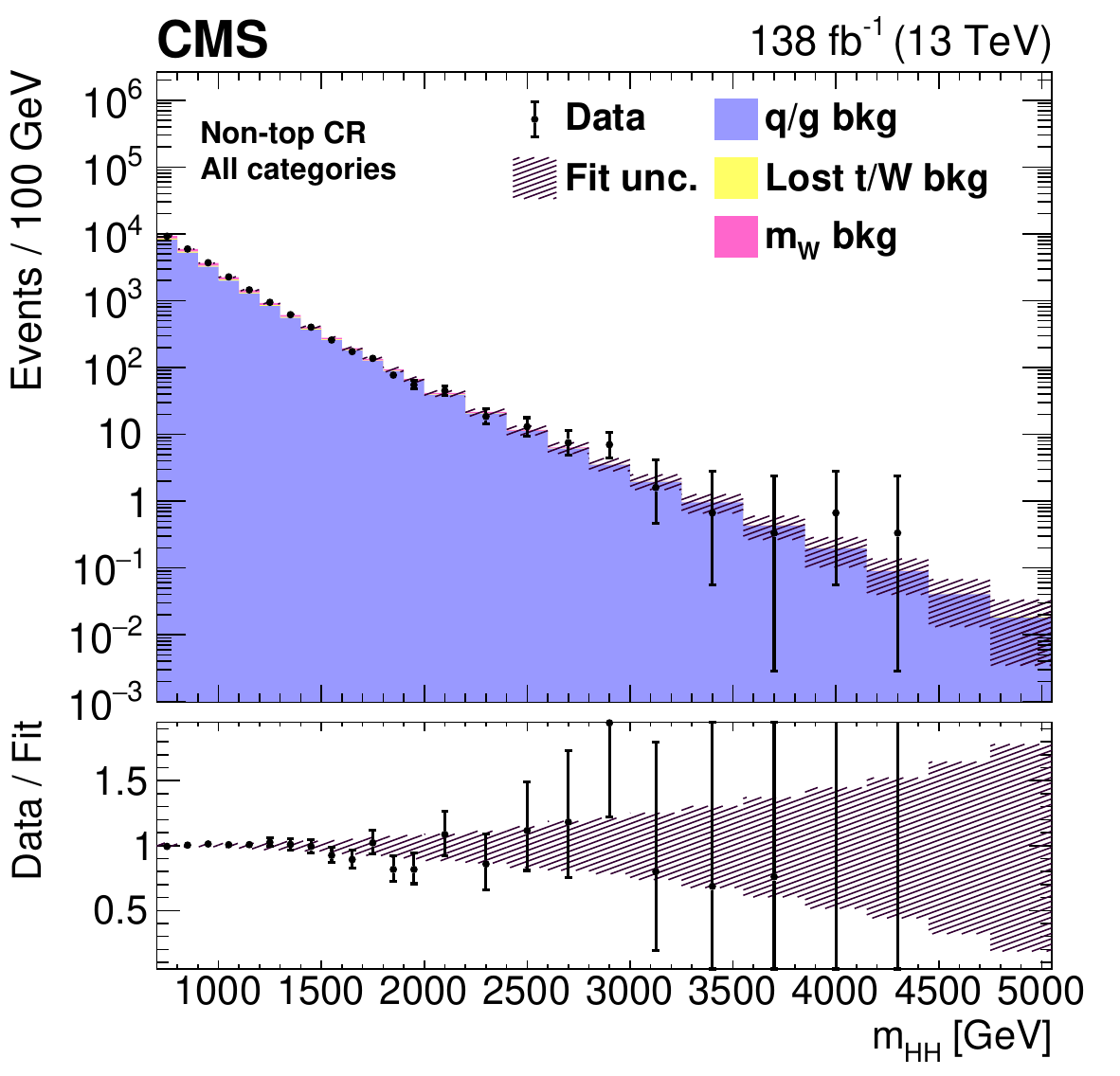}
\caption{ The post-fit model compared to data in the top~CR (upper plots) and non-top~CR (lower plots), projected into \mbb (left) and \mhh (right). Events from all categories are combined. The fit result is the filled histogram, with the different colors indicating different background components. The background shape uncertainty is shown as the hatched band. The lower panels of each plot show the ratio of the data to the fit result.}
\label{fig:modeling_cr}
\end{figure}

The background models are validated in the top~CR and non-top~CR data samples. For both CRs, background templates are constructed in the same way as for the search region, except using the CR selection. The background templates are then fit to the CR data with the same nuisance parameters that are used in the standard 2D maximum likelihood fit. In the non-top CR, the \mtn background is negligible and not included in the modeling. The result of the simultaneous fit is shown in Fig.~\ref{fig:modeling_cr} for both CRs.
To improve visualization, the displayed binning in these and subsequent histograms is coarser than the binning used in the maximum likelihood fit.
The projections in both mass dimensions are shown for the combination of all event categories. In both CRs, the fit results model the data well in all categories, indicating that the shape uncertainties can account sufficiently for potential differences between data and simulation.

\section{Systematic uncertainties\label{sec:syst}}

Systematic uncertainties that affect the normalization and shape of the signal and background are modeled with nuisance parameters in the 2D maximum likelihood fit to data. Nuisance parameters for shape uncertainties have Gaussian constraints, while normalization uncertainties have log-normal constraints. In certain cases a single nuisance parameter may affect both the normalization and the shape of a resonance, in which case the nuisance parameter constraint is Gaussian. Detailed methods of parameterizing the background and signal uncertainties are described in Sections~\ref{sec:bkgUncs} and~\ref{sec:signalUncs}, respectively.

To implement nonresonant mass shape uncertainties, templates are first generated with modified event weights that include multiplicative parameters proportional to \mbb, \mhh, 1/\mbb, and 1/\mhh. Each of these four modifications produces two alternative templates that represent an upward and downward shift from the nominal model. The 2D fit then interpolates between these two alternative templates to constrain the magnitudes of these parameters.
Resonant mass shape uncertainties are implemented as uncertainties in the mean and width parameters of a double Crystal Ball function. In most cases, different nuisance parameters are used for the background shape uncertainties from those used for the signal uncertainties.

All background and signal uncertainties are listed in Tables~\ref{tab:bkg_uncertainties} and~\ref{tab:sig_uncertainties}, respectively, with their initial sizes. A single uncertainty type can be applied to multiple event categories with independent nuisance parameters for each category. The background model contains 104 total nuisance parameters, while the signal model contains 27, with two parameters shared between signal and background. The descriptions of all uncertainties and their correlations are also described in the rest of this section.

\subsection{Background uncertainties \label{sec:bkgUncs}}

Background uncertainty parameters are chosen by considering possible discrepancies between data and simulation, such as in the relative background composition or in the jet energy scale. Studies of the two CRs are used to verify that the chosen uncertainties cover such differences. The fitted values and the sensitivity to signal do not depend strongly on the sizes of the pre-fit uncertainties because they serve as loose constraints on the fit. We verify this by inflating all pre-fit background uncertainties by a factor of two and observing that the final result does not change. Therefore, the pre-fit uncertainties are sufficiently large to account for discrepancies between data and simulation in the CRs. More complex background models, such as those with more nuisance parameters or higher-order shape distortions, were studied following the same approach as in Ref.~\cite{B2G-18-008} and not found to be necessary.

\begin{table}[ht]
\centering
\topcaption{\label{tab:bkg_uncertainties} Background systematic uncertainties included in the maximum likelihood fit. The uncertainty types with "normalization" correspond to uncertainties in the background yield, while all others are uncertainties in the background shape. The $N_{\text{p}}$ column indicates the number of nuisance parameters used to model the uncertainty. In the last two columns, $\sigma_{\text{I}}$ refers to the initial estimate of the uncertainty, and $\sigma_{\text{C}}$ refers to the constrained uncertainty obtained post-fit. For the \qgn, \ttbar, and \losttwn shape uncertainties, ``scale'' uncertainties are those implemented with alternative templates with multiplicative parameters proportional to mass $m$, and ``inverse scale'' uncertainties are those implemented with parameters proportional to $1/m$.}
\begin{tabular}{lllll}
Uncertainty type & Processes & $N_{\text{p}}$ & $\sigma_{\text{I}}$ & $\sigma_{\text{C}}/\sigma_{\text{I}}$ \\
\hline
SD jet \mbb scale & \mwn, \mtn, signal & 2 & 0.54\%, 2.0\% (\mtn) & 98\%, 19\% (\mtn) \\
SD jet \mbb resolution & \mwn, \mtn, signal & 2 & 8.6\%, 17.2\% (\mtn) & 95\%, 25\% (\mtn) \\
\qgn normalization & \qgn & 12 & 50\% (1\Pell), 100\% (2\Pell) & 37--78\% \\
\qgn \mhh scale & \qgn & 10 & $\pm0.5 ~\mhh/\TeV$ & 78--99\% \\
\qgn \mhh inverse scale & \qgn & 10 & $\pm1.4  \TeV/\mhh$ & 64--99\% \\
\qgn \mbb scale & \qgn & 4 & $\pm0.00375~\mbb/\GeV$ & 81--99\% \\
\qgn \mbb inverse scale & \qgn & 4 & $\pm15  \GeV/\mbb$ & 77--99\% \\
\Losttwn \mbb scale & \losttwn & 4 & $\pm0.003 ~ \mbb/\GeV$ & 71--99\% \\
\Losttwn \mbb inverse scale & \losttwn & 4 & $\pm18  \GeV/\mbb$ & 88--99\% \\
\ttbar normalization & \losttwn, \mwn, \mtn & 12 & 35\% (1\Pell), 70\% (2\Pell) & 19--68\% \\
\ttbar relative normalization & \losttwn, \mwn, \mtn & 8 & 35\% (1\Pell), 70\% (2\Pell) & 9--96\% \\
\ttbar \mhh scale & \losttwn, \mwn, \mtn & 12 & $\pm0.25 ~ \mhh/\TeV$ & 84--99\% \\
\ttbar \mhh relative scale & \losttwn, \mwn, \mtn & 8 & $\pm0.25 ~ \mhh/\TeV$ & 74--99\% \\
\ttbar \mhh inverse scale & \losttwn, \mwn, \mtn & 12 & $\pm0.7  \TeV/\mhh$ & 61--99\% \\
\end{tabular}
\end{table}

In the following subsections, we detail the parameterization of the different uncertainties for the background.

\subsubsection{Background normalization uncertainties \label{sec:normUncs_bkg}}

The \mwn, \mtn, and \losttwn backgrounds all primarily arise from \ttbar production. Consequently, some uncertainties are applied by treating these three backgrounds together, referred to collectively as the \ttbar background in Table~\ref{tab:bkg_uncertainties}. We account for differences between data and simulation in the \ttbar normalization by including independent nuisance parameters for each category that allow the normalizations of these backgrounds to vary in a correlated manner (``\ttbar normalization"). However, the three \ttbar-dominated background components exhibit differences in the \PQb~tagging efficiency and the \hbbjet \pt spectrum, so we include additional nuisance parameters (``\ttbar relative normalization") that allow the relative normalizations of each of these to vary within the absolute normalization, which itself also varies. Separate nuisance parameters are used to control the \qgn background normalization, as this is the only background component to arise primarily from non-\ttbar processes.

\subsubsection{Background shape uncertainties \label{sec:shapeUncs_bkg}}

The shape uncertainties for the backgrounds are modeled differently depending on whether or not the shape is resonant in the \mbb dimension. All backgrounds are nonresonant in the \mhh dimension, and mismodeling of the background \pt spectrum can manifest as an incorrect \mhh scale. To account for this, the \mhh shape uncertainties are implemented with alternative background templates built with parameters proportional to \mhh (``scale'') and $1/\mhh$ (``inverse scale''), as described in the beginning of Section~\ref{sec:syst}.
For the \qgn background, a pair of these nuisance parameters is included for each category in the SL channel and for each \PQb tagging category in the DL channel. For the \ttbar-dominated backgrounds, we include a pair of these nuisance parameters for each search category. Furthermore, to allow the \ttbar-dominated backgrounds to be anticorrelated, we include nuisance parameters for the relative \mhh scale (alternative templates built with factors proportional to \mhh) for each \PQb tagging category, separately for the SL and DL channels.

The \qgn and \losttwn backgrounds are nonresonant in \mbb, and thus alternative templates are also used to encode the shape uncertainties for the \mbb dimension with factors proportional to \mbb or $1/\mbb$. The uncertainties account for mismodeling in the simulated jet energy scale and resolution. For both of these nonresonant backgrounds, the \mbb shape does not depend on the lepton flavor or the \hww~purity, and so there is a pair of nuisance parameters for each background and each \hbbjet tagging category, separately for the SL and DL channels.

For the \mwn and \mtn backgrounds in the \mbb dimension, where resonances are constructed using AK8 SD jets, the jet mass uncertainties are dependent on the jet substructure. Because of this, the jet mass uncertainties for the signal and the \mwn background, respectively from the two-prong decays \hbb and \wqq, are correlated. This is the only such instance where signal and background are correlated, sharing nuisance parameters. Uncertainties that have been measured in data for \PW boson decays into merged jets in \ttbar events are found to cover discrepancies between our simulation and data for the \mwn background but not for the \mtn background. We do not expect these uncertainties to cover discrepancies in the \mtn background because the SD algorithm behaves differently for the three-prong top quark jets ($\PQt \to \PQb\qqpr$) in this background. Thus, these uncertainties are larger than for two-prong jets and are not correlated with the \mwn jet mass shape uncertainties, as seen in the upper two rows of Table~\ref{tab:bkg_uncertainties}.

\subsection{Signal uncertainties \label{sec:signalUncs}}

As shown in Table~\ref{tab:sig_uncertainties}, uncertainties are applied to the normalization of the signal to account for mismeasurements in the total integrated luminosity~\cite{CMS:2021xjt,CMS-PAS-LUM-17-004,CMS-PAS-LUM-18-002}, the pileup profile, the trigger efficiency, the lepton selection efficiencies, and other detector effects. Signal acceptance uncertainties from the choices of the PDFs and also the factorization and renormalization scales are also applied. The scale uncertainties are obtained following Refs.~\cite{Cacciari:2003fi,Catani:2003zt}, and the PDF uncertainty is evaluated using the NNPDF 3.1 PDF set~\cite{Ball:2017mu}.

\begin{table}[ht]
\centering
\topcaption{\label{tab:sig_uncertainties} Signal systematic uncertainties included in the maximum likelihood fit. The $N_{\text{p}}$ column indicates the number of nuisance parameters used to model the uncertainty. In the ``Uncertainty values'' column, some uncertainties are noted as affecting both the yield ($Y$) and \mhh shape ($S$ for scale, $R$ for resolution) of the signal. All other uncertainties, except the SD jet mass uncertainties, are uncertainties in the signal yield alone.}
\begin{tabular}{lll}

Uncertainty type & $N_{\text{p}}$ & Uncertainty values \\
\hline
SD jet \mbb scale & 1 & $S(\mbb)$: 0.54\% \\
SD jet \mbb resolution & 1 & $R(\mbb)$: 8.6\% \\
Integrated luminosity & 1 & 1.6\% \\
PDFs+scales & 1 & spin-0: 2.0\%, spin-2: 2.5\%  \\
Trigger & 6 & 1\Pell: 2.0\%, 2\Pell: 3.0\%  \\
Pileup & 1 & 1\Pell: 1.0\%, 2\Pell: 0.6\% \\
Electron reconstruction \quad & 1 & 1\Pell: 0.5\%, 2\Pell: $<$0.8\%  \\
Electron identification & 2 & 1\Pell: 4.2\%, 2\Pell: $<$2.6\%  \\
Muon identification & 2 & 1\Pell: 2.3\%, 2\Pell: $<$2.3\%  \\
Electron isolation & 1 & 1\Pell: 6\%, 2\Pell: 3\% for each electron  \\
Muon isolation & 1 & 1\Pell: 6\%, 2\Pell: 2\% for each muon  \\
Jet energy scale & 1 & $Y$: 2\%, $S(\mhh)$: 0.8\%, $R(\mhh)$: 3\% \\
Jet energy resolution & 1 & $Y$: 0.5\%. $S(\mhh)$: 0.3\%, $R(\mhh)$: 4\%  \\
Unclustered energy & 1 & $Y$: 0.5\%, $S(\mhh)$: 0.1\%, $R(\mhh)$: 1.5\%  \\
Other detector effects & 2 & $Y$: 0.6\%, $R(\mhh)$: 1.0\%  \\
AK4 \PQb tag efficiency & 1 & $<$4.0\%  \\
AK4 \PQb tag misidentification rate & 1 & $<$2.5\%  \\
\hbbjet tagging & 1 & \bL: 8.5\%, \bT: 11.5\% \\
\wqqjet \tauTO efficiency & 1 & LP: 26\% HP: 6.7\% \\
\end{tabular}
\end{table}

The signal acceptance and the \mhh resonance scale and resolution all have uncertainties due to the jet energy scale and resolution, the unclustered energy resolution, and other detector effects. The same \mbb resonance scale and resolution uncertainties that are applied for the \mwn background are applied to the signal because they are both SD jets with two-prong substructure.

The \wqqjet \tauTO selection efficiency is measured in a \ttbar data sample enriched with hadronically decaying \PW bosons. The uncertainties in this measurement are included as normalization uncertainties in the \hww decay purity categories, and the LP and HP uncertainties are anticorrelated. Normalization uncertainties are also applied to account for the efficiency and misidentification rate of AK4 jet \PQb tagging used to identify and reject jets from \ttbar production. The uncertainty in the \hbbjet tagging efficiency is included as a single nuisance parameter that varies the signal normalization and is dependent on both the \hbbjet tagging category and \mx. These \hbbjet tagging uncertainties are the dominant systematic uncertainties associated with the signal normalization, followed by the uncertainties in the \tauTO efficiencies.

\section{Results\label{sec:results}}

\begin{figure}[htp]
\centering
\includegraphics[width=0.32\textwidth]{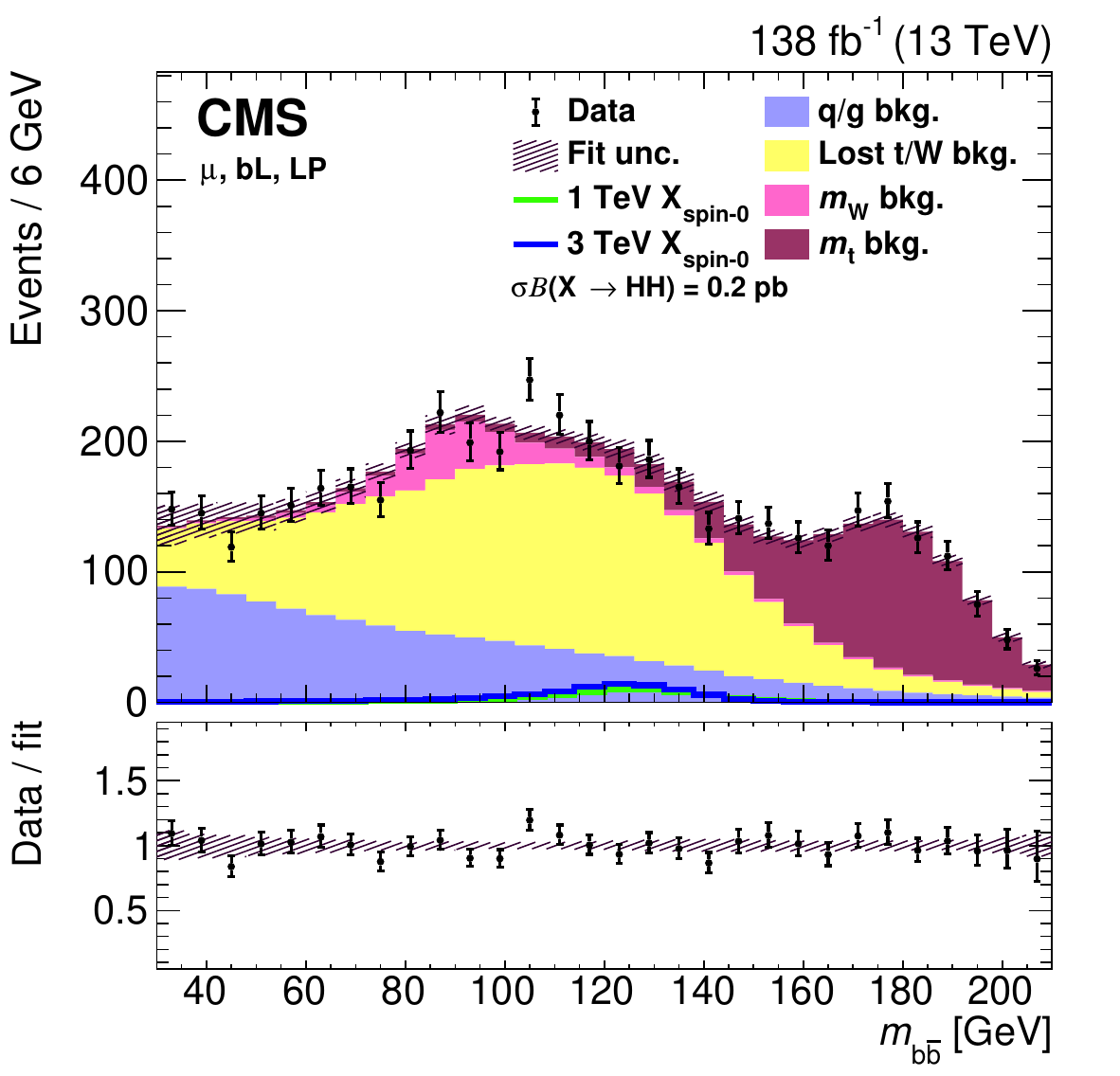}
\includegraphics[width=0.32\textwidth]{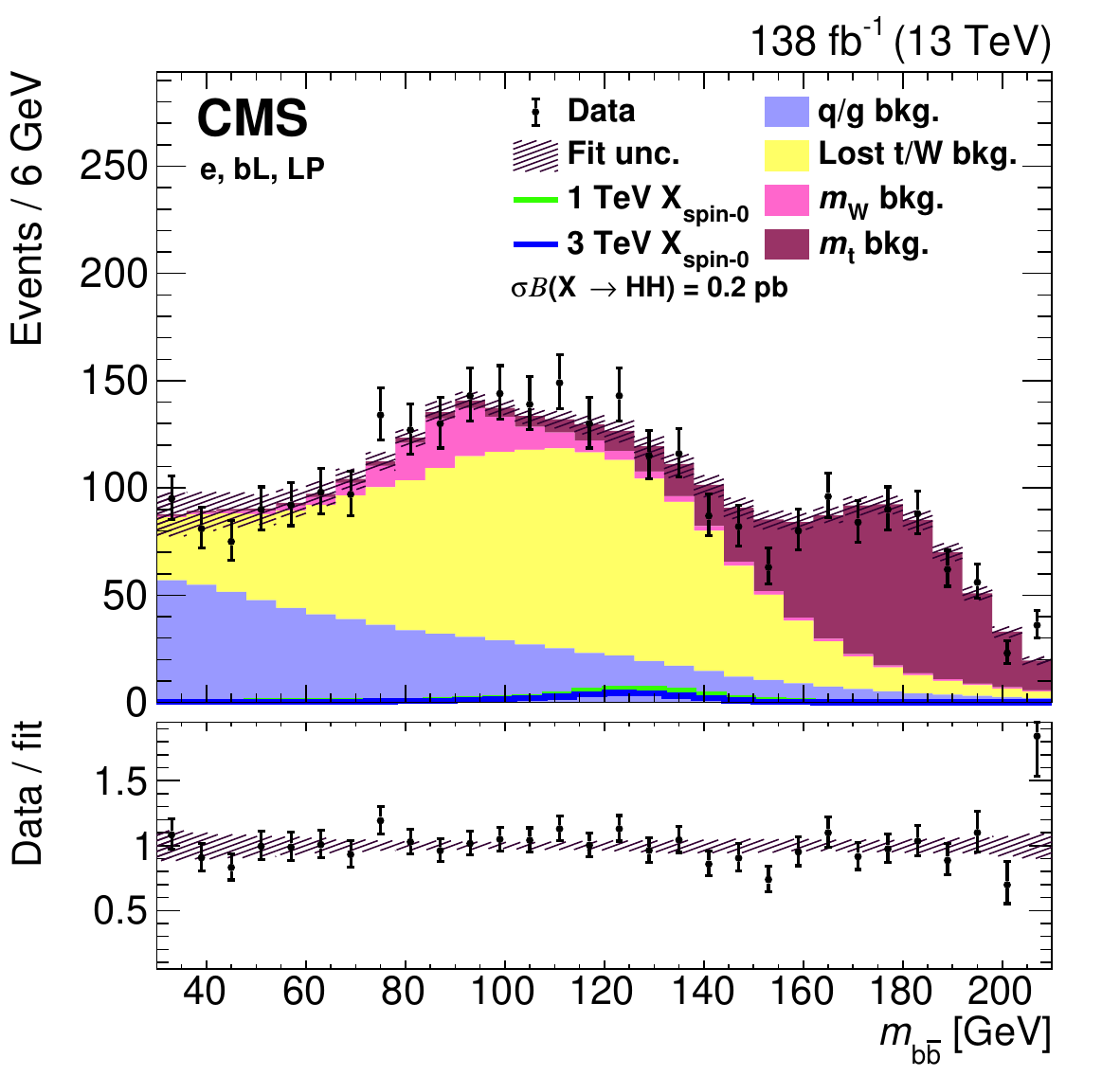}
\includegraphics[width=0.32\textwidth]{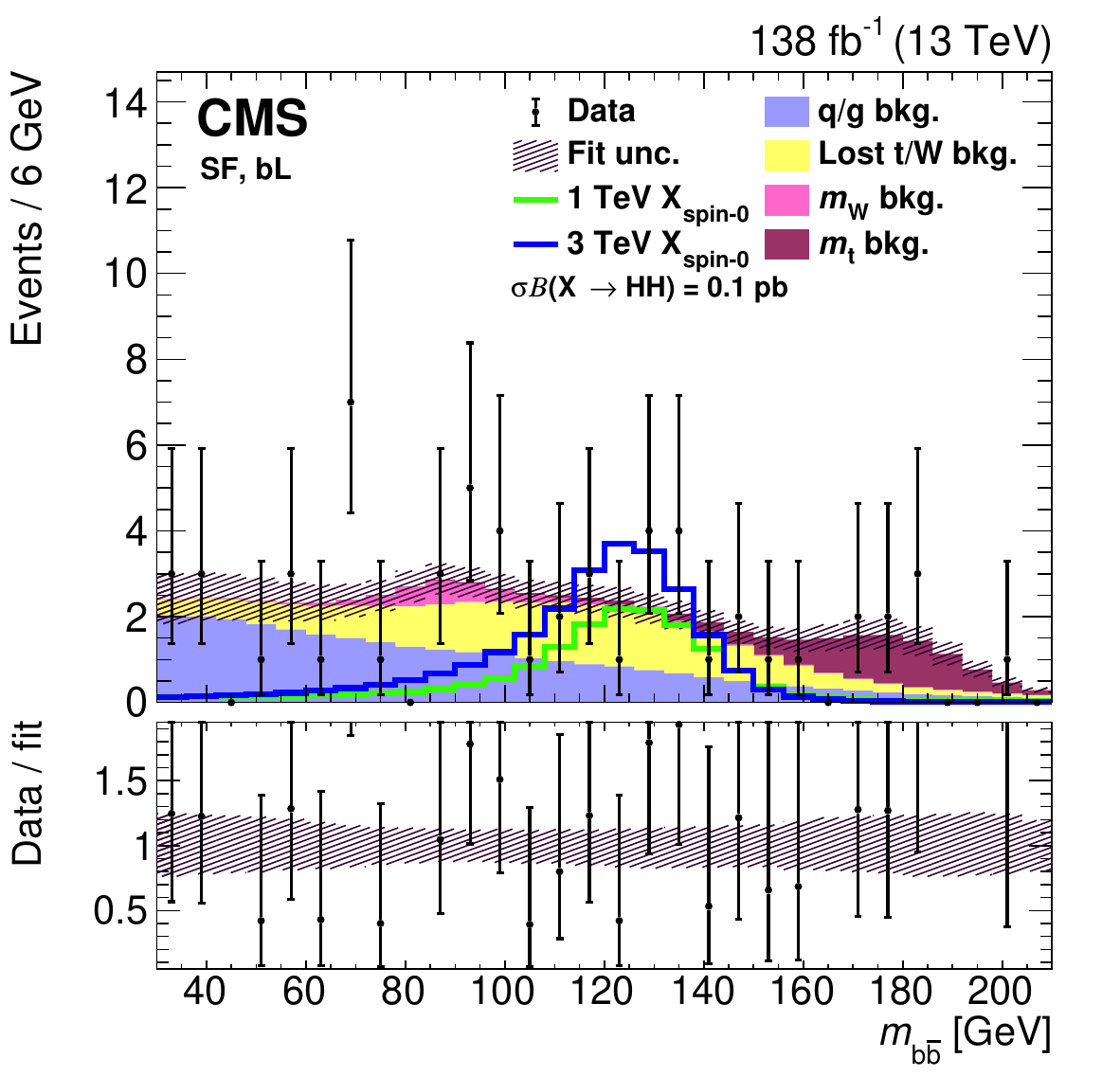}
\includegraphics[width=0.32\textwidth]{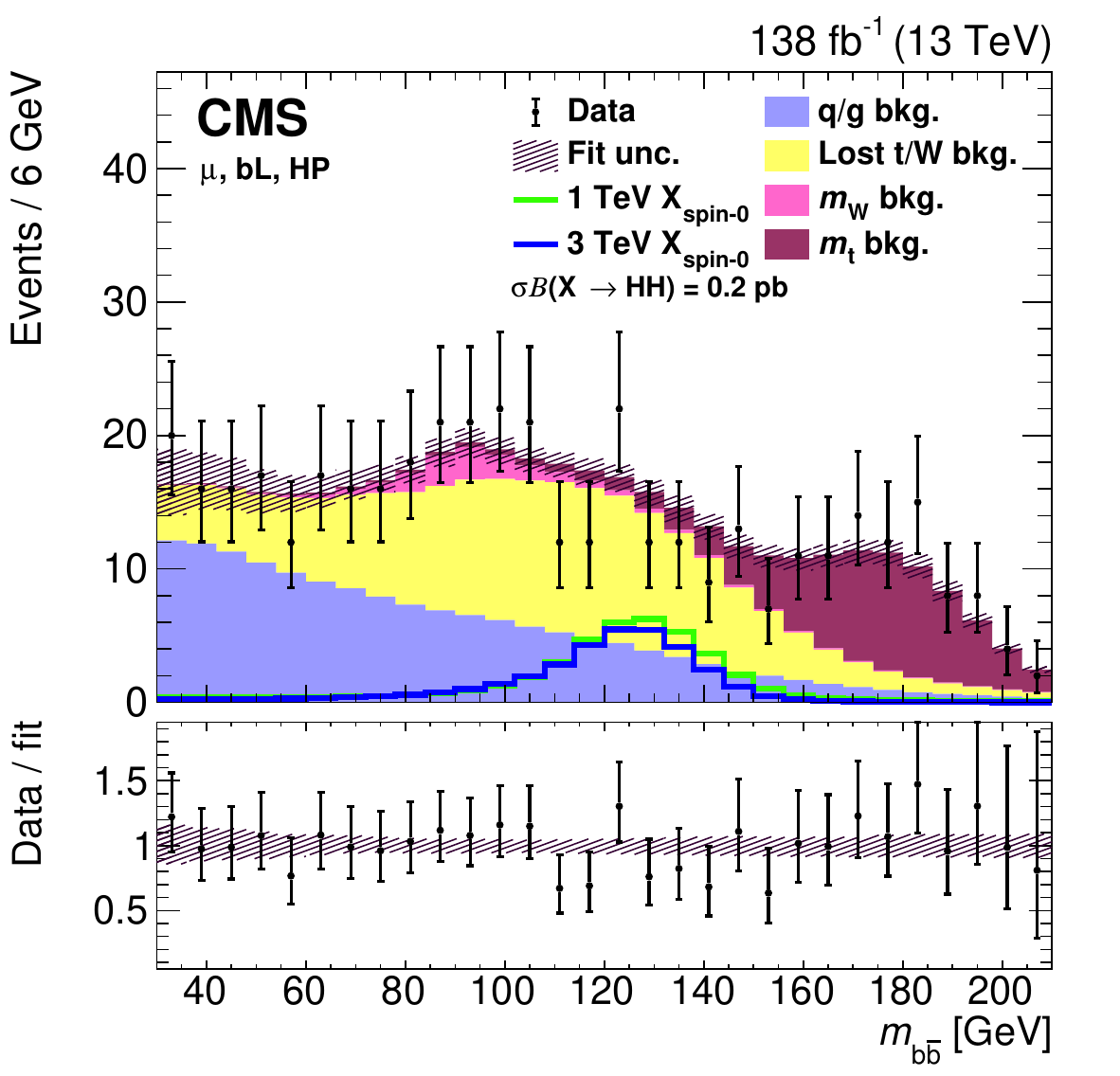}
\includegraphics[width=0.32\textwidth]{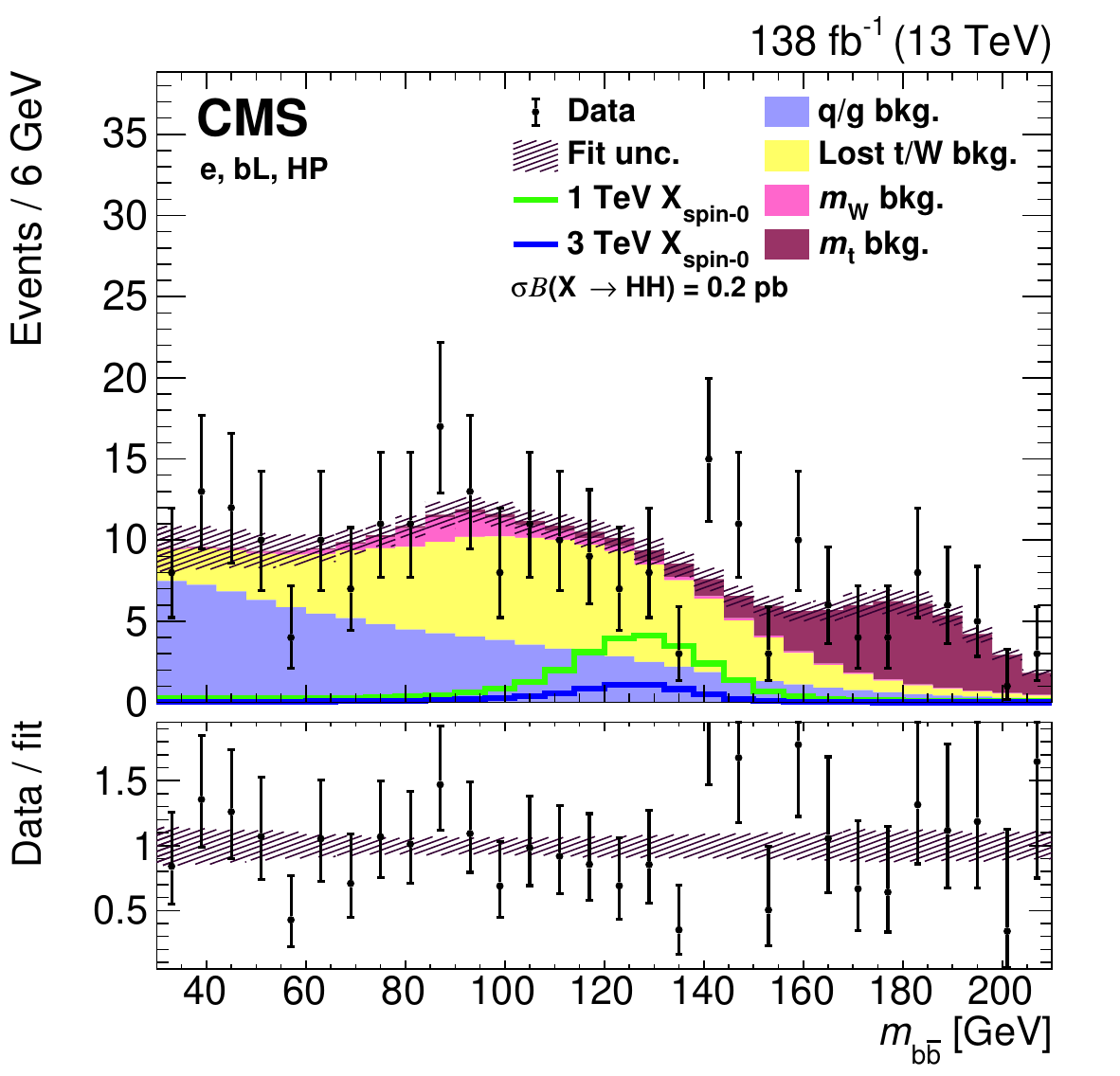}
\includegraphics[width=0.32\textwidth]{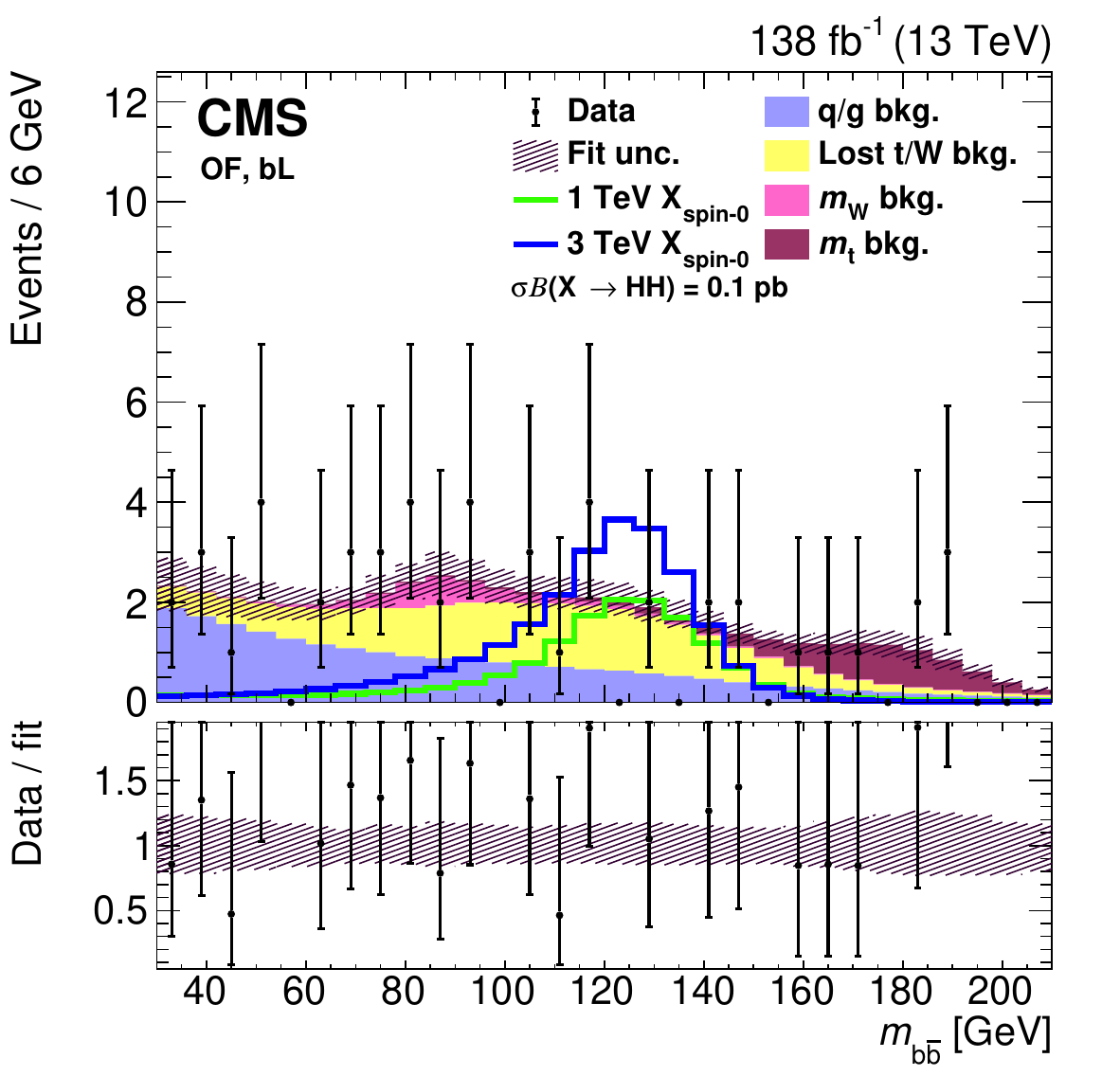}
\includegraphics[width=0.32\textwidth]{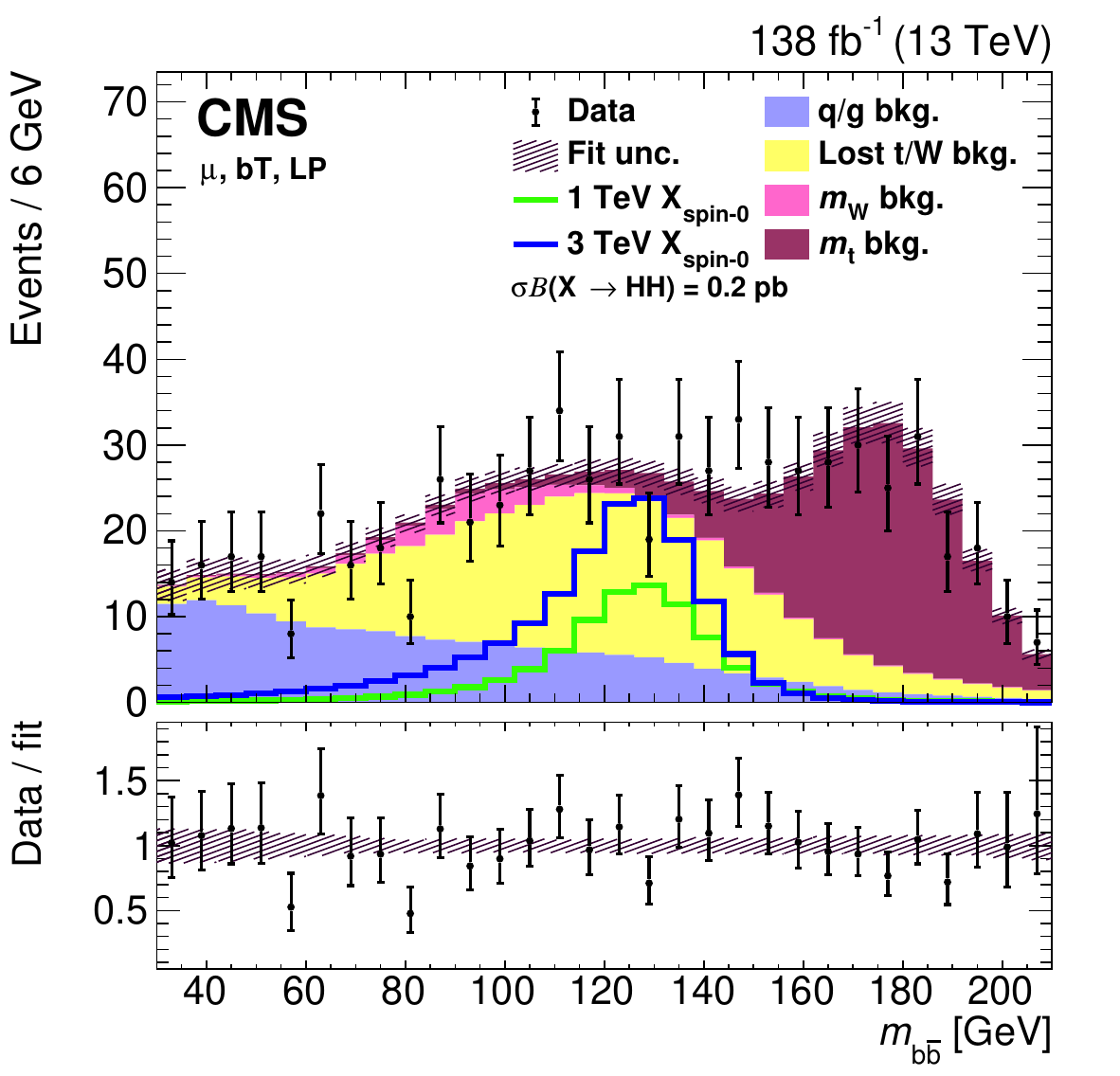}
\includegraphics[width=0.32\textwidth]{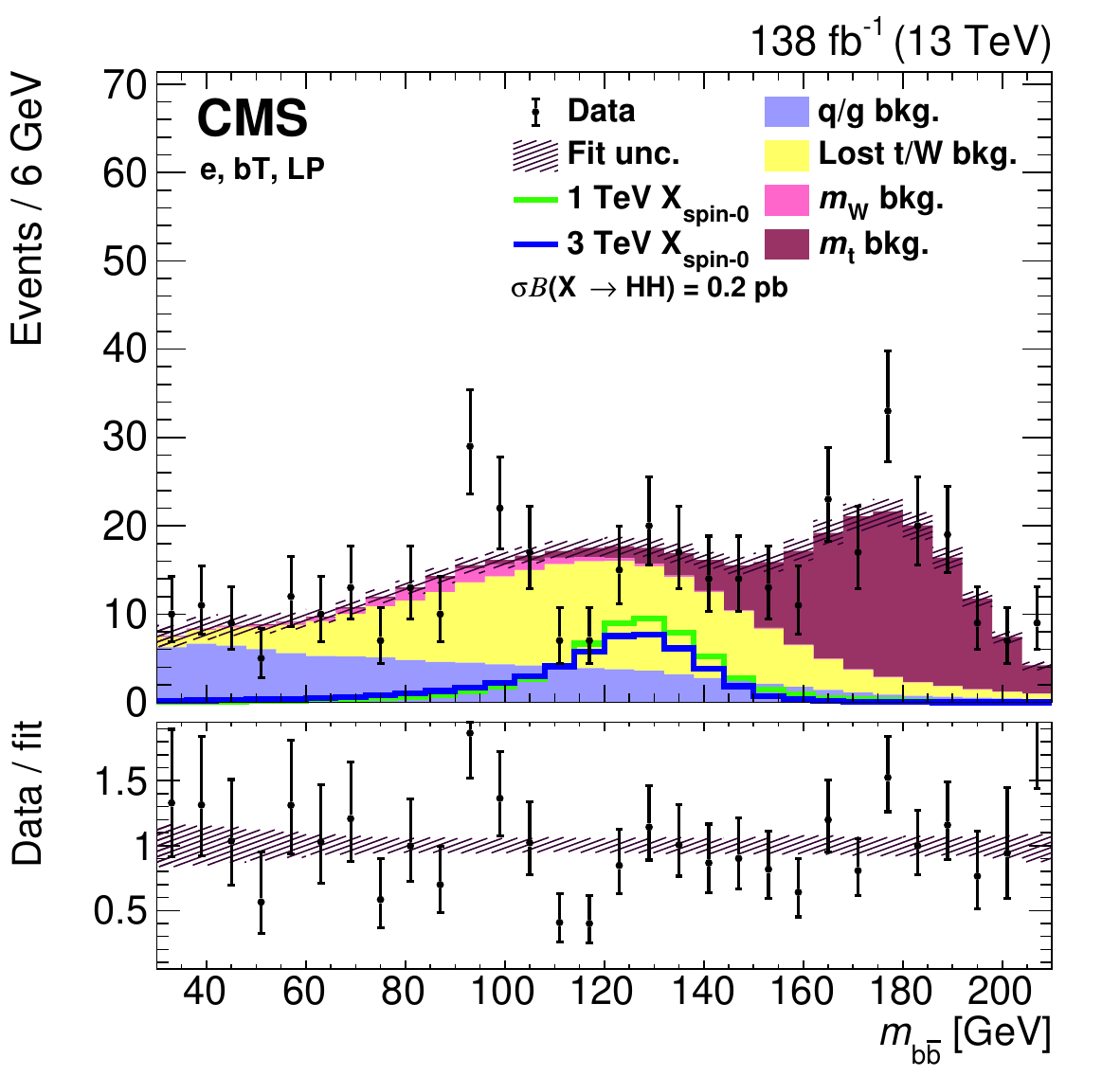}
\includegraphics[width=0.32\textwidth]{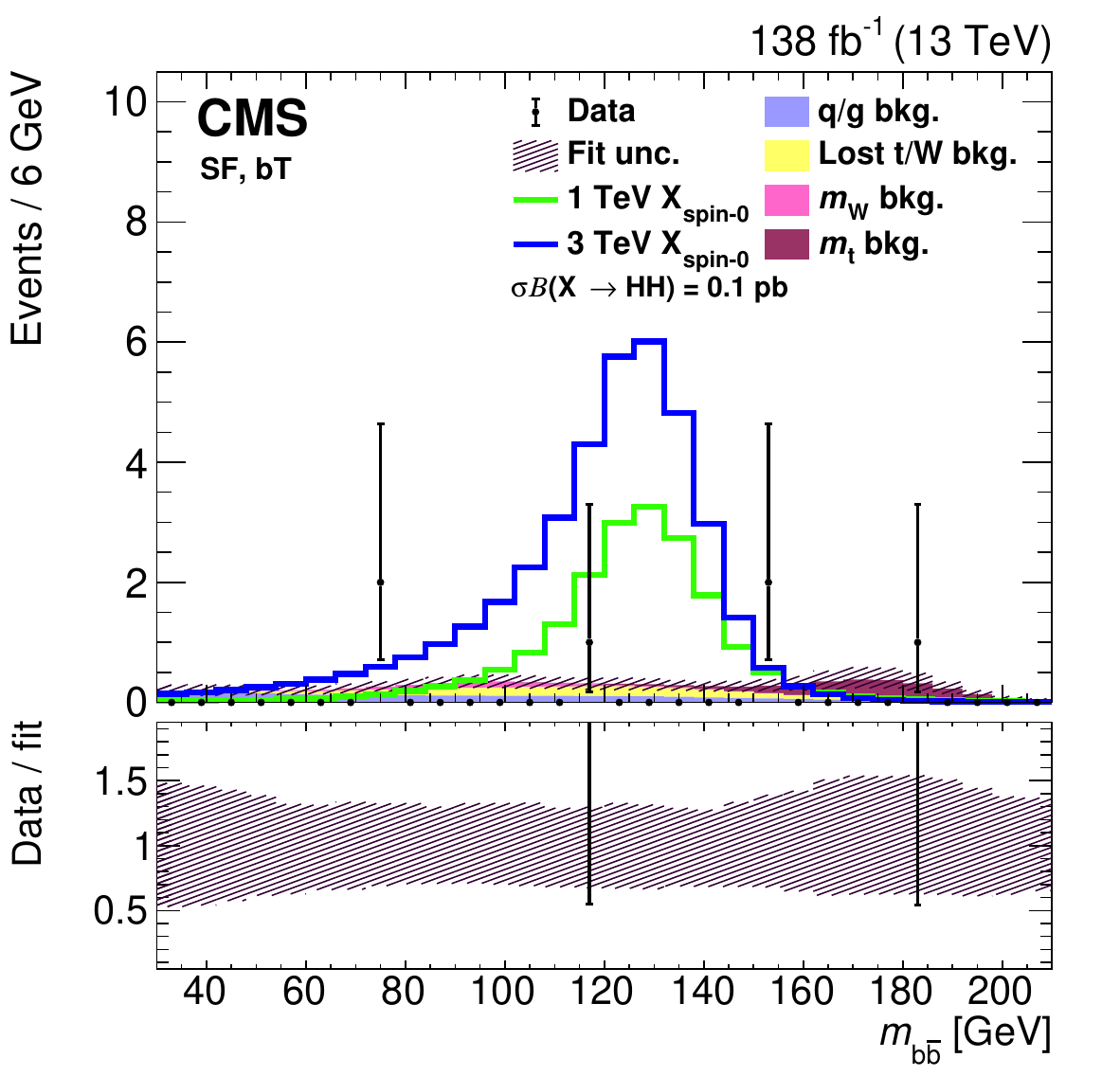}
\includegraphics[width=0.32\textwidth]{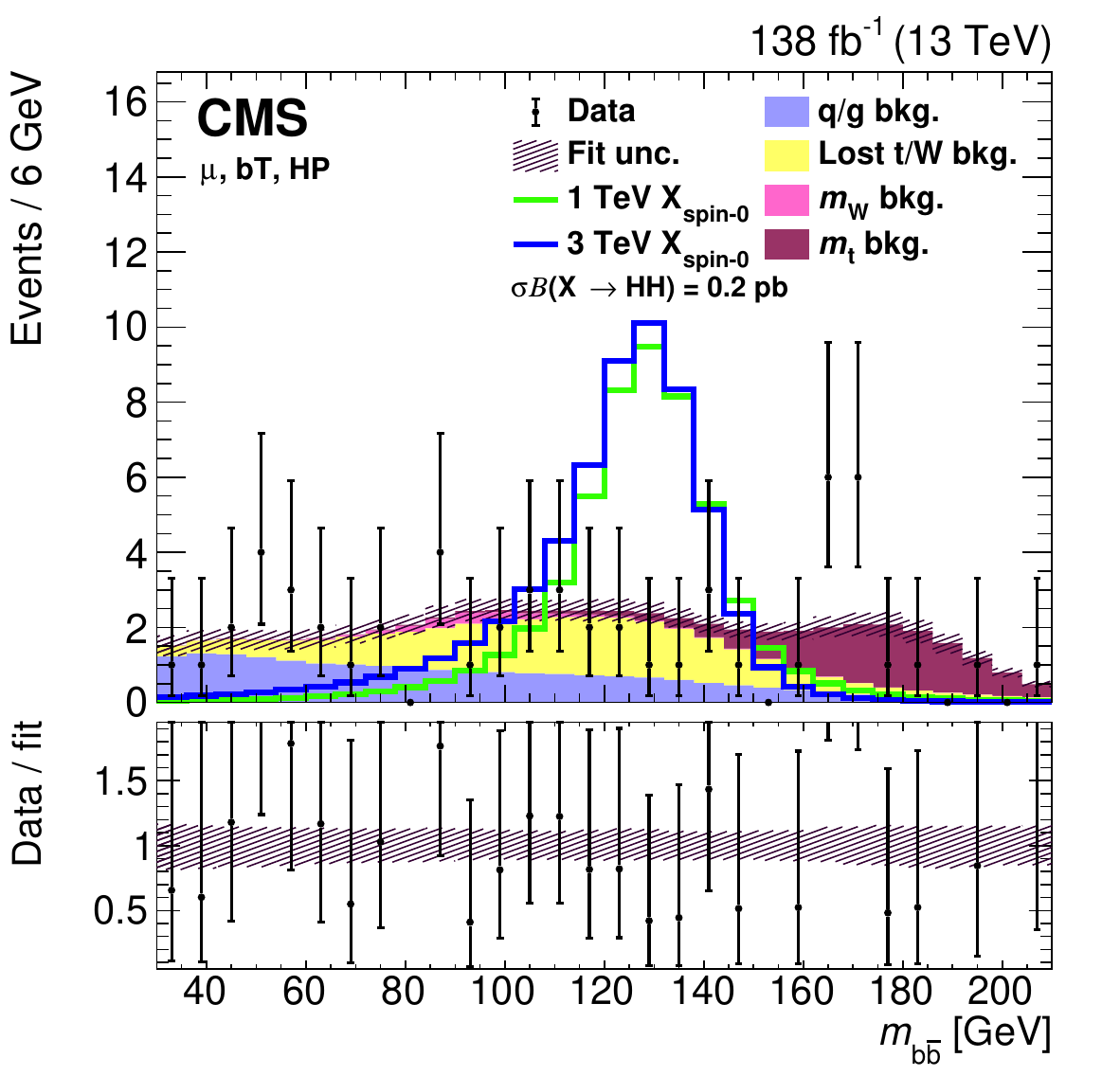}
\includegraphics[width=0.32\textwidth]{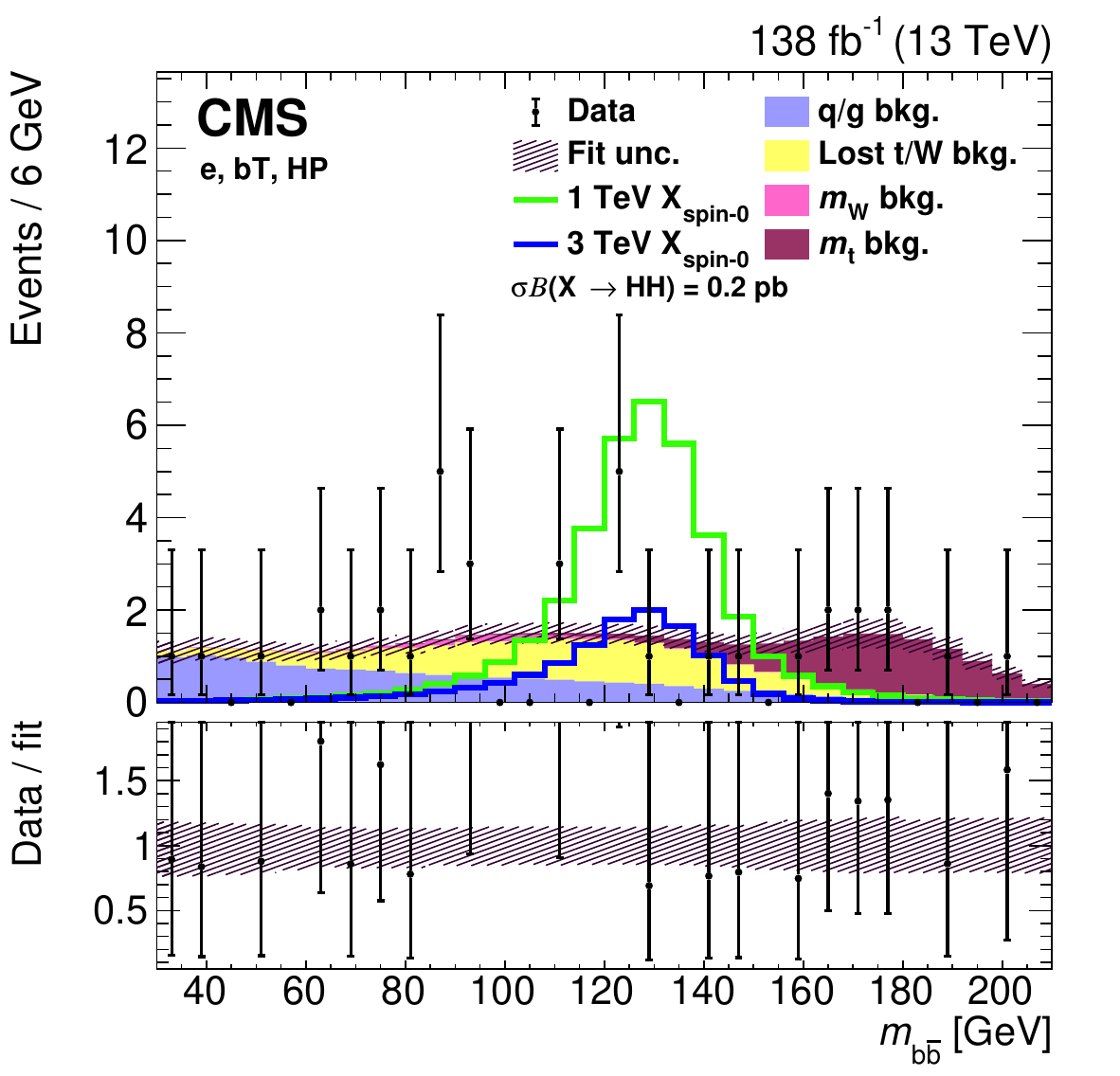}
\includegraphics[width=0.32\textwidth]{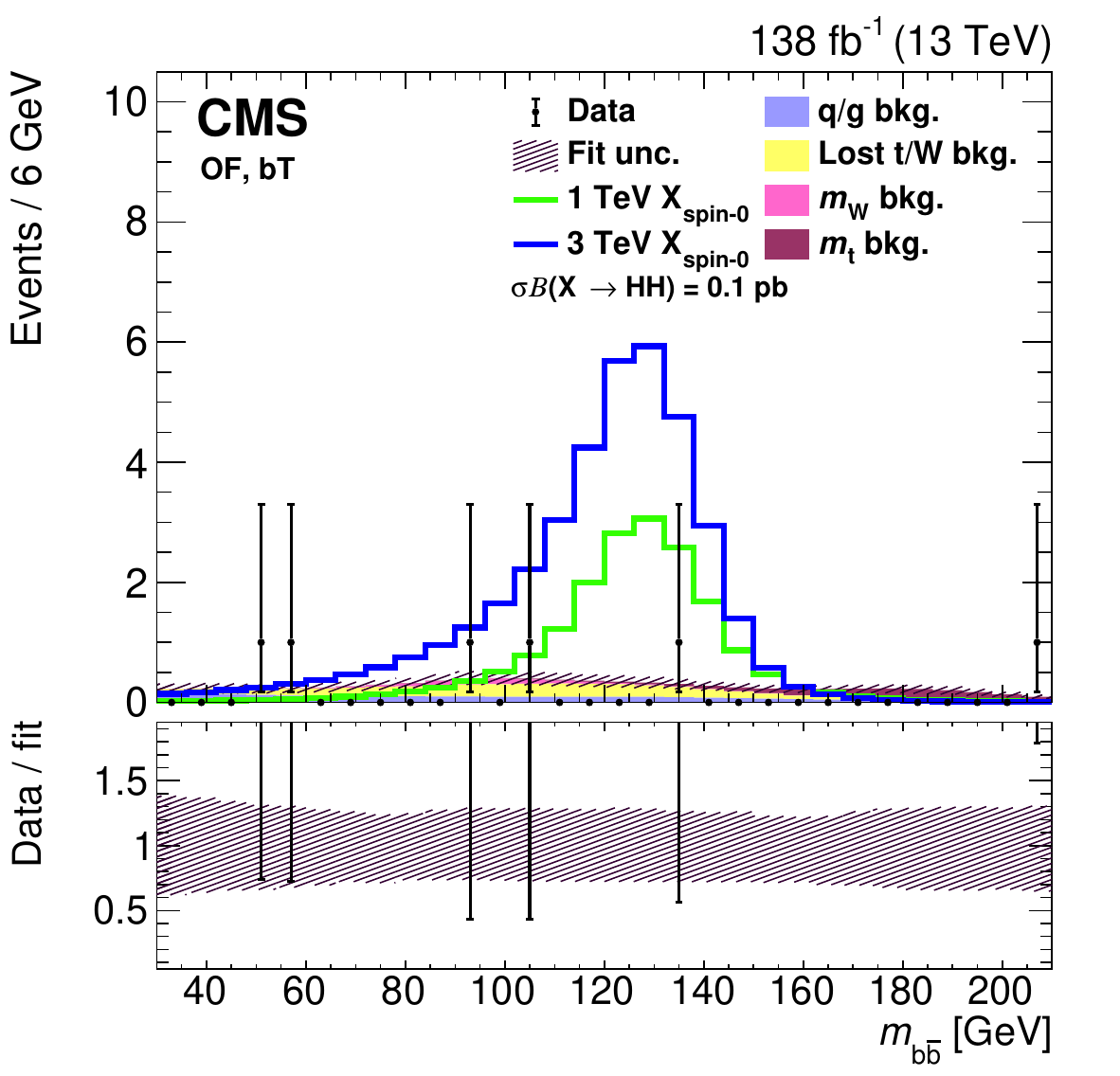}
\caption{The background-only 2D fit result compared to data projected onto the \mbb axis for both the SL and DL channels. The label for each search category is in the upper left of each plot. The fit result is the filled histogram, with the different colors indicating different background components. The background shape uncertainty from the fit is shown as the hatched band. Example spin-0 signal distributions for $\mx = 1.0$ and 3.0\TeV are shown as solid lines, with $\sigma \mathcal{B} (\PX\to\hh)$ set to 0.2 and 0.1\unit{pb} for the SL and DL channels, respectively. The lower panels show the ratio of the data to the fit result. Only nonzero data entries are shown in the interest of clarity.}
\label{fig:results_mbb}
\end{figure}

\begin{figure}[htp]
\centering
\includegraphics[width=0.32\textwidth]{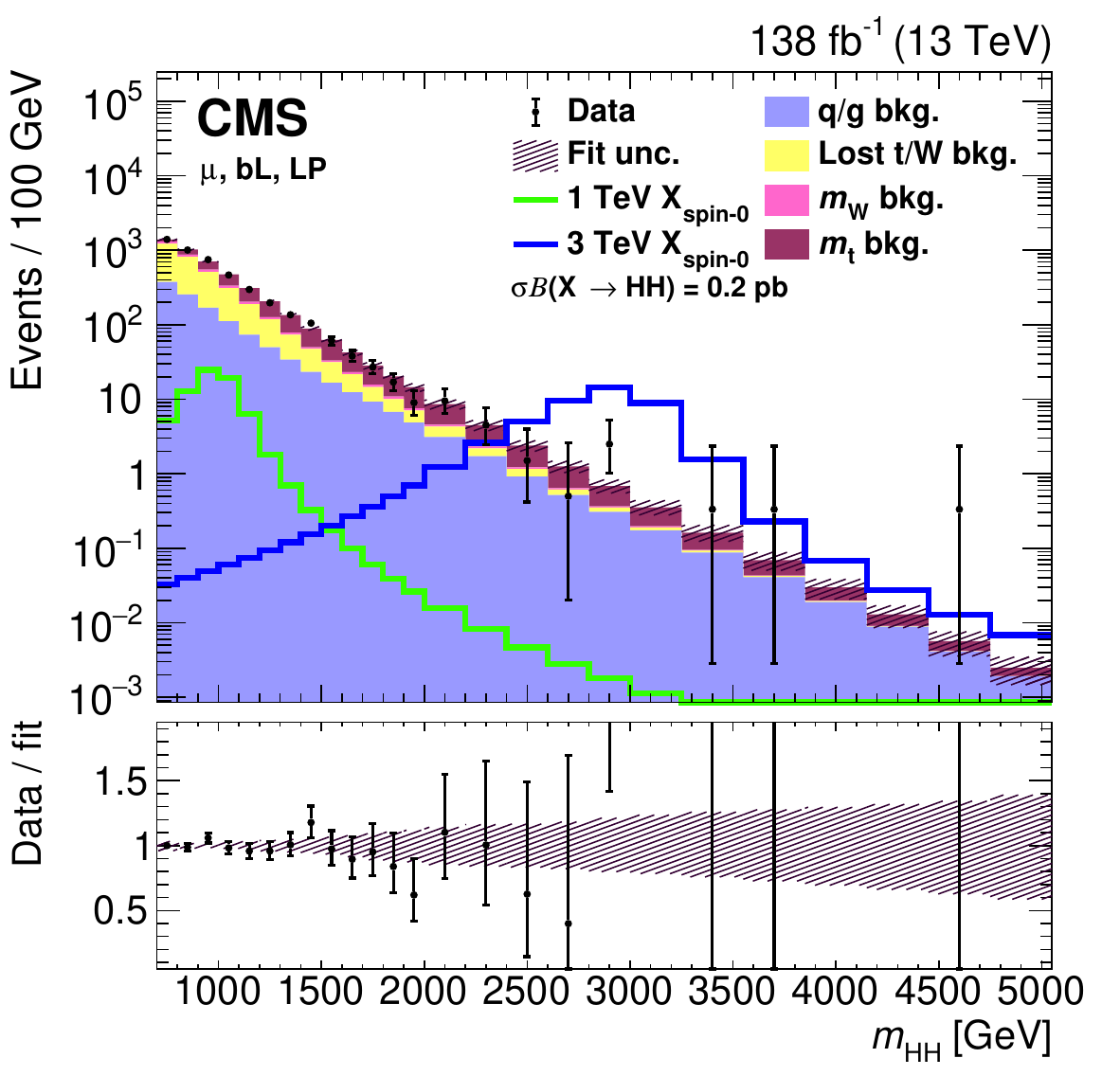}
\includegraphics[width=0.32\textwidth]{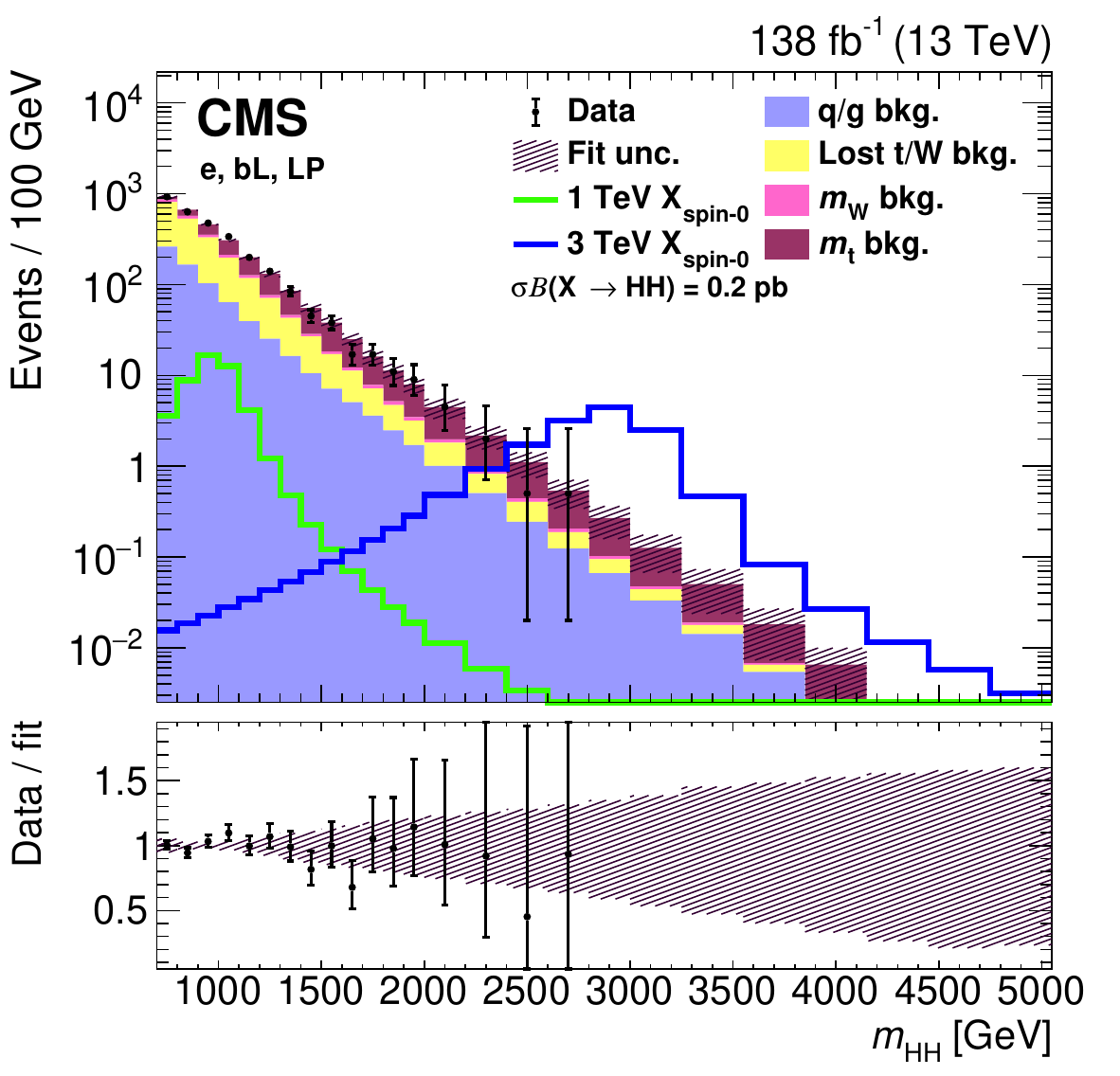}
\includegraphics[width=0.32\textwidth]{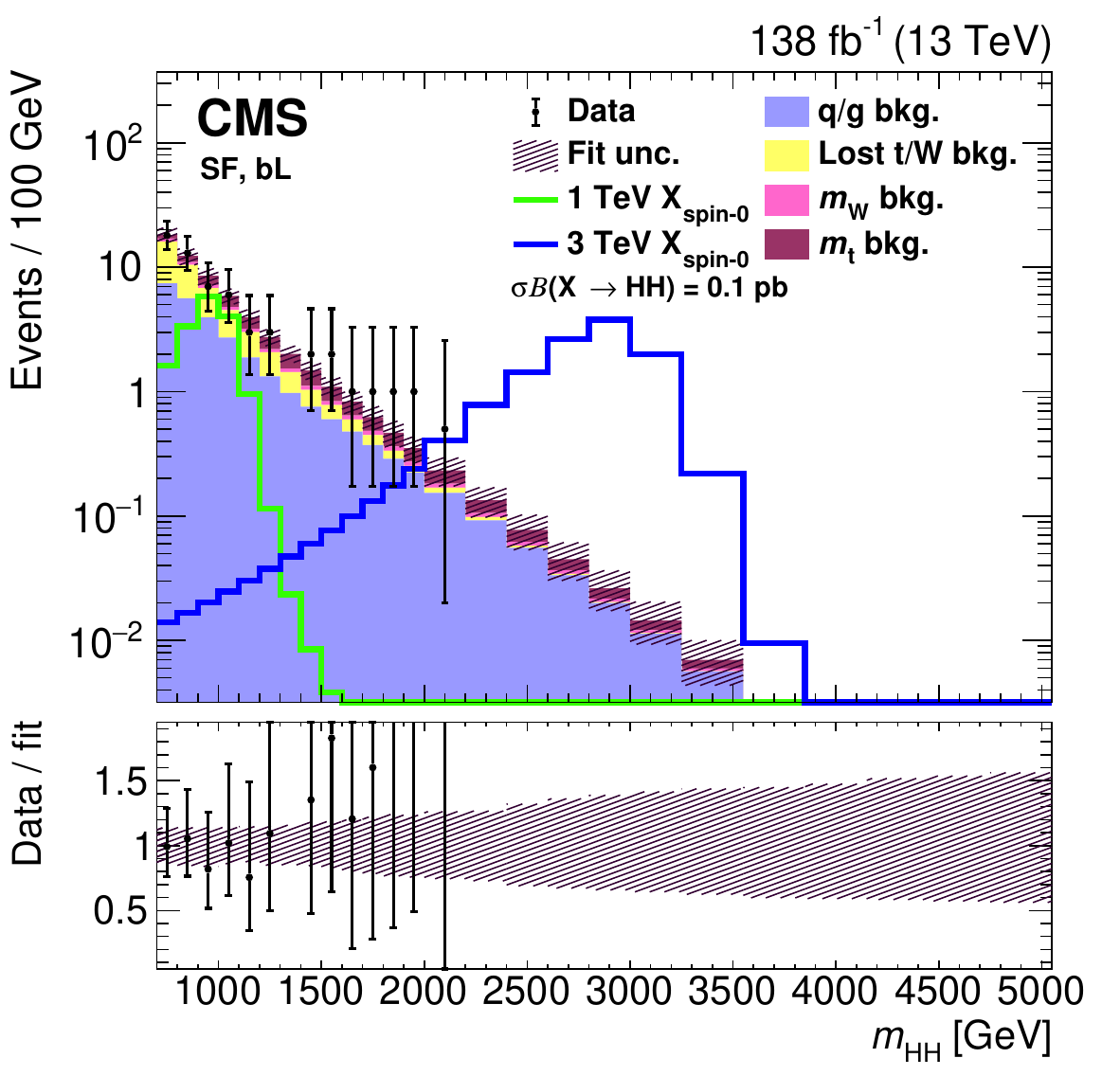}
\includegraphics[width=0.32\textwidth]{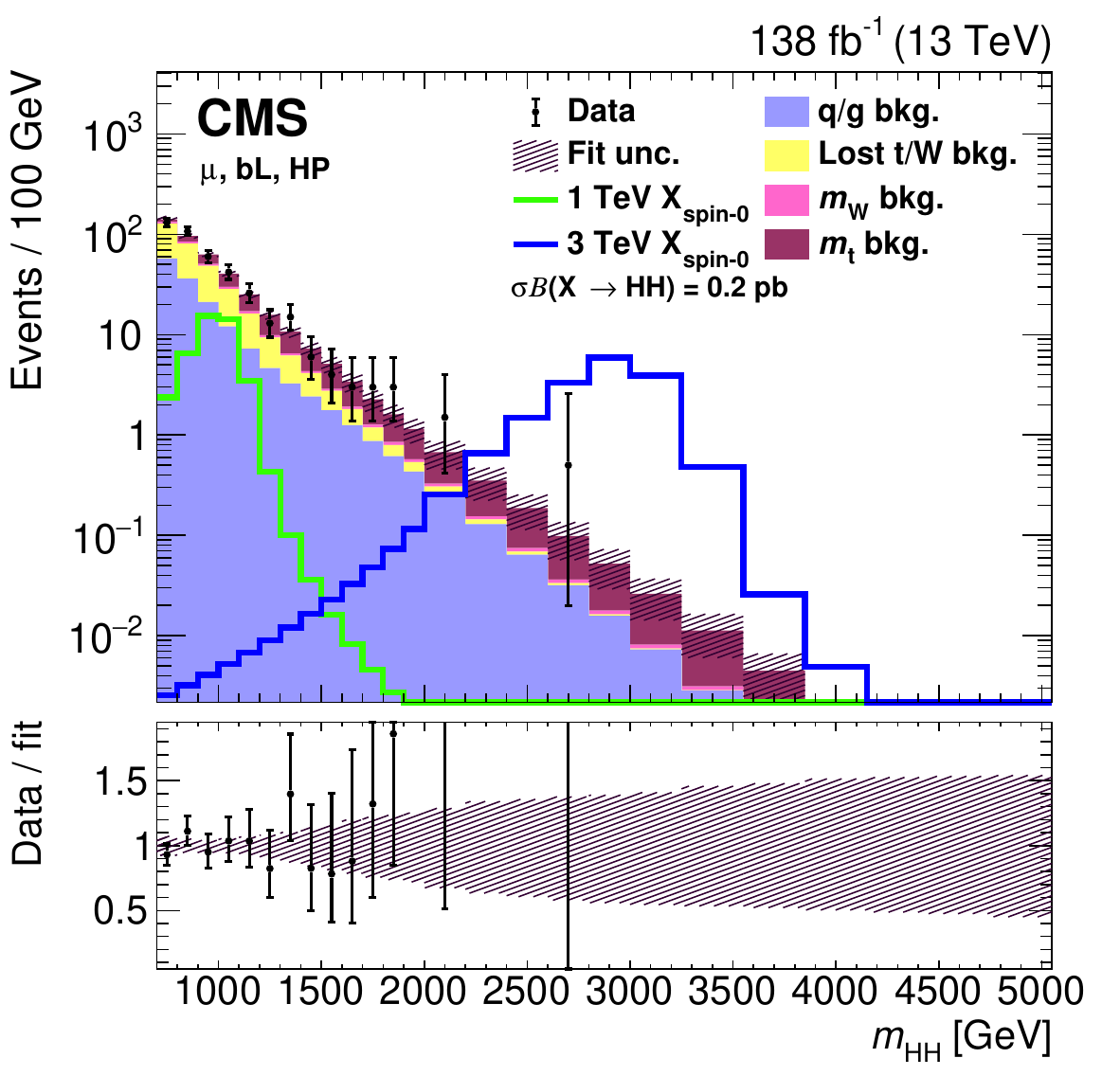}
\includegraphics[width=0.32\textwidth]{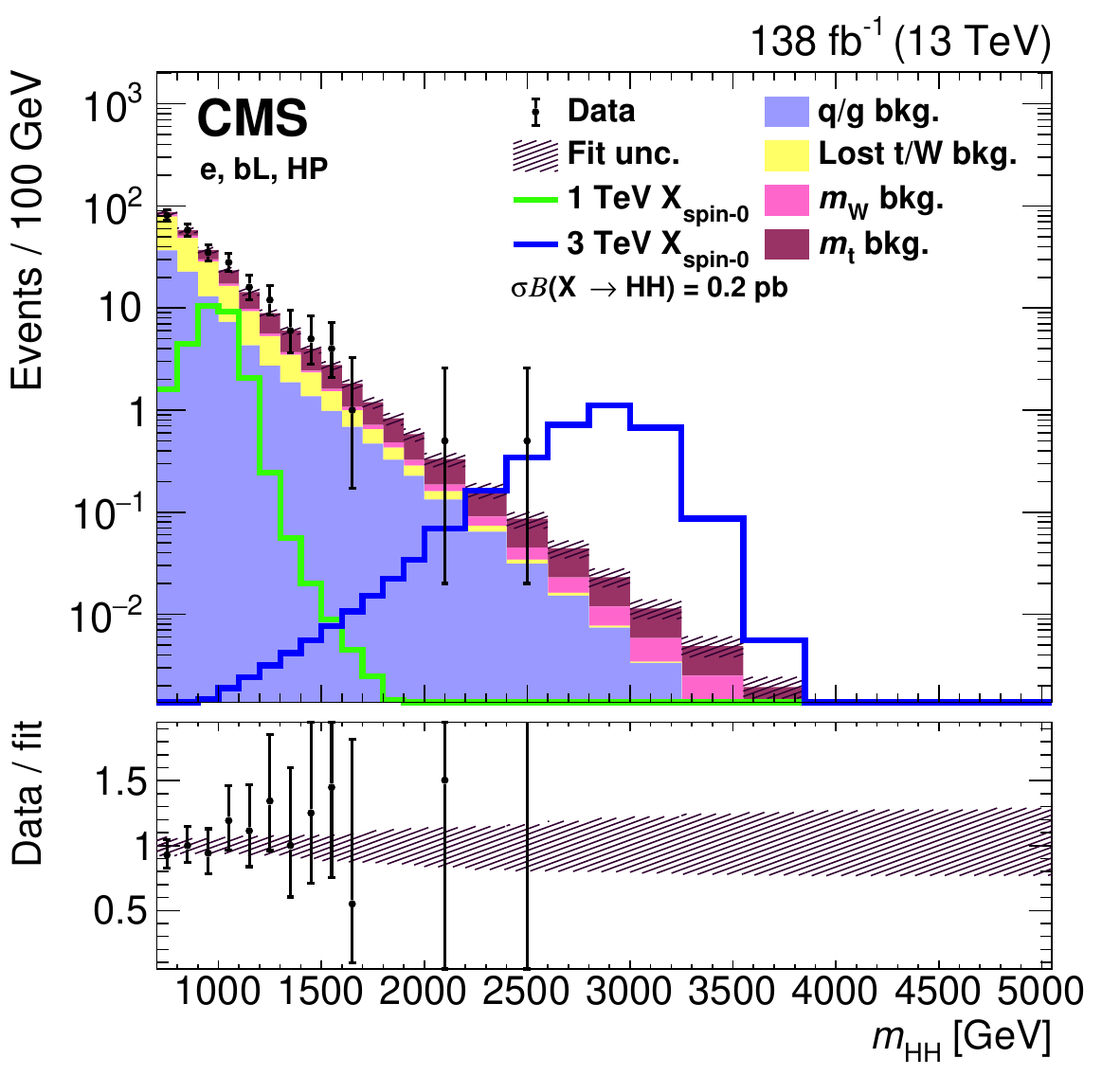}
\includegraphics[width=0.32\textwidth]{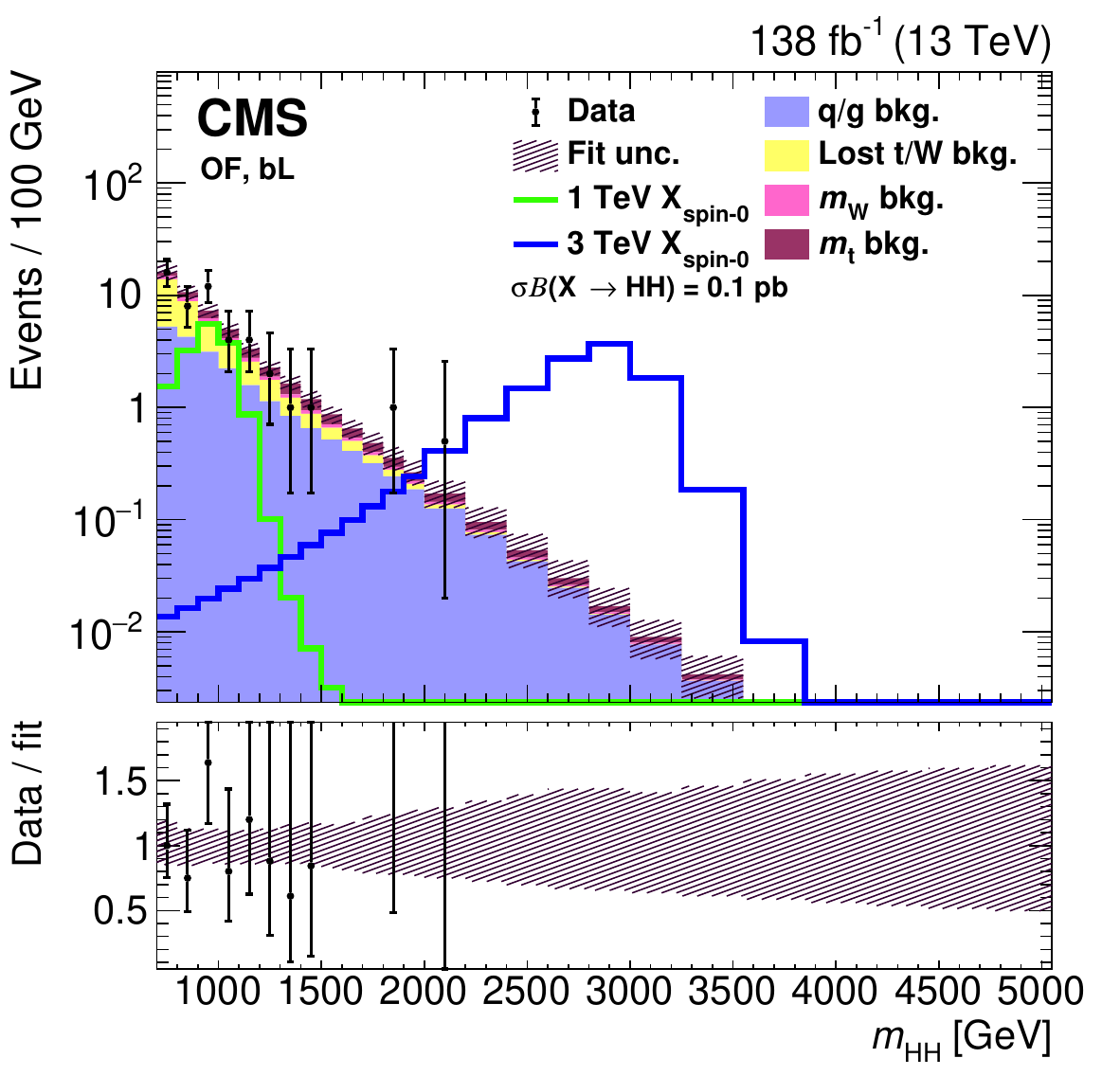}
\includegraphics[width=0.32\textwidth]{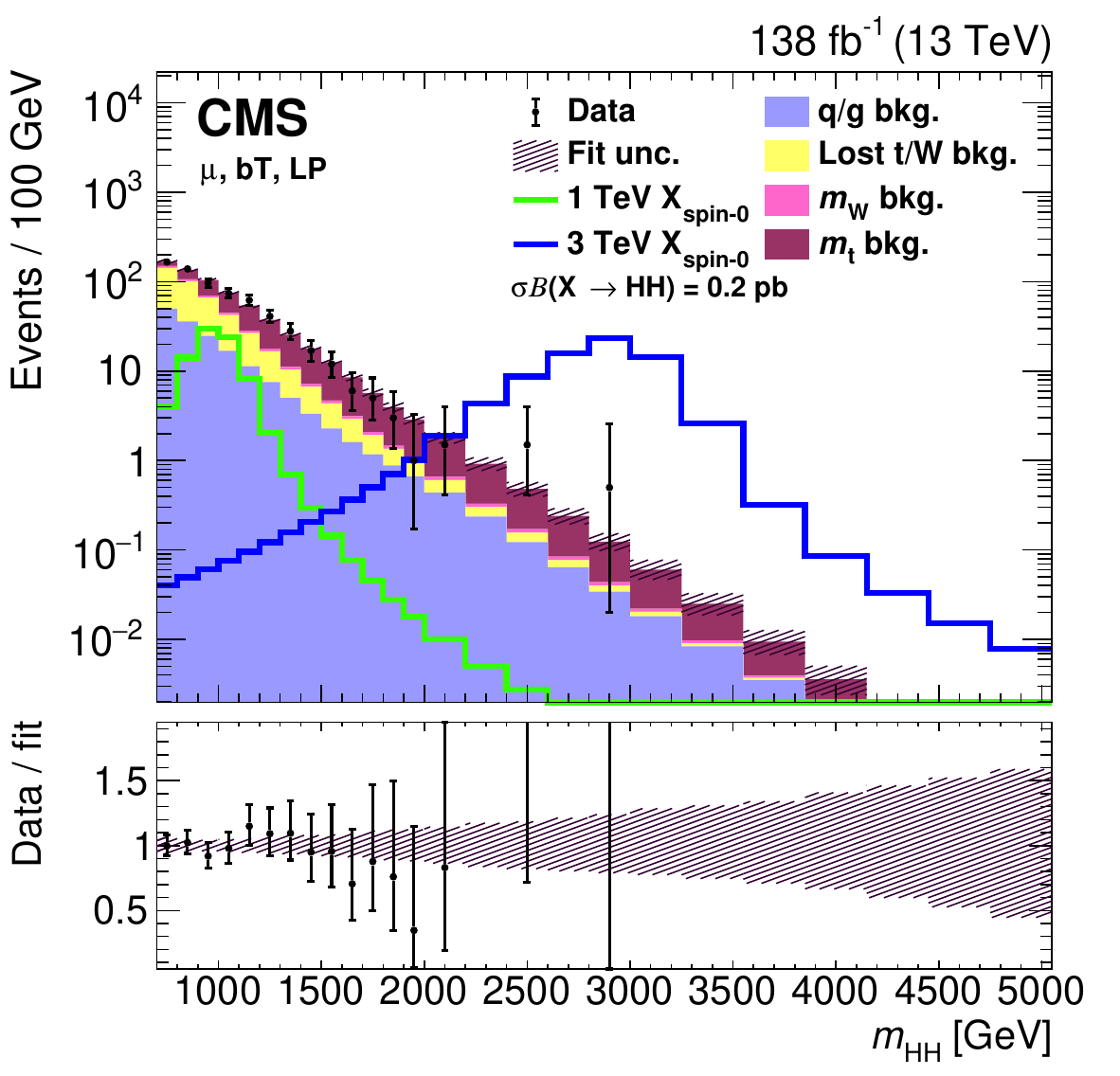}
\includegraphics[width=0.32\textwidth]{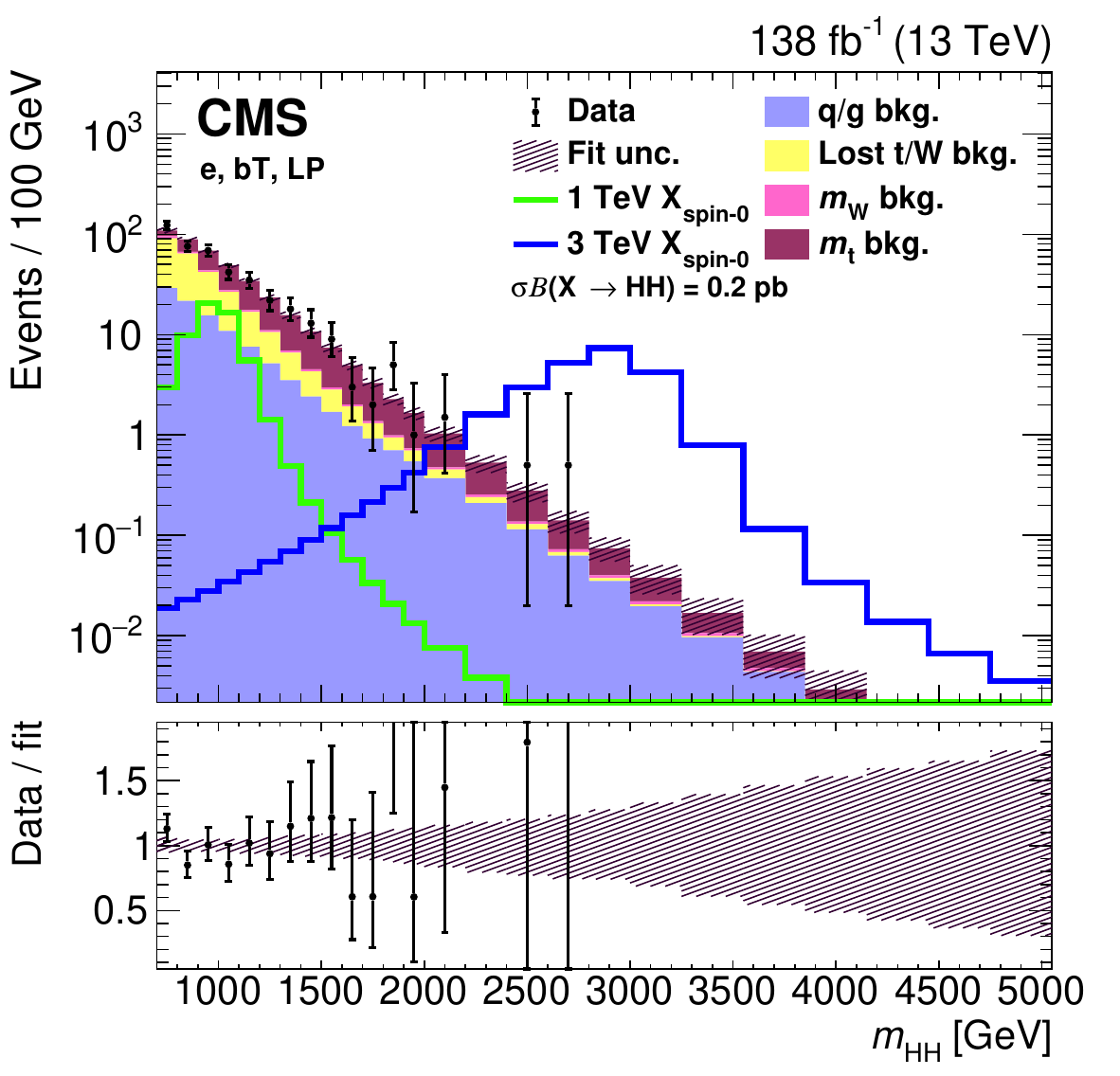}
\includegraphics[width=0.32\textwidth]{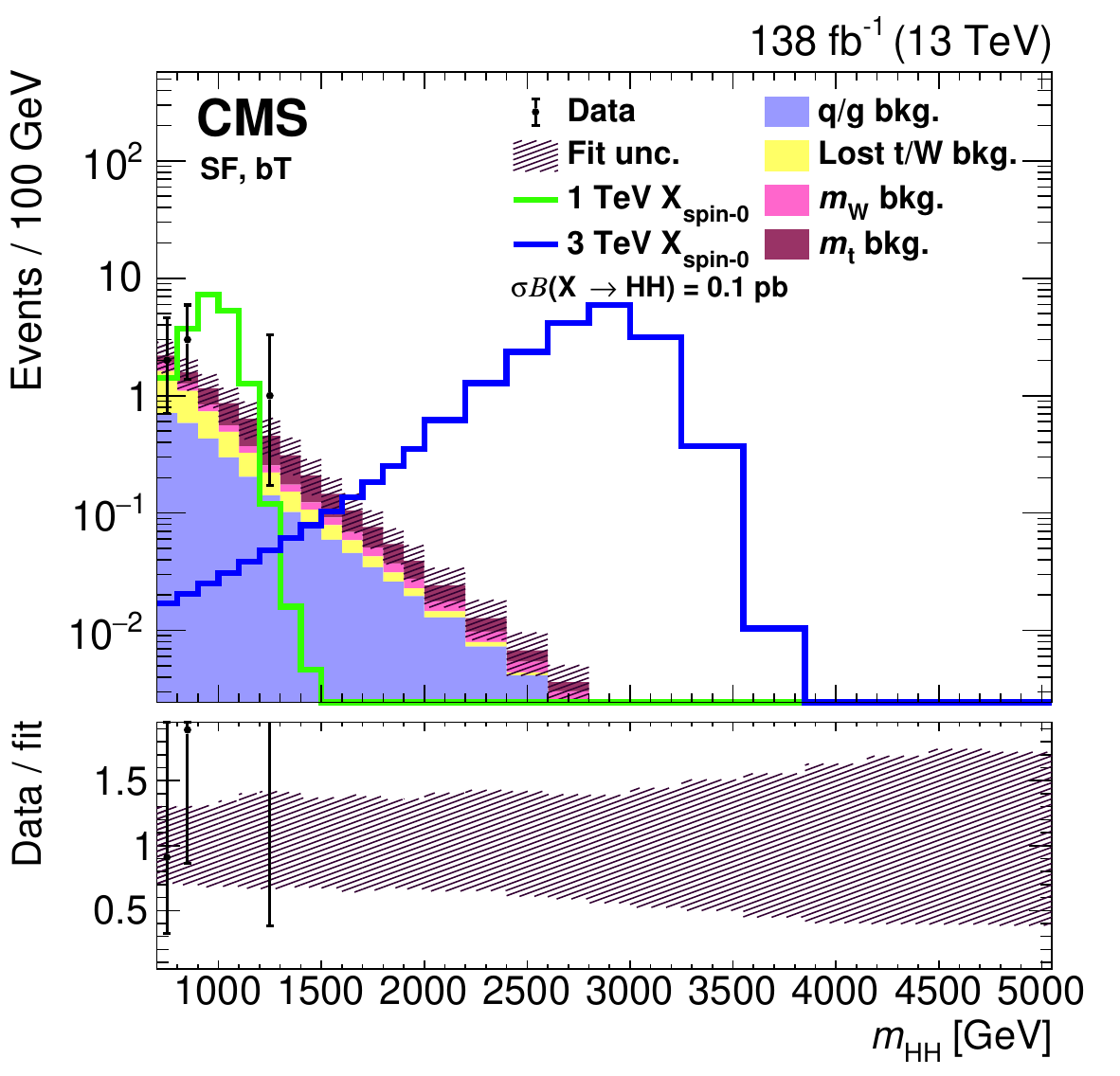}
\includegraphics[width=0.32\textwidth]{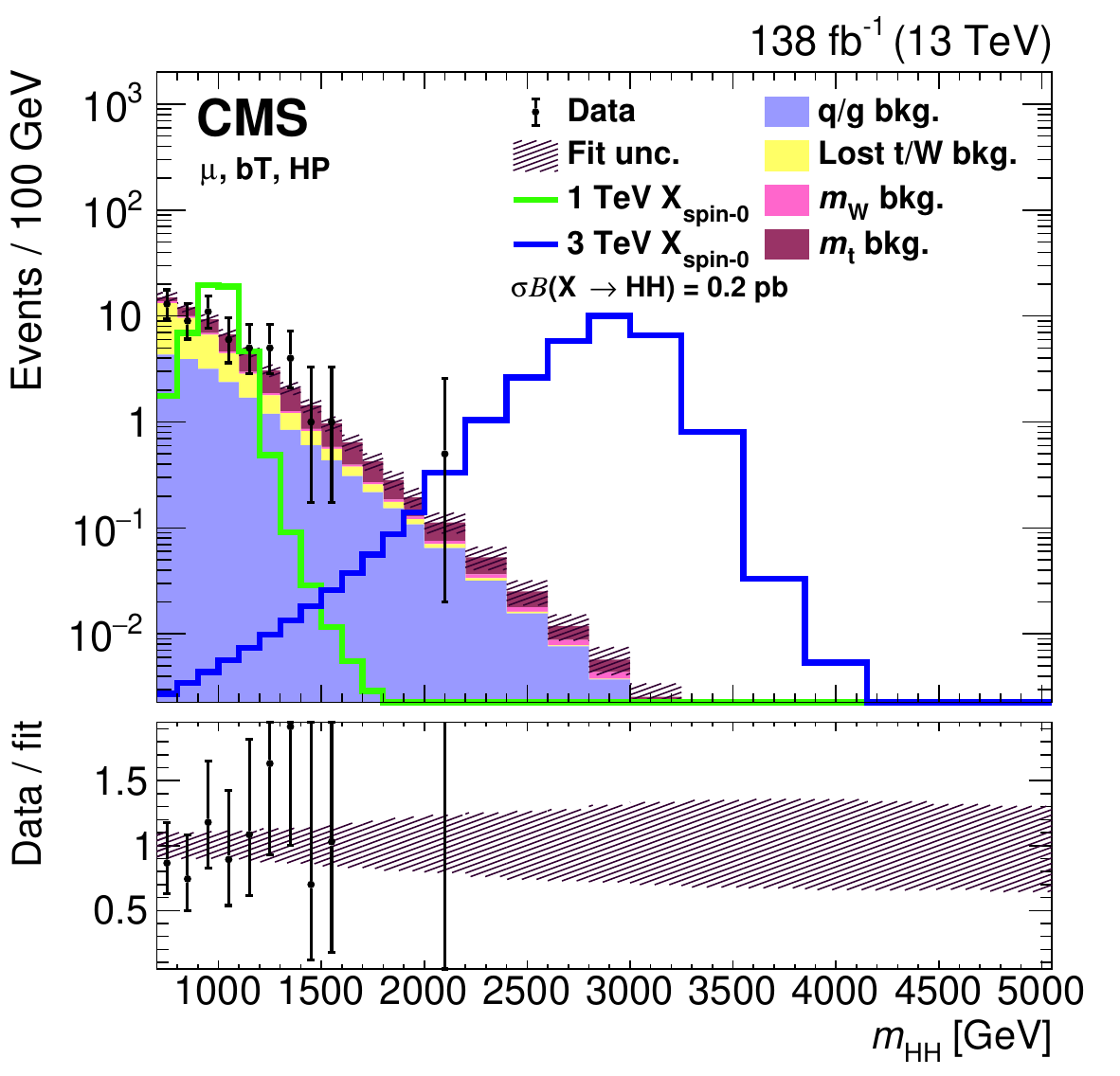}
\includegraphics[width=0.32\textwidth]{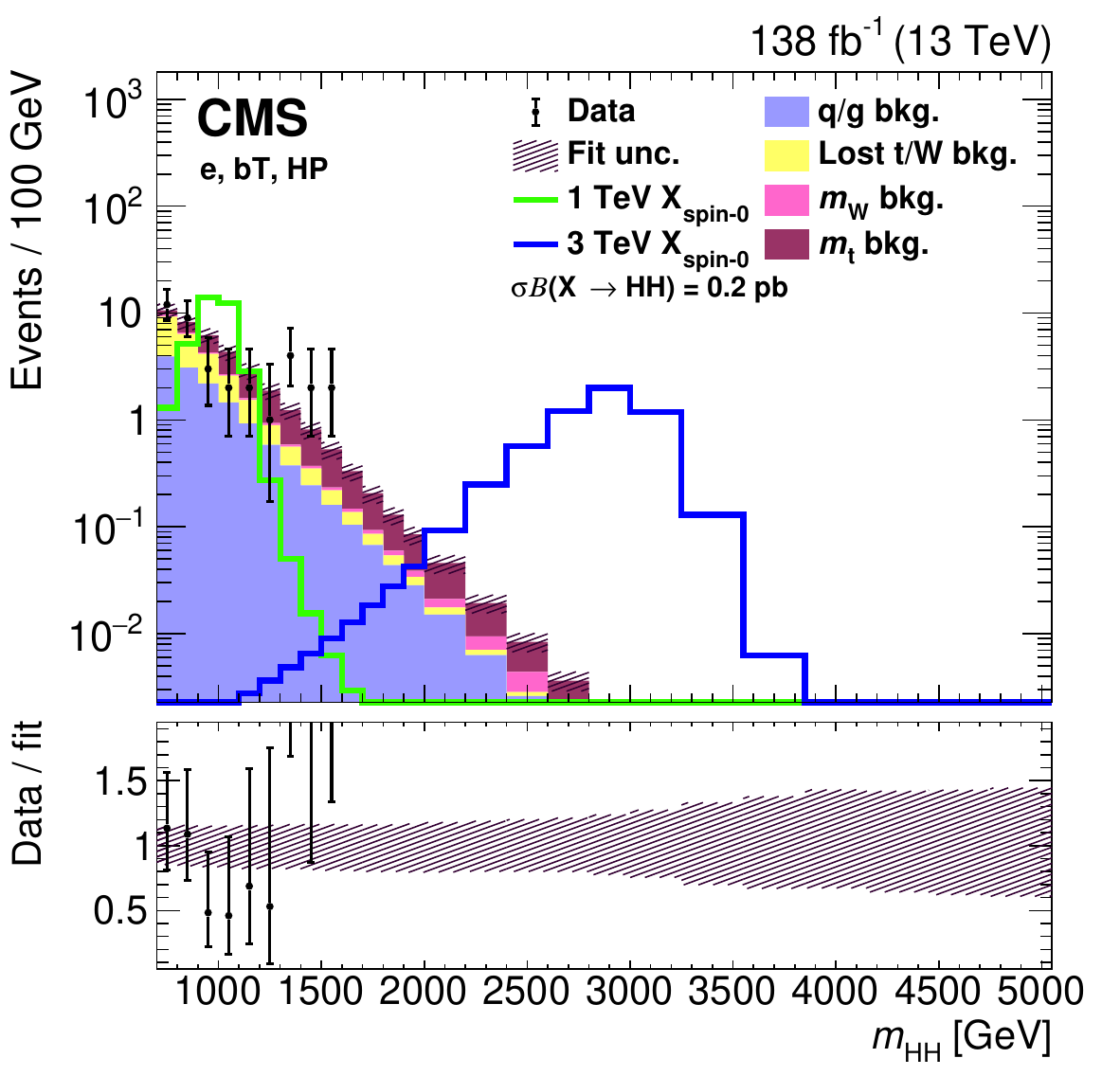}
\includegraphics[width=0.32\textwidth]{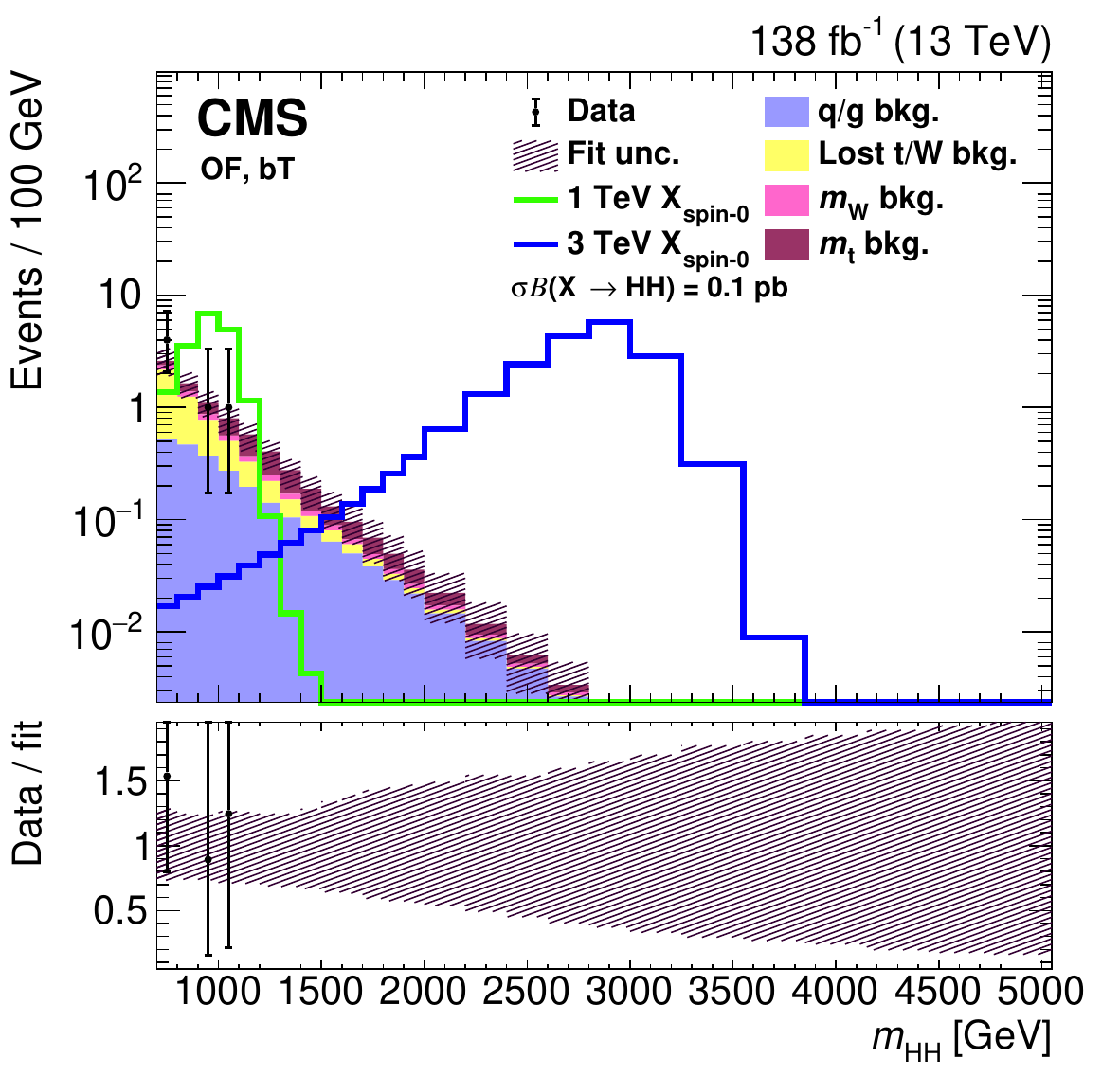}
\caption{The background-only 2D fit result compared to data projected onto the \mhh axis for both the SL and DL channels. The label for each search category is in the upper left of each plot. The fit result is the filled histogram, with the different colors indicating different background components. The background shape uncertainty from the fit is shown as the hatched band. Example spin-0 signal distributions for $\mx = 1.0$ and 3.0\TeV are shown as solid lines, with $\sigma \mathcal{B} (\PX\to\hh)$ set to 0.2 and 0.1\unit{pb} for the SL and DL channels, respectively. The lower panels of each plot show the ratio of the data to the fit result. Only nonzero data entries are shown in the interest of clarity.}
\label{fig:results_mhh}
\end{figure}

The data are interpreted by performing a maximum likelihood fit in the 2D (\mbb, \mhh) mass plane using one model containing only background processes and using one containing both background and signal processes. We find that the background-only model fits the data well. We interpret the results as upper limits at 95\% confidence level (CL) on $\sigma \mathcal{B} (\PX\to\hh)$.

The quality of the fit is quantified with a likelihood ratio goodness-of-fit test using the saturated model~\cite{Baker:1983tu}. The probability distribution function of the test statistic is obtained with pseudo-experiments, and the observed value is within the central 68\% interval of expected results. The best-fit values of the nuisance parameters are consistent with the initial 1 standard deviation range of uncertainty.

The fit result and the data are projected in \mbb for each event category in Fig.~\ref{fig:results_mbb}. The \mbb shape is modeled well by the background-only model, and each background component is important in at least some subspace of the mass range. Particularly, the resonant peaks corresponding to the \PW boson and top quark are correctly modeled by the fit. Similarly, the \mhh projections of the fit are shown in Fig.~\ref{fig:results_mhh}. There is good agreement for the full \mhh mass range in these figures as well.

\begin{figure}[ht]
\centering
\includegraphics[width=0.45\textwidth]{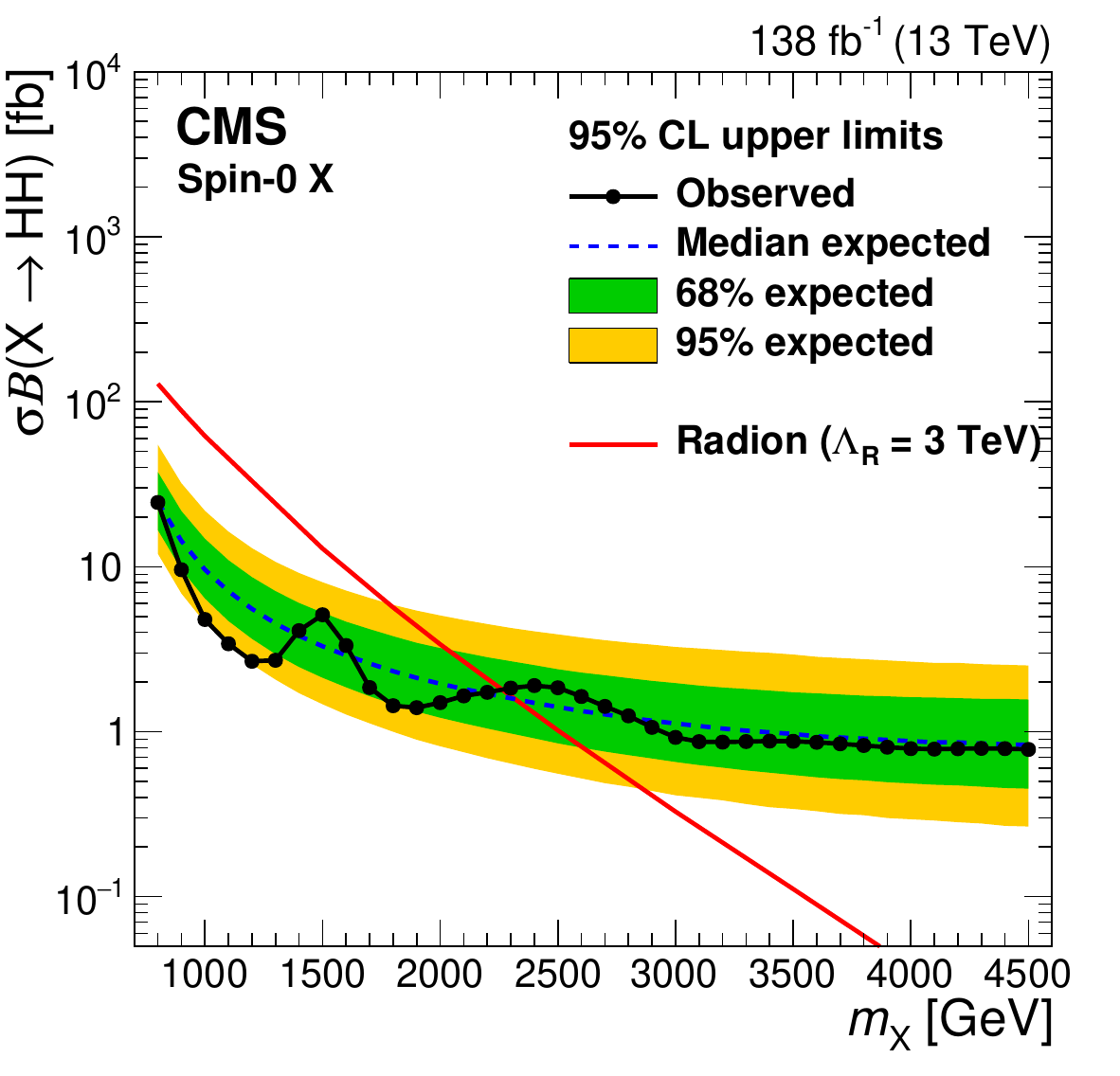}
\includegraphics[width=0.45\textwidth]{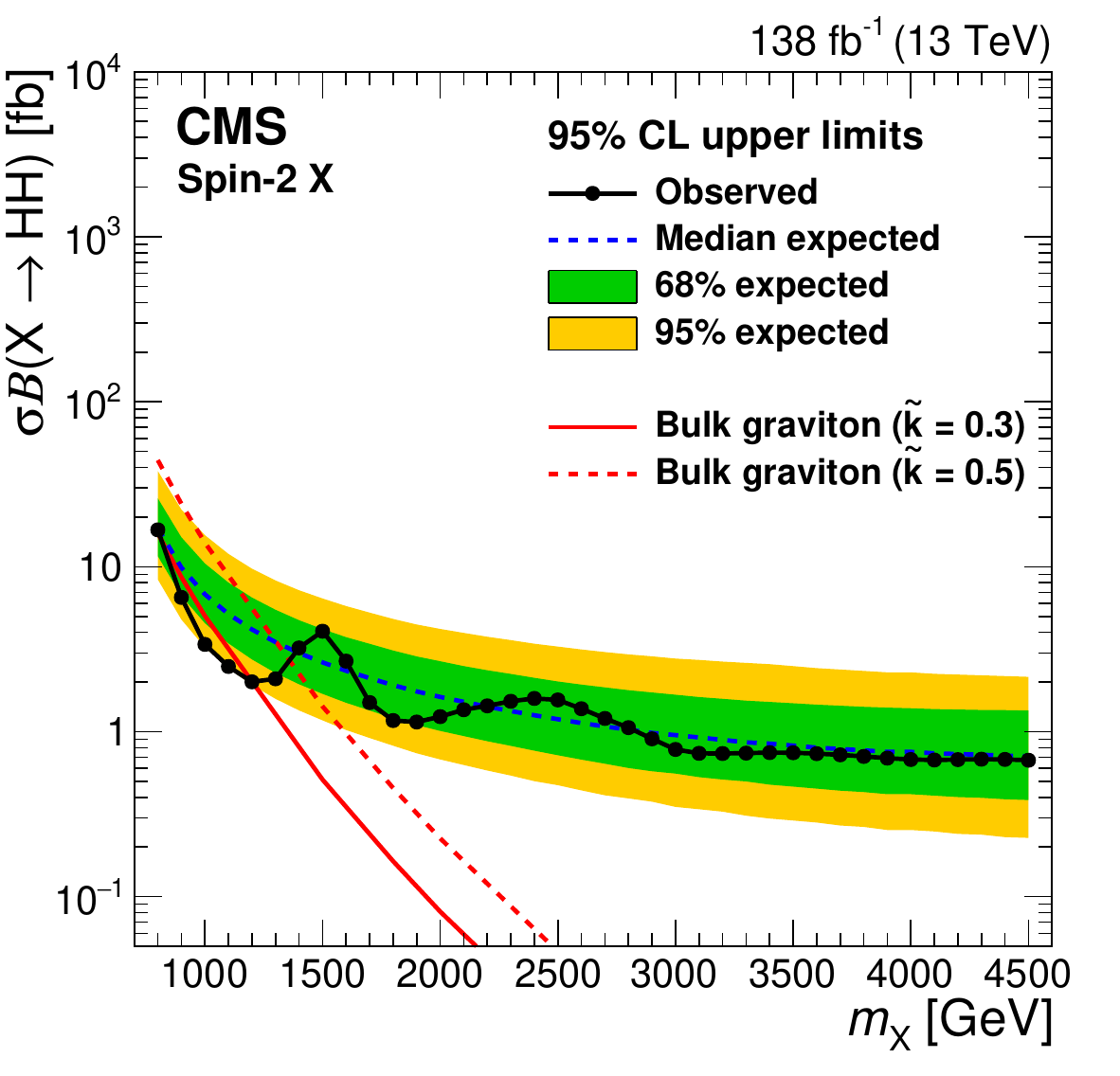}
\caption{Observed and expected 95\% \CL upper limits on the product of the cross section and branching fraction to \hh for a generic spin-0 (left) and spin-2 (right) boson \PX, as functions of mass. Example radion and bulk graviton predictions are also shown. The \hh branching fraction is assumed to be 25\% for radions and 10\% for bulk gravitons. }
\label{fig:results_limits}
\end{figure}

Upper limits are shown at 95\% \CL in Fig.~\ref{fig:results_limits} for both the spin-0 and spin-2 boson scenarios. The limits are evaluated using the asymptotic approximation~\cite{Cowan:2010js} of the \CLs method~\cite{Junk:1999kv,Read:2002hq}, and the validity of this approximation was confirmed by calculating limits with pseudo-experiments. The difference in the limits calculated with pseudo-experiments versus the asymptotic approximation is significantly smaller than 1 standard deviation in the expected limit. The observed exclusion limits are consistent with the expected limits.
A spin-0 signal at $\mx=0.8\TeV$ is excluded for $\sigma \mathcal{B} > 24.5\unit{fb}$, and the exclusion limits strengthen over the full mass range to $\sigma \mathcal{B} > 0.78\unit{fb}$ at $\mx=4.5\TeV$. Spin-2 signals have larger acceptance, and so the exclusion limits on these signals are stronger: at $\mx=0.8\TeV$, we exclude $\sigma \mathcal{B} > 16.7\unit{fb}$, and at $\mx=4.5\TeV$ we exclude $\sigma \mathcal{B} > 0.67\unit{fb}$.

\begin{figure}[htp]
\centering
\includegraphics[width=0.75\textwidth]{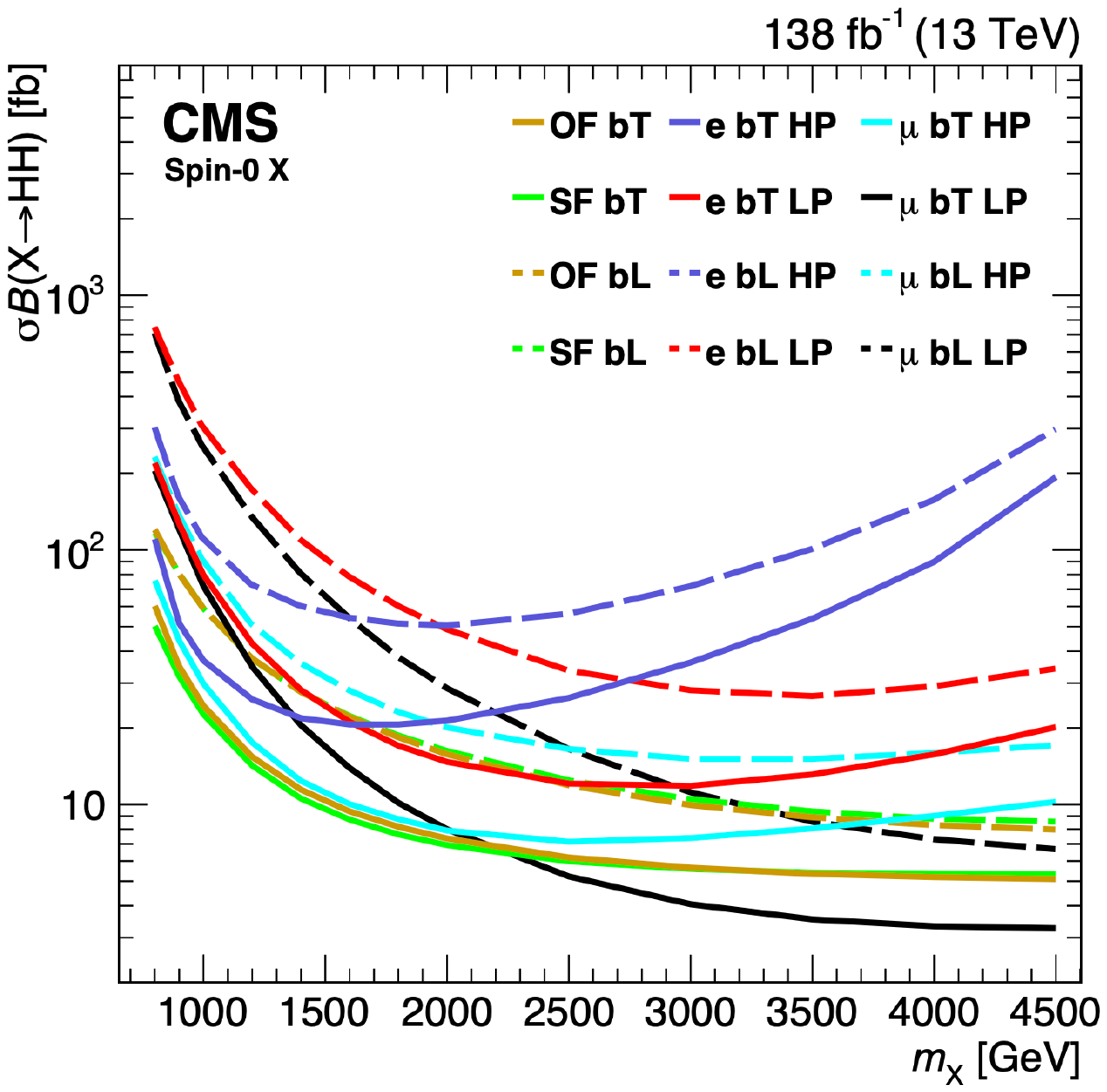}
\caption{Median expected upper limits at 95\% confidence level for each of the 12 search categories individually.}
\label{fig:catLimits}
\end{figure}

Table~\ref{tab:evt_counts} shows the event yields for each search category that are observed in data and are expected before and after a background-only fit, along with the associated post-fit uncertainty in the total background yield in each category. Figure~\ref{fig:catLimits} shows the expected exclusion limit at 95\% \CL for each search category alone. In general, the tight (\bT) \hbbjet tagging categories are the most sensitive over the full range of \mx, since these contain the most signal and the least background. The DL categories are generally more sensitive than most SL categories since the background yields are much lower in the DL channel. A notable exception to this trend is the \PGm \bT LP category, which is the most sensitive above approximately 2.5\TeV. At high \mx, the electron categories in the SL channel are the least sensitive because the electron reconstruction efficiency is degraded.

\begin{table}[!ht]
\centering
\topcaption{\label{tab:evt_counts} Event yields broken down by search category. For each category, shown are the event yields observed in data, expected before and after a fit of the background-only model, and the corresponding relative uncertainty.}
\begin{tabular}{lrrrr}
Search category & Observed & Expected (pre-fit) & Expected (post-fit) & Post-fit uncertainty \\
\hline
\PGm \bL LP & 4542 & 4362.2 & 4540.9 & 1.5\% \\
\PGm \bL HP & 417 & 402.4 & 416.1 & 4.8\% \\
\PGm \bT LP & 657 & 731.8 & 658.5 & 4.2\% \\
\PGm \bT HP & 56 & 67.0 & 57.3 & 10.0\% \\
\Pe \bL LP & 2945 & 2973.7 & 2945.4 & 1.9\% \\
\Pe \bL HP & 248 & 246.1 & 247.7 & 5.7\% \\
\Pe \bT LP & 423 & 443.0 & 423.9 & 4.2\%\\
\Pe \bT HP & 37 & 41.0 & 37.7 & 14.6\% \\
SF \bL & 59 & 70.2 & 59.6 & 14.2\% \\
OF \bL & 50 & 61.1 & 50.8 & 13.5\% \\
SF \bT & 6 & 11.3 & 7.9 & 31.6\% \\
OF \bT & 6 & 11.6 & 8.1 & 25.8\% \\
\end{tabular}
\end{table}

The total uncertainty in the signal sensitivity is dominated by the statistical uncertainty of the data in the analysis. As mentioned in Section~\ref{sec:signalUncs}, the dominant systematic uncertainty for all \mx comes from the \hbbjet tagging efficiency for the signal. Most of the background systematic uncertainties do not have an impact on the signal sensitivity. For high \mx signals, none of the background systematic uncertainties have a significant impact, and only the signal systematic uncertainties have an effect. For low \mx signals, however, the background normalization uncertainties in the most sensitive categories (those with the least background) have an impact.

Relative to the $\PX\to\bbbar\Pell\Pgn\qqpr$ search in Ref.~\cite{B2G-18-008}, this analysis ranges from 6 times more sensitive at low \mx to 14 times more sensitive at high \mx. The improvements in sensitivity arise primarily from three developments. First, an improvement in the expected upper limits by a factor of $\approx$3.5 is achieved because of the larger integrated luminosity alone. This level of improvement is expected because the number of background events is much smaller than the number of signal events under a typical signal peak, even at low \mx.
Second, because of improved techniques in the SL channel alone, we achieve similar sensitivity at $\mx = 0.8\TeV$ and up to a $\approx$2 times improvement at $\mx = 4.5\TeV$. Finally, the addition of the DL channel provides significant improvement in sensitivity. At low \mx, the DL channel is $\approx$70\% more sensitive than the SL channel, largely because the background level is over an order of magnitude smaller. At high \mx, where there is virtually no background in any channel, the DL channel has similar sensitivity to the SL channel. This occurs because the dilepton signal efficiency is largest at high mass, and in the SL channel, despite the larger branching fraction, the lepton efficiency (particularly for electrons) degrades at high mass because of the nearby \wqqjet.

This search yields the most sensitive upper limits for $\PX\to\hh$ production with leptons in the final state. The only $\PX\to\hh$ searches that are more sensitive in any subspace of the considered \mx and for any spin hypothesis are those in the $\bbbar\bbbar$ final state from ATLAS and CMS, each of which extend only up to 3.0\TeV. From 0.8 to 3.0\TeV, the sensitivity of this search is mostly comparable to and in some places stronger than the $\bbbar\bbbar$ searches.

Predicted radion and bulk graviton cross sections~\cite{Oliveira:2014kla} are also shown in Fig.~\ref{fig:results_limits} in the context of Randall--Sundrum models that allow the SM fields to propagate through an extra dimension. Typical model parameters are chosen as proposed in Ref.~\cite{Gouzevitch:2013qca}. A branching fraction of 25\% to \hh and an ultraviolet cutoff $\Lambda_R=3\TeV$ are assumed for the radion, which is excluded for $\mx < 2.25\TeV$. A 10\% branching fraction is assumed for the bulk graviton, which occurs in scenarios that include significant coupling between the bulk graviton and top quarks. Bulk graviton production cross sections depend on the dimensionless quantity $\widetilde{k}=\sqrt{8\pi}k/\Mpl$, where $k$ is the curvature of the extra dimension and \Mpl is the Planck mass. For this interpretation, we choose $\widetilde{k}=0.3$ and 0.5. The bulk gravitons with $\widetilde{k}=0.3$ and 0.5 are excluded for $\mx<1.20$ and 1.35\TeV, respectively. For these particular signal parameters, the radion and bulk graviton decay widths are larger than the 1\MeV width chosen for signal sample generation but much smaller than the detector resolution.

\section{Summary\label{sec:summary}}
A search has been performed for new bosons (narrow resonances) decaying to a pair of Higgs bosons (\hh) where one decays into a bottom quark pair ($\bbbar$) and the other via one of three different modes into final states with leptons. The large Lorentz boost of the Higgs bosons produces a distinct experimental signature with one jet that has substructure consistent with the decay $\PH\to\bbbar$. For the Higgs boson that does not decay to $\bbbar$, the single-lepton decay $\PH\to\wwlnuqq$ and the dilepton decays $\PH\to\wwlnulnu$ and $\PH\to\tautaulnunulnunu$ are considered. In the single-lepton channel, the experimental signature is characterized by a second large jet with a nearby lepton, which is consistent with the decay of \hww. In the dilepton channel, the experimental signature contains two leptons and significant missing transverse momentum. This search uses a sample of proton-proton collisions at $\sqrt{s}=13\TeV$, corresponding to an integrated luminosity of 138\fbinv, collected by the CMS detector at the LHC. The primary standard model backgrounds---production of top quark pairs and of vector bosons in association with jets---are suppressed by reconstructing the \hh decay chain and applying selections to discriminate signal from background. The signal and background yields are estimated by a two-dimensional template fit in the plane of the \hbbjet mass and the \hh resonance mass. The templates are validated in a variety of data control regions and are shown to model the data well. The data are consistent with the expected standard model background. Upper limits are set on the product of the cross section and branching fraction for new bosons decaying to \hh. The observed limit at 95\% confidence level for a spin-0 (spin-2) boson ranges from 24.5 (16.7)\unit{fb} at 0.8\TeV to 0.78 (0.67)\unit{fb} at 4.5\TeV. The results of this search provide the most stringent exclusion limits to date for $\PX\to\hh$ signatures with leptons in the final state and are among the most stringent of all $\PX\to\hh$ searches, at certain mass points the most sensitive.

\begin{acknowledgments}

We congratulate our colleagues in the CERN accelerator departments for the excellent performance of the LHC and thank the technical and administrative staffs at CERN and at other CMS institutes for their contributions to the success of the CMS effort. In addition, we gratefully acknowledge the computing centers and personnel of the Worldwide LHC Computing Grid and other centers for delivering so effectively the computing infrastructure essential to our analyses. Finally, we acknowledge the enduring support for the construction and operation of the LHC, the CMS detector, and the supporting computing infrastructure provided by the following funding agencies: BMBWF and FWF (Austria); FNRS and FWO (Belgium); CNPq, CAPES, FAPERJ, FAPERGS, and FAPESP (Brazil); MES and BNSF (Bulgaria); CERN; CAS, MoST, and NSFC (China); MINCIENCIAS (Colombia); MSES and CSF (Croatia); RIF (Cyprus); SENESCYT (Ecuador); MoER, ERC PUT and ERDF (Estonia); Academy of Finland, MEC, and HIP (Finland); CEA and CNRS/IN2P3 (France); BMBF, DFG, and HGF (Germany); GSRI (Greece); NKFIA (Hungary); DAE and DST (India); IPM (Iran); SFI (Ireland); INFN (Italy); MSIP and NRF (Republic of Korea); MES (Latvia); LAS (Lithuania); MOE and UM (Malaysia); BUAP, CINVESTAV, CONACYT, LNS, SEP, and UASLP-FAI (Mexico); MOS (Montenegro); MBIE (New Zealand); PAEC (Pakistan); MSHE and NSC (Poland); FCT (Portugal); JINR (Dubna); MON, RosAtom, RAS, RFBR, and NRC KI (Russia); MESTD (Serbia); MCIN/AE and PCTI (Spain); MOSTR (Sri Lanka); Swiss Funding Agencies (Switzerland); MST (Taipei); ThEPCenter, IPST, STAR, and NSTDA (Thailand); TUBITAK and TAEK (Turkey); NASU (Ukraine); STFC (United Kingdom); DOE and NSF (USA).

\hyphenation{Rachada-pisek} Individuals have received support from the Marie-Curie program and the European Research Council and Horizon 2020 Grant, contract Nos.\ 675440, 724704, 752730, 758316, 765710, 824093, 884104, and COST Action CA16108 (European Union); the Leventis Foundation; the Alfred P.\ Sloan Foundation; the Alexander von Humboldt Foundation; the Belgian Federal Science Policy Office; the Fonds pour la Formation \`a la Recherche dans l'Industrie et dans l'Agriculture (FRIA-Belgium); the Agentschap voor Innovatie door Wetenschap en Technologie (IWT-Belgium); the F.R.S.-FNRS and FWO (Belgium) under the ``Excellence of Science -- EOS" -- be.h project n.\ 30820817; the Beijing Municipal Science \& Technology Commission, No. Z191100007219010; the Ministry of Education, Youth and Sports (MEYS) of the Czech Republic; the Deutsche Forschungsgemeinschaft (DFG), under Germany's Excellence Strategy -- EXC 2121 ``Quantum Universe" -- 390833306, and under project number 400140256 - GRK2497; the Lend\"ulet (``Momentum") Program and the J\'anos Bolyai Research Scholarship of the Hungarian Academy of Sciences, the New National Excellence Program \'UNKP, the NKFIA research grants 123842, 123959, 124845, 124850, 125105, 128713, 128786, and 129058 (Hungary); the Council of Science and Industrial Research, India; the Latvian Council of Science; the Ministry of Science and Higher Education and the National Science Center, contracts Opus 2014/15/B/ST2/03998 and 2015/19/B/ST2/02861 (Poland); the Funda\c{c}\~ao para a Ci\^encia e a Tecnologia, grant CEECIND/01334/2018 (Portugal); the National Priorities Research Program by Qatar National Research Fund; the Ministry of Science and Higher Education, projects no. 14.W03.31.0026 and no. FSWW-2020-0008, and the Russian Foundation for Basic Research, project No.19-42-703014 (Russia); MCIN/AEI/10.13039/501100011033, ERDF ``a way of making Europe", and the Programa Estatal de Fomento de la Investigaci{\'o}n Cient{\'i}fica y T{\'e}cnica de Excelencia Mar\'{\i}a de Maeztu, grant MDM-2017-0765 and Programa Severo Ochoa del Principado de Asturias (Spain); the Stavros Niarchos Foundation (Greece); the Rachadapisek Sompot Fund for Postdoctoral Fellowship, Chulalongkorn University and the Chulalongkorn Academic into Its 2nd Century Project Advancement Project (Thailand); the Kavli Foundation; the Nvidia Corporation; the SuperMicro Corporation; the Welch Foundation, contract C-1845; and the Weston Havens Foundation (USA).

\end{acknowledgments}

\bibliography{auto_generated}
\cleardoublepage \appendix\section{The CMS Collaboration \label{app:collab}}\begin{sloppypar}\hyphenpenalty=5000\widowpenalty=500\clubpenalty=5000\input{B2G-20-007-authorlist.tex}\end{sloppypar}
\end{document}

%% file: B2G-20-007-authorlist.tex
\cmsinstitute{Yerevan~Physics~Institute, Yerevan, Armenia}
A.~Tumasyan
\cmsinstitute{Institut~f\"{u}r~Hochenergiephysik, Vienna, Austria}
W.~Adam\cmsorcid{0000-0001-9099-4341}, J.W.~Andrejkovic, T.~Bergauer\cmsorcid{0000-0002-5786-0293}, S.~Chatterjee\cmsorcid{0000-0003-2660-0349}, K.~Damanakis, M.~Dragicevic\cmsorcid{0000-0003-1967-6783}, A.~Escalante~Del~Valle\cmsorcid{0000-0002-9702-6359}, R.~Fr\"{u}hwirth\cmsAuthorMark{1}, M.~Jeitler\cmsAuthorMark{1}\cmsorcid{0000-0002-5141-9560}, N.~Krammer, L.~Lechner\cmsorcid{0000-0002-3065-1141}, D.~Liko, I.~Mikulec, P.~Paulitsch, F.M.~Pitters, J.~Schieck\cmsAuthorMark{1}\cmsorcid{0000-0002-1058-8093}, R.~Sch\"{o}fbeck\cmsorcid{0000-0002-2332-8784}, D.~Schwarz, S.~Templ\cmsorcid{0000-0003-3137-5692}, W.~Waltenberger\cmsorcid{0000-0002-6215-7228}, C.-E.~Wulz\cmsAuthorMark{1}\cmsorcid{0000-0001-9226-5812}
% ...
\vskip\cmsinstskip
% ...
1:~Also at TU~Wien, Wien, Austria\\
% ...


%% file: B2G-20-007_temp.bbl
\providecommand{\href}[2]{#2}\begingroup\raggedright\begin{thebibliography}{100}%
\makeatletter
\providecommand{\hrefCMSnoop }[0]{\@secondoftwo}%
\makeatother
\providecommand{\doi}{\texttt{doi:}\begingroup \urlstyle{tt}\Url}

\bibitem{HiggsDiscoveryAtlas}
\hrefCMSnoop {}{{ATLAS Collaboration}, ``Observation of a new particle in the
  search for the standard model {Higgs} boson with the {ATLAS} detector at the
  {LHC}'',} \textit{ Phys. Lett. B} \textbf{ 716} (2012) 01,
  \href{http://dx.doi.org/10.1016/j.physletb.2012.08.020}{\doi{10.1016/j.physletb.2012.08.020}},
\href{http://www.arXiv.org/abs/1207.7214}{\texttt{arXiv:1207.7214}}.

\bibitem{HiggsDiscoveryCMS}
\hrefCMSnoop {}{{{CMS}} Collaboration, ``Observation of a new boson at a mass
  of 125 {\GeV} with the {CMS} experiment at the {LHC}'',} \textit{ Phys. Lett.
  B} \textbf{ 716} (2012) 30,
  \href{http://dx.doi.org/10.1016/j.physletb.2012.08.021}{\doi{10.1016/j.physletb.2012.08.021}},
\href{http://www.arXiv.org/abs/1207.7235}{\texttt{arXiv:1207.7235}}.

\bibitem{Chatrchyan:2013lba}
\hrefCMSnoop {}{{CMS Collaboration}, ``Observation of a new boson with mass
  near {125 GeV} in pp collisions at {$\sqrt{s}$ = 7 and 8~TeV}'',} \textit{
  JHEP} \textbf{ 06} (2013) 081,
  \href{http://dx.doi.org/10.1007/JHEP06(2013)081}{\doi{10.1007/JHEP06(2013)081}},
\href{http://www.arXiv.org/abs/1303.4571}{\texttt{arXiv:1303.4571}}.

\bibitem{Englert:1964et}
\hrefCMSnoop {}{F.~Englert and R.~Brout, ``Broken symmetry and the mass of
  gauge vector mesons'',} \textit{ Phys. Rev. Lett.} \textbf{ 13} (1964) 321,
  \href{http://dx.doi.org/10.1103/PhysRevLett.13.321}{\doi{10.1103/PhysRevLett.13.321}}.

\bibitem{Higgs}
\hrefCMSnoop {}{P.~W. Higgs, ``Broken symmetries and the masses of gauge
  bosons'',} \textit{ Phys. Rev. Lett.} \textbf{ 13} (1964) 508,
  \href{http://dx.doi.org/10.1103/PhysRevLett.13.508}{\doi{10.1103/PhysRevLett.13.508}}.

\bibitem{Branco:2011iw}
G.~C. Branco\hrefCMSnoop {}{ {et~al.}, ``Theory and phenomenology of
  {two-Higgs-doublet models}'',} \textit{ Phys. Rept.} \textbf{ 516} (2012) 1,
  \href{http://dx.doi.org/10.1016/j.physrep.2012.02.002}{\doi{10.1016/j.physrep.2012.02.002}},
\href{http://www.arXiv.org/abs/1106.0034}{\texttt{arXiv:1106.0034}}.

\bibitem{Ramond:1971gb}
\hrefCMSnoop {}{P.~Ramond, ``{Dual theory for free fermions}'',} \textit{ Phys.
  Rev. D} \textbf{ 3} (1971) 2415,
\href{http://dx.doi.org/10.1103/PhysRevD.3.2415}{\doi{10.1103/PhysRevD.3.2415}}.

\bibitem{Golfand:1971iw}
\href {http://www.jetpletters.ru/ps/1584/article_24309.shtml}{Y.~A. Golfand and
  E.~P. Likhtman, ``{Extension of the algebra of {P}oincar\'{e} group
  generators and violation of {P} invariance}'',} \textit{ JETP Lett.} \textbf{
  13} (1971)
323.

\bibitem{Neveu:1971rx}
\hrefCMSnoop {}{A.~Neveu and J.~H. Schwarz, ``{Factorizable dual model of
  pions}'',} \textit{ Nucl. Phys. B} \textbf{ 31} (1971) 86,
\href{http://dx.doi.org/10.1016/0550-3213(71)90448-2}{\doi{10.1016/0550-3213(71)90448-2}}.

\bibitem{Volkov:1972jx}
\href {http://www.jetpletters.ru/ps/1766/article_26864.shtml}{D.~V. Volkov and
  V.~P. Akulov, ``{Possible universal neutrino interaction}'',} \textit{ JETP
  Lett.} \textbf{ 16} (1972)
438.

\bibitem{Wess:1973kz}
\hrefCMSnoop {}{J.~Wess and B.~Zumino, ``{A {L}agrangian model invariant under
  supergauge transformations}'',} \textit{ Phys. Lett. B} \textbf{ 49} (1974)
  52,
\href{http://dx.doi.org/10.1016/0370-2693(74)90578-4}{\doi{10.1016/0370-2693(74)90578-4}}.

\bibitem{Wess:1974tw}
\hrefCMSnoop {}{J.~Wess and B.~Zumino, ``{Supergauge transformations in four
  dimensions}'',} \textit{ Nucl. Phys. B} \textbf{ 70} (1974) 39,
\href{http://dx.doi.org/10.1016/0550-3213(74)90355-1}{\doi{10.1016/0550-3213(74)90355-1}}.

\bibitem{Fayet:1974pd}
\hrefCMSnoop {}{P.~Fayet, ``{Supergauge invariant extension of the {H}iggs
  mechanism and a model for the electron and its neutrino}'',} \textit{ Nucl.
  Phys. B} \textbf{ 90} (1975) 104,
\href{http://dx.doi.org/10.1016/0550-3213(75)90636-7}{\doi{10.1016/0550-3213(75)90636-7}}.

\bibitem{Nilles:1983ge}
\hrefCMSnoop {}{H.~P. Nilles, ``{Supersymmetry, supergravity and particle
  physics}'',} \textit{ Phys. Rep.} \textbf{ 110} (1984) 1,
\href{http://dx.doi.org/10.1016/0370-1573(84)90008-5}{\doi{10.1016/0370-1573(84)90008-5}}.

\bibitem{Randall:1999ee}
\hrefCMSnoop {}{L.~Randall and R.~Sundrum, ``A large mass hierarchy from a
  small extra dimension'',} \textit{ Phys. Rev. Lett.} \textbf{ 83} (1999)
  3370,
  \href{http://dx.doi.org/10.1103/PhysRevLett.83.3370}{\doi{10.1103/PhysRevLett.83.3370}},
\href{http://www.arXiv.org/abs/hep-ph/9905221}{\texttt{arXiv:hep-ph/9905221}}.

\bibitem{Goldberger:1999uk}
\hrefCMSnoop {}{W.~D. Goldberger and M.~B. Wise, ``Modulus stabilization with
  bulk fields'',} \textit{ Phys. Rev. Lett.} \textbf{ 83} (1999) 4922,
  \href{http://dx.doi.org/10.1103/PhysRevLett.83.4922}{\doi{10.1103/PhysRevLett.83.4922}},
\href{http://www.arXiv.org/abs/hep-ph/9907447}{\texttt{arXiv:hep-ph/9907447}}.

\bibitem{DeWolfe:1999cp}
\hrefCMSnoop {}{O.~DeWolfe, D.~Z. Freedman, S.~S. Gubser, and A.~Karch,
  ``Modeling the fifth dimension with scalars and gravity'',} \textit{ Phys.
  Rev. D} \textbf{ 62} (2000) 046008,
  \href{http://dx.doi.org/10.1103/PhysRevD.62.046008}{\doi{10.1103/PhysRevD.62.046008}},
\href{http://www.arXiv.org/abs/hep-th/9909134}{\texttt{arXiv:hep-th/9909134}}.

\bibitem{Csaki:1999mp}
\hrefCMSnoop {}{C.~Csaki, M.~Graesser, L.~Randall, and J.~Terning, ``Cosmology
  of brane models with radion stabilization'',} \textit{ Phys. Rev. D} \textbf{
  62} (2000) 045015,
  \href{http://dx.doi.org/10.1103/PhysRevD.62.045015}{\doi{10.1103/PhysRevD.62.045015}},
\href{http://www.arXiv.org/abs/hep-ph/9911406}{\texttt{arXiv:hep-ph/9911406}}.

\bibitem{Csaki:2000zn}
\hrefCMSnoop {}{C.~Csaki, M.~L. Graesser, and G.~D. Kribs, ``Radion dynamics
  and electroweak physics'',} \textit{ Phys. Rev. D} \textbf{ 63} (2001)
  065002,
  \href{http://dx.doi.org/10.1103/PhysRevD.63.065002}{\doi{10.1103/PhysRevD.63.065002}},
\href{http://www.arXiv.org/abs/hep-th/0008151}{\texttt{arXiv:hep-th/0008151}}.

\bibitem{Davoudiasl:1999jd}
\hrefCMSnoop {}{H.~Davoudiasl, J.~L. Hewett, and T.~G. Rizzo, ``Phenomenology
  of the {Randall-Sundrum} gauge hierarchy model'',} \textit{ Phys. Rev. Lett.}
  \textbf{ 84} (2000) 2080,
  \href{http://dx.doi.org/10.1103/PhysRevLett.84.2080}{\doi{10.1103/PhysRevLett.84.2080}},
\href{http://www.arXiv.org/abs/hep-ph/9909255}{\texttt{arXiv:hep-ph/9909255}}.

\bibitem{Agashe:2007zd}
\hrefCMSnoop {}{K.~Agashe, H.~Davoudiasl, G.~Perez, and A.~Soni, ``Warped
  gravitons at the {LHC} and beyond'',} \textit{ Phys. Rev. D} \textbf{ 76}
  (2007) 036006,
  \href{http://dx.doi.org/10.1103/PhysRevD.76.036006}{\doi{10.1103/PhysRevD.76.036006}},
\href{http://www.arXiv.org/abs/hep-ph/0701186}{\texttt{arXiv:hep-ph/0701186}}.

\bibitem{Fitzpatrick:2007qr}
\hrefCMSnoop {}{L.~Fitzpatrick, J.~Kaplan, L.~Randall, and L.-T. Wang,
  ``Searching for the {Kaluza-Klein} graviton in bulk {RS} models'',} \textit{
  JHEP} \textbf{ 09} (2007) 013,
  \href{http://dx.doi.org/10.1088/1126-6708/2007/09/013}{\doi{10.1088/1126-6708/2007/09/013}},
\href{http://www.arXiv.org/abs/hep-ph/0701150}{\texttt{arXiv:hep-ph/0701150}}.

\bibitem{Aaboud:2018ohp}
\hrefCMSnoop {}{{ATLAS Collaboration}, ``Search for resonant {$WZ$} production
  in the fully leptonic final state in proton-proton collisions at $\sqrt{s} =
  13$ {TeV} with the {ATLAS} detector'',} \textit{ Phys. Lett. B} \textbf{ 787}
  (2018) 68,
  \href{http://dx.doi.org/10.1016/j.physletb.2018.10.021}{\doi{10.1016/j.physletb.2018.10.021}},
\href{http://www.arXiv.org/abs/1806.01532}{\texttt{arXiv:1806.01532}}.

\bibitem{Aaboud:2017gsl}
\hrefCMSnoop {}{{ATLAS Collaboration}, ``Search for heavy resonances decaying
  into {$WW$} in the $e\nu\mu\nu$ final state in $pp$ collisions at
  $\sqrt{s}=13$ {TeV} with the {ATLAS} detector'',} \textit{ Eur. Phys. J. C}
  \textbf{ 78} (2018) 24,
  \href{http://dx.doi.org/10.1140/epjc/s10052-017-5491-4}{\doi{10.1140/epjc/s10052-017-5491-4}},
\href{http://www.arXiv.org/abs/1710.01123}{\texttt{arXiv:1710.01123}}.

\bibitem{Aaboud:2017rel}
\hrefCMSnoop {}{{ATLAS Collaboration}, ``Search for heavy {ZZ} resonances in
  the $\ell ^+\ell ^-\ell ^+\ell ^-$ and $\ell ^+\ell ^-\nu \bar{\nu }$ final
  states using proton-proton collisions at $\sqrt{s}= 13$ $\text {TeV}$ with
  the {ATLAS} detector'',} \textit{ Eur. Phys. J. C} \textbf{ 78} (2018) 293,
  \href{http://dx.doi.org/10.1140/epjc/s10052-018-5686-3}{\doi{10.1140/epjc/s10052-018-5686-3}},
\href{http://www.arXiv.org/abs/1712.06386}{\texttt{arXiv:1712.06386}}.

\bibitem{Aaboud:2017cxo}
\hrefCMSnoop {}{{ATLAS Collaboration}, ``Search for heavy resonances decaying
  into a {$W$ or $Z$} boson and a {Higgs} boson in final states with leptons
  and $b$-jets in 36 fb$^{-1}$ of $\sqrt s = 13$ {TeV} $pp$ collisions with the
  {ATLAS} detector'',} \textit{ JHEP} \textbf{ 03} (2018) 174,
  \href{http://dx.doi.org/10.1007/JHEP03(2018)174}{\doi{10.1007/JHEP03(2018)174}},
\href{http://www.arXiv.org/abs/1712.06518}{\texttt{arXiv:1712.06518}}.

\bibitem{Aaboud:2018zhh}
\hrefCMSnoop {}{{ATLAS Collaboration}, ``Search for {Higgs} boson pair
  production in the {$b\bar{b} WW^{*}$} decay mode at {$\sqrt{s}=13$~TeV} with
  the {ATLAS} detector'',} \textit{ JHEP} \textbf{ 04} (2019) 092,
  \href{http://dx.doi.org/10.1007/JHEP04(2019)092}{\doi{10.1007/JHEP04(2019)092}},
\href{http://www.arXiv.org/abs/1811.04671}{\texttt{arXiv:1811.04671}}.

\bibitem{ATLAS:2019nat}
\hrefCMSnoop {}{{ATLAS Collaboration}, ``{Search for diboson resonances in
  hadronic final states in 139 fb$^{-1}$ of $pp$ collisions at $\sqrt{s} = 13$
  TeV with the ATLAS detector}'',} \textit{ JHEP} \textbf{ 09} (2019) 091,
  \href{http://dx.doi.org/10.1007/JHEP09(2019)091}{\doi{10.1007/JHEP09(2019)091}},
  \href{http://www.arXiv.org/abs/1906.08589}{\texttt{arXiv:1906.08589}}.
  [Erratum: \DOI{10.1007/JHEP06(2020)042}].

\bibitem{ATLAS:2020jgy}
\hrefCMSnoop {}{{ATLAS Collaboration}, ``{Search for the $HH \rightarrow b
  \bar{b} b \bar{b}$ process via vector-boson fusion production using
  proton-proton collisions at $\sqrt{s} = 13$ TeV with the ATLAS detector}'',}
  \textit{ JHEP} \textbf{ 07} (2020) 108,
  \href{http://dx.doi.org/10.1007/JHEP07(2020)108}{\doi{10.1007/JHEP07(2020)108}},
  \href{http://www.arXiv.org/abs/2001.05178}{\texttt{arXiv:2001.05178}}.
  [Errata: \DOI{10.1007/JHEP01(2021)145}, \DOI{10.1007/JHEP05(2021)207}].

\bibitem{ATLAS:2020fry}
\hrefCMSnoop {}{{ATLAS Collaboration}, ``{Search for heavy diboson resonances
  in semileptonic final states in pp collisions at $\sqrt{s}=13$ TeV with the
  ATLAS detector}'',} \textit{ Eur. Phys. J. C} \textbf{ 80} (2020) 1165,
  \href{http://dx.doi.org/10.1140/epjc/s10052-020-08554-y}{\doi{10.1140/epjc/s10052-020-08554-y}},
  \href{http://www.arXiv.org/abs/2004.14636}{\texttt{arXiv:2004.14636}}.

\bibitem{ATLAS:2020qiz}
\hrefCMSnoop {}{{ATLAS Collaboration}, ``{Search for resonances decaying into a
  weak vector boson and a Higgs boson in the fully hadronic final state
  produced in proton$-$proton collisions at $\sqrt{s} = 13$ TeV with the ATLAS
  detector}'',} \textit{ Phys. Rev. D} \textbf{ 102} (2020) 112008,
  \href{http://dx.doi.org/10.1103/PhysRevD.102.112008}{\doi{10.1103/PhysRevD.102.112008}},
  \href{http://www.arXiv.org/abs/2007.05293}{\texttt{arXiv:2007.05293}}.

\bibitem{ATLAS:2020azv}
\hrefCMSnoop {}{{ATLAS Collaboration}, ``{Reconstruction and identification of
  boosted di-$\tau$ systems in a search for Higgs boson pairs using 13 TeV
  proton-proton collision data in ATLAS}'',} \textit{ JHEP} \textbf{ 11} (2020)
  163,
  \href{http://dx.doi.org/10.1007/JHEP11(2020)163}{\doi{10.1007/JHEP11(2020)163}},
  \href{http://www.arXiv.org/abs/2007.14811}{\texttt{arXiv:2007.14811}}.

\bibitem{ATLAS:2022hwc}
\hrefCMSnoop {}{{ATLAS Collaboration}, ``{Search for resonant pair production
  of Higgs bosons in the $b\bar{b}b\bar{b}$ final state using $pp$ collisions
  at $\sqrt{s}$ = 13 TeV with the ATLAS detector}'',} 2022.
  \href{http://www.arXiv.org/abs/2202.07288}{\texttt{arXiv:2202.07288}}.
  Submitted to \textit{Phys. Rev. D.}

\bibitem{ATLAS:2018sbw}
\hrefCMSnoop {}{{ATLAS Collaboration}, ``{Combination of searches for heavy
  resonances decaying into bosonic and leptonic final states using 36 fb$^{-1}$
  of proton-proton collision data at $\sqrt{s} = 13$ TeV with the ATLAS
  detector}'',} \textit{ Phys. Rev. D} \textbf{ 98} (2018) 052008,
  \href{http://dx.doi.org/10.1103/PhysRevD.98.052008}{\doi{10.1103/PhysRevD.98.052008}},
  \href{http://www.arXiv.org/abs/1808.02380}{\texttt{arXiv:1808.02380}}.

\bibitem{Sirunyan:2017nrt}
\hrefCMSnoop {}{{CMS Collaboration}, ``Combination of searches for heavy
  resonances decaying to {WW}, {WZ}, {ZZ}, {WH}, and {ZH} boson pairs in
  proton-proton collisions at {$\sqrt{s}=8$} and 13 {TeV}'',} \textit{ Phys.
  Lett. B} \textbf{ 774} (2017) 533,
  \href{http://dx.doi.org/10.1016/j.physletb.2017.09.083}{\doi{10.1016/j.physletb.2017.09.083}},
\href{http://www.arXiv.org/abs/1705.09171}{\texttt{arXiv:1705.09171}}.

\bibitem{Sirunyan:2018fuh}
\hrefCMSnoop {}{{CMS Collaboration}, ``Search for heavy resonances decaying
  into two {Higgs} bosons or into a {Higgs} boson and a {W} or {Z} boson in
  proton-proton collisions at 13 {TeV}'',} \textit{ JHEP} \textbf{ 01} (2019)
  051,
  \href{http://dx.doi.org/10.1007/JHEP01(2019)051}{\doi{10.1007/JHEP01(2019)051}},
\href{http://www.arXiv.org/abs/1808.01365}{\texttt{arXiv:1808.01365}}.

\bibitem{Sirunyan:2018hsl}
\hrefCMSnoop {}{{CMS Collaboration}, ``Search for a heavy resonance decaying
  into a {Z} boson and a {Z or W} boson in 2$\ell$2q final states at $
  \sqrt{s}=13 $ {TeV}'',} \textit{ JHEP} \textbf{ 09} (2018) 101,
  \href{http://dx.doi.org/10.1007/JHEP09(2018)101}{\doi{10.1007/JHEP09(2018)101}},
\href{http://www.arXiv.org/abs/1803.10093}{\texttt{arXiv:1803.10093}}.

\bibitem{Sirunyan:2018qca}
\hrefCMSnoop {}{{CMS Collaboration}, ``Search for production of {Higgs} boson
  pairs in the four b quark final state using large-area jets in proton-proton
  collisions at $\sqrt{s}=$ 13 {TeV}'',} \textit{ JHEP} \textbf{ 01} (2019)
  040,
  \href{http://dx.doi.org/10.1007/JHEP01(2019)040}{\doi{10.1007/JHEP01(2019)040}},
\href{http://www.arXiv.org/abs/1808.01473}{\texttt{arXiv:1808.01473}}.

\bibitem{Sirunyan:2017acf}
\hrefCMSnoop {}{{CMS Collaboration}, ``Search for massive resonances decaying
  into {$WW$, $WZ$, $ZZ$, $qW$, and $qZ$} with dijet final states at
  $\sqrt{s}=13$ {TeV}'',} \textit{ Phys. Rev. D} \textbf{ 97} (2018) 072006,
  \href{http://dx.doi.org/10.1103/PhysRevD.97.072006}{\doi{10.1103/PhysRevD.97.072006}},
\href{http://www.arXiv.org/abs/1708.05379}{\texttt{arXiv:1708.05379}}.

\bibitem{Sirunyan:2017wto}
\hrefCMSnoop {}{{CMS Collaboration}, ``Search for heavy resonances that decay
  into a vector boson and a {Higgs} boson in hadronic final states at $\sqrt{s}
  = 13$ {TeV}'',} \textit{ Eur. Phys. J. C} \textbf{ 77} (2017) 636,
  \href{http://dx.doi.org/10.1140/epjc/s10052-017-5192-z}{\doi{10.1140/epjc/s10052-017-5192-z}},
\href{http://www.arXiv.org/abs/1707.01303}{\texttt{arXiv:1707.01303}}.

\bibitem{Sirunyan:2018qob}
\hrefCMSnoop {}{{CMS Collaboration}, ``Search for heavy resonances decaying
  into a vector boson and a {Higgs} boson in final states with charged leptons,
  neutrinos and b quarks at $\sqrt{s}=13$ {TeV}'',} \textit{ JHEP} \textbf{ 11}
  (2018) 172,
  \href{http://dx.doi.org/10.1007/JHEP11(2018)172}{\doi{10.1007/JHEP11(2018)172}},
\href{http://www.arXiv.org/abs/1807.02826}{\texttt{arXiv:1807.02826}}.

\bibitem{Sirunyan:2017isc}
\hrefCMSnoop {}{{CMS Collaboration}, ``Search for a massive resonance decaying
  to a pair of {Higgs} bosons in the four b quark final state in proton-proton
  collisions at $\sqrt{s}=$ 13 {TeV}'',} \textit{ Phys. Lett. B} \textbf{ 781}
  (2018) 244,
  \href{http://dx.doi.org/10.1016/j.physletb.2018.03.084}{\doi{10.1016/j.physletb.2018.03.084}},
\href{http://www.arXiv.org/abs/1710.04960}{\texttt{arXiv:1710.04960}}.

\bibitem{Sirunyan:2016cao}
\hrefCMSnoop {}{{CMS Collaboration}, ``Search for massive resonances decaying
  into {WW}, {WZ} or {ZZ} bosons in proton-proton collisions at $\sqrt{s} =
  13{\TeV}$'',} \textit{ JHEP} \textbf{ 03} (2017) 162,
  \href{http://dx.doi.org/10.1007/JHEP03(2017)162}{\doi{10.1007/JHEP03(2017)162}},
\href{http://www.arXiv.org/abs/1612.09159}{\texttt{arXiv:1612.09159}}.

\bibitem{Sirunyan:2017jtu}
\hrefCMSnoop {}{{CMS Collaboration}, ``Search for {ZZ} resonances in the
  2$\ell$2$\nu$ final state in proton-proton collisions at 13 {TeV}'',}
  \textit{ JHEP} \textbf{ 03} (2018) 003,
  \href{http://dx.doi.org/10.1007/JHEP03(2018)003}{\doi{10.1007/JHEP03(2018)003}},
\href{http://www.arXiv.org/abs/1711.04370}{\texttt{arXiv:1711.04370}}.

\bibitem{Khachatryan:2014xja}
\hrefCMSnoop {}{{CMS Collaboration}, ``Search for new resonances decaying via
  {WZ} to leptons in proton-proton collisions at $\sqrt{s} = 13$ {TeV}'',}
  \textit{ Phys. Lett. B} \textbf{ 740} (2015) 83,
  \href{http://dx.doi.org/10.1016/j.physletb.2014.11.026}{\doi{10.1016/j.physletb.2014.11.026}},
\href{http://www.arXiv.org/abs/1407.3476}{\texttt{arXiv:1407.3476}}.

\bibitem{Khachatryan:2015ywa}
\hrefCMSnoop {}{{CMS Collaboration}, ``Search for narrow high-mass resonances
  in proton-proton collisions at {$\sqrt{s} = 8\TeV $} decaying to a {Z} and a
  {Higgs} boson'',} \textit{ Phys. Lett. B} \textbf{ 748} (2015) 255,
  \href{http://dx.doi.org/10.1016/j.physletb.2015.07.011}{\doi{10.1016/j.physletb.2015.07.011}},
\href{http://www.arXiv.org/abs/1502.04994}{\texttt{arXiv:1502.04994}}.

\bibitem{CMS:2019kaf}
\hrefCMSnoop {}{{CMS Collaboration}, ``{Combination of CMS searches for heavy
  resonances decaying to pairs of bosons or leptons}'',} \textit{ Phys. Lett.
  B} \textbf{ 798} (2019) 134952,
  \href{http://dx.doi.org/10.1016/j.physletb.2019.134952}{\doi{10.1016/j.physletb.2019.134952}},
  \href{http://www.arXiv.org/abs/1906.00057}{\texttt{arXiv:1906.00057}}.

\bibitem{CMS:2021fyk}
\hrefCMSnoop {}{{CMS Collaboration}, ``{Search for a heavy vector resonance
  decaying to a ${\mathrm{Z}}_{\mathrm{}}^{\mathrm{}}$ ~boson and a Higgs boson
  in proton-proton collisions at $\sqrt{s} = 13\,\text {Te}\text {V} $}'',}
  \textit{ Eur. Phys. J. C} \textbf{ 81} (2021) 688,
  \href{http://dx.doi.org/10.1140/epjc/s10052-021-09348-6}{\doi{10.1140/epjc/s10052-021-09348-6}},
  \href{http://www.arXiv.org/abs/2102.08198}{\texttt{arXiv:2102.08198}}.

\bibitem{CMS:2021klu}
\hrefCMSnoop {}{{CMS Collaboration}, ``{Search for heavy resonances decaying to
  $WW$, $WZ$, or $WH$ boson pairs in a final state consisting of a lepton and a
  large-radius jet in proton-proton collisions at $\sqrt{s}=13\text{ }\text{
  }\mathrm{TeV}$}'',} \textit{ Phys. Rev. D} \textbf{ 105} (2022) 032008,
  \href{http://dx.doi.org/10.1103/PhysRevD.105.032008}{\doi{10.1103/PhysRevD.105.032008}},
  \href{http://www.arXiv.org/abs/2109.06055}{\texttt{arXiv:2109.06055}}.

\bibitem{CMS:2021itu}
\hrefCMSnoop {}{{CMS Collaboration}, ``{Search for heavy resonances decaying to
  Z($\nu\bar{\nu}$)V(q$\bar{\mathrm{q}}$') in proton-proton collisions at
  $\sqrt{s}$ = 13 TeV}'',} 2021.
  \href{http://www.arXiv.org/abs/2109.08268}{\texttt{arXiv:2109.08268}}.
  Submitted to \textit{Phys. Rev. D.}

\bibitem{CMS:2021xor}
\hrefCMSnoop {}{{CMS Collaboration}, ``{Search for heavy resonances decaying to
  ZZ or ZW and axion-like particles mediating nonresonant ZZ or ZH production
  at $\sqrt{s}$ = 13 TeV}'',} 2021.
  \href{http://www.arXiv.org/abs/2111.13669}{\texttt{arXiv:2111.13669}}.
  Submitted to \textit{JHEP}.

\bibitem{B2G-18-008}
\hrefCMSnoop {}{{CMS Collaboration}, ``Search for resonances decaying to a pair
  of {Higgs} bosons in the $\bbbar \qqbar' \ell \nu$ final state in
  proton-proton collisions at $\sqrt{s}=13~{\TeV}$'',} \textit{ JHEP} \textbf{
  10} (2019) 125,
  \href{http://dx.doi.org/10.1007/JHEP10(2019)125}{\doi{10.1007/JHEP10(2019)125}},
\href{http://www.arXiv.org/abs/1904.04193}{\texttt{arXiv:1904.04193}}.

\bibitem{hepdata}
\hrefCMSnoop {}{}{HEPD}ata record for this analysis, 2021.
\newblock
  \href{http://dx.doi.org/10.17182/hepdata.115024}{\doi{10.17182/hepdata.115024}}.

\bibitem{Chatrchyan:2008zzk}
\hrefCMSnoop {}{{CMS Collaboration}, ``The {CMS} experiment at the {CERN}
  {LHC}'',} \textit{ JINST} \textbf{ 3} (2008) S08004,
  \href{http://dx.doi.org/10.1088/1748-0221/3/08/S08004}{\doi{10.1088/1748-0221/3/08/S08004}}.

\bibitem{Sirunyan:2020zal}
\hrefCMSnoop {}{{CMS Collaboration}, ``{Performance of the CMS Level-1 trigger
  in proton-proton collisions at $\sqrt{s} = 13$\,TeV}'',} \textit{ JINST}
  \textbf{ 15} (2020) P10017,
  \href{http://dx.doi.org/10.1088/1748-0221/15/10/P10017}{\doi{10.1088/1748-0221/15/10/P10017}},
  \href{http://www.arXiv.org/abs/2006.10165}{\texttt{arXiv:2006.10165}}.

\bibitem{Khachatryan:2016bia}
\hrefCMSnoop {}{{CMS Collaboration}, ``The {CMS} trigger system'',} \textit{
  JINST} \textbf{ 12} (2017) P01020,
  \href{http://dx.doi.org/10.1088/1748-0221/12/01/P01020}{\doi{10.1088/1748-0221/12/01/P01020}},
\href{http://www.arXiv.org/abs/1609.02366}{\texttt{arXiv:1609.02366}}.

\bibitem{CMS-PRF-14-001}
\hrefCMSnoop {}{{CMS Collaboration}, ``Particle-flow reconstruction and global
  event description with the {CMS} detector'',} \textit{ JINST} \textbf{ 12}
  (2017) P10003,
  \href{http://dx.doi.org/10.1088/1748-0221/12/10/P10003}{\doi{10.1088/1748-0221/12/10/P10003}},
\href{http://www.arXiv.org/abs/1706.04965}{\texttt{arXiv:1706.04965}}.

\bibitem{Sirunyan:2019kia}
\hrefCMSnoop {}{{CMS Collaboration}, ``Performance of missing transverse
  momentum reconstruction in proton-proton collisions at $\sqrt{s} = 13$\,{TeV}
  using the {CMS} detector'',} \textit{ JINST} \textbf{ 14} (2019) P07004,
  \href{http://dx.doi.org/10.1088/1748-0221/14/07/P07004}{\doi{10.1088/1748-0221/14/07/P07004}},
\href{http://www.arXiv.org/abs/1903.06078}{\texttt{arXiv:1903.06078}}.

\bibitem{Cacciari:2008gp}
\hrefCMSnoop {}{M.~Cacciari, G.~P. Salam, and G.~Soyez, ``The anti-\kt jet
  clustering algorithm'',} \textit{ JHEP} \textbf{ 04} (2008) 063,
  \href{http://dx.doi.org/10.1088/1126-6708/2008/04/063}{\doi{10.1088/1126-6708/2008/04/063}},
  \href{http://www.arXiv.org/abs/0802.1189}{\texttt{arXiv:0802.1189}}.

\bibitem{Cacciari:2011ma}
\hrefCMSnoop {}{M.~Cacciari, G.~P. Salam, and G.~Soyez, ``{FastJet} user
  manual'',} \textit{ Eur. Phys. J. C} \textbf{ 72} (2012) 1896,
  \href{http://dx.doi.org/10.1140/epjc/s10052-012-1896-2}{\doi{10.1140/epjc/s10052-012-1896-2}},
\href{http://www.arXiv.org/abs/1111.6097}{\texttt{arXiv:1111.6097}}.

\bibitem{Sirunyan:2020foa}
\hrefCMSnoop {}{{CMS Collaboration}, ``{Pileup mitigation at CMS in 13~TeV
  data}'',} \textit{ JINST} \textbf{ 15} (2020) P09018,
  \href{http://dx.doi.org/10.1088/1748-0221/15/09/p09018}{\doi{10.1088/1748-0221/15/09/p09018}},
  \href{http://www.arXiv.org/abs/2003.00503}{\texttt{arXiv:2003.00503}}.

\bibitem{Bertolini:2014bba}
\hrefCMSnoop {}{D.~Bertolini, P.~Harris, M.~Low, and N.~Tran, ``Pileup per
  particle identification'',} \textit{ JHEP} \textbf{ 10} (2014) 059,
  \href{http://dx.doi.org/10.1007/JHEP10(2014)059}{\doi{10.1007/JHEP10(2014)059}},
\href{http://www.arXiv.org/abs/1407.6013}{\texttt{arXiv:1407.6013}}.

\bibitem{Khachatryan:2016kdb}
\hrefCMSnoop {}{{CMS Collaboration}, ``Jet energy scale and resolution in the
  {CMS} experiment in pp collisions at 8 {TeV}'',} \textit{ JINST} \textbf{ 12}
  (2017) P02014,
  \href{http://dx.doi.org/10.1088/1748-0221/12/02/P02014}{\doi{10.1088/1748-0221/12/02/P02014}},
\href{http://www.arXiv.org/abs/1607.03663}{\texttt{arXiv:1607.03663}}.

\bibitem{Alwall:2014hca}
J.~Alwall\hrefCMSnoop {}{ {et~al.}, ``The automated computation of tree-level
  and next-to-leading order differential cross sections, and their matching to
  parton shower simulations'',} \textit{ JHEP} \textbf{ 07} (2014) 079,
  \href{http://dx.doi.org/10.1007/JHEP07(2014)079}{\doi{10.1007/JHEP07(2014)079}},
\href{http://www.arXiv.org/abs/1405.0301}{\texttt{arXiv:1405.0301}}.

\bibitem{Alwall:2007fs}
J.~Alwall\hrefCMSnoop {}{ {et~al.}, ``Comparative study of various algorithms
  for the merging of parton showers and matrix elements in hadronic
  collisions'',} \textit{ Eur. Phys. J. C} \textbf{ 53} (2008) 473,
  \href{http://dx.doi.org/10.1140/epjc/s10052-007-0490-5}{\doi{10.1140/epjc/s10052-007-0490-5}},
\href{http://www.arXiv.org/abs/0706.2569}{\texttt{arXiv:0706.2569}}.

\bibitem{Li:2012wna}
\hrefCMSnoop {}{Y.~Li and F.~Petriello, ``Combining {QCD} and electroweak
  corrections to dilepton production in the framework of the {FEWZ} simulation
  code'',} \textit{ Phys. Rev. D} \textbf{ 86} (2012) 094034,
  \href{http://dx.doi.org/10.1103/PhysRevD.86.094034}{\doi{10.1103/PhysRevD.86.094034}},
\href{http://www.arXiv.org/abs/1208.5967}{\texttt{arXiv:1208.5967}}.

\bibitem{Frederix2012}
\hrefCMSnoop {}{R.~Frederix and S.~Frixione, ``Merging meets matching in
  {MC@NLO}'',} \textit{ JHEP} \textbf{ 12} (2012) 61,
  \href{http://dx.doi.org/10.1007/JHEP12(2012)061}{\doi{10.1007/JHEP12(2012)061}},
\href{http://www.arXiv.org/abs/1209.6215}{\texttt{arXiv:1209.6215}}.

\bibitem{Nason:2004rx}
\hrefCMSnoop {}{P.~Nason, ``A new method for combining {NLO QCD} with shower
  {M}onte {C}arlo algorithms'',} \textit{ JHEP} \textbf{ 11} (2004) 040,
  \href{http://dx.doi.org/10.1088/1126-6708/2004/11/040}{\doi{10.1088/1126-6708/2004/11/040}},
\href{http://www.arXiv.org/abs/hep-ph/0409146}{\texttt{arXiv:hep-ph/0409146}}.

\bibitem{Frixione:2007vw}
\hrefCMSnoop {}{S.~Frixione, P.~Nason, and C.~Oleari, ``Matching {NLO} {QCD}
  computations with parton shower simulations: the {POWHEG} method'',} \textit{
  JHEP} \textbf{ 11} (2007) 070,
  \href{http://dx.doi.org/10.1088/1126-6708/2007/11/070}{\doi{10.1088/1126-6708/2007/11/070}},
\href{http://www.arXiv.org/abs/0709.2092}{\texttt{arXiv:0709.2092}}.

\bibitem{Alioli:2010xd}
\hrefCMSnoop {}{S.~Alioli, P.~Nason, C.~Oleari, and E.~Re, ``A general
  framework for implementing {NLO} calculations in shower {Monte Carlo}
  programs: the {POWHEG BOX}'',} \textit{ JHEP} \textbf{ 06} (2010) 043,
  \href{http://dx.doi.org/10.1007/JHEP06(2010)043}{\doi{10.1007/JHEP06(2010)043}},
\href{http://www.arXiv.org/abs/1002.2581}{\texttt{arXiv:1002.2581}}.

\bibitem{Re:2010bp}
\hrefCMSnoop {}{E.~Re, ``Single-top {$Wt$-channel} production matched with
  parton showers using the {POWHEG} method'',} \textit{ Eur. Phys. J. C}
  \textbf{ 71} (2011) 1547,
  \href{http://dx.doi.org/10.1140/epjc/s10052-011-1547-z}{\doi{10.1140/epjc/s10052-011-1547-z}},
\href{http://www.arXiv.org/abs/1009.2450}{\texttt{arXiv:1009.2450}}.

\bibitem{Melia:2011tj}
\hrefCMSnoop {}{T.~Melia, P.~Nason, R.~R{\"o}ntsch, and G.~Zanderighi,
  ``{$\PW^+\PW^-$, $\PW\PZ$ and $\PZ\PZ$} production in the {POWHEG BOX}'',}
  \textit{ JHEP} \textbf{ 11} (2011) 078,
  \href{http://dx.doi.org/10.1007/JHEP11(2011)078}{\doi{10.1007/JHEP11(2011)078}},
\href{http://www.arXiv.org/abs/1107.5051}{\texttt{arXiv:1107.5051}}.

\bibitem{Nason:2013ydw}
\hrefCMSnoop {}{P.~Nason and G.~Zanderighi, ``{$W^+ W^-$ , $W Z$ and $Z Z$}
  production in the {POWHEG-BOX-V2}'',} \textit{ Eur. Phys. J. C} \textbf{ 74}
  (2014) 2702,
  \href{http://dx.doi.org/10.1140/epjc/s10052-013-2702-5}{\doi{10.1140/epjc/s10052-013-2702-5}},
\href{http://www.arXiv.org/abs/1311.1365}{\texttt{arXiv:1311.1365}}.

\bibitem{Frederix:2012dh}
\hrefCMSnoop {}{R.~Frederix, E.~Re, and P.~Torrielli, ``{Single-top}
  $t$-channel hadroproduction in the four-flavour scheme with {POWHEG} and
  {aMC@NLO}'',} \textit{ JHEP} \textbf{ 09} (2012) 130,
  \href{http://dx.doi.org/10.1007/JHEP09(2012)130}{\doi{10.1007/JHEP09(2012)130}},
\href{http://www.arXiv.org/abs/1207.5391}{\texttt{arXiv:1207.5391}}.

\bibitem{Hartanto:2015uka}
\hrefCMSnoop {}{H.~B. Hartanto, B.~J{\"a}ger, L.~Reina, and D.~Wackeroth,
  ``{Higgs} boson production in association with top quarks in the {POWHEG
  BOX}'',} \textit{ Phys. Rev. D} \textbf{ 91} (2015) 094003,
  \href{http://dx.doi.org/10.1103/PhysRevD.91.094003}{\doi{10.1103/PhysRevD.91.094003}},
\href{http://www.arXiv.org/abs/1501.04498}{\texttt{arXiv:1501.04498}}.

\bibitem{Czakon:2011xx}
\hrefCMSnoop {}{M.~Czakon and A.~Mitov, ``{Top++}: A program for the
  calculation of the top-pair cross-section at hadron colliders'',} \textit{
  Comput. Phys. Commun.} \textbf{ 185} (2014) 2930,
  \href{http://dx.doi.org/10.1016/j.cpc.2014.06.021}{\doi{10.1016/j.cpc.2014.06.021}},
\href{http://www.arXiv.org/abs/1112.5675}{\texttt{arXiv:1112.5675}}.

\bibitem{Sjostrand:2014zea}
T.~Sj{\"o}strand\hrefCMSnoop {}{ {et~al.}, ``An introduction to {PYTHIA}
  8.2'',} \textit{ Comput. Phys. Commun.} \textbf{ 191} (2015) 159,
  \href{http://dx.doi.org/10.1016/j.cpc.2015.01.024}{\doi{10.1016/j.cpc.2015.01.024}},
\href{http://www.arXiv.org/abs/1410.3012}{\texttt{arXiv:1410.3012}}.

\bibitem{Khachatryan:2015pea}
\hrefCMSnoop {}{{CMS Collaboration}, ``Event generator tunes obtained from
  underlying event and multiparton scattering measurements'',} \textit{ Eur.
  Phys. J. C} \textbf{ 76} (2016) 155,
  \href{http://dx.doi.org/10.1140/epjc/s10052-016-3988-x}{\doi{10.1140/epjc/s10052-016-3988-x}},
\href{http://www.arXiv.org/abs/1512.00815}{\texttt{arXiv:1512.00815}}.

\bibitem{Sirunyan:2020pea}
\hrefCMSnoop {}{{CMS Collaboration}, ``Extraction and validation of a new set
  of {CMS PYTHIA8} tunes from underlying-event measurements'',} \textit{ Eur.
  Phys. J. C} \textbf{ 80} (2020) 4,
  \href{http://dx.doi.org/10.1140/epjc/s10052-019-7499-4}{\doi{10.1140/epjc/s10052-019-7499-4}},
\href{http://www.arXiv.org/abs/1903.12179}{\texttt{arXiv:1903.12179}}.

\bibitem{Ball:2015mu}
\hrefCMSnoop {}{{NNPDF} Collaboration, ``Parton distributions for the {LHC Run
  II}'',} \textit{ JHEP} \textbf{ 04} (2015) 040,
  \href{http://dx.doi.org/10.1007/JHEP04(2015)040}{\doi{10.1007/JHEP04(2015)040}},
\href{http://www.arXiv.org/abs/1410.8849}{\texttt{arXiv:1410.8849}}.

\bibitem{Ball:2017mu}
\hrefCMSnoop {}{{{NNPDF}} Collaboration, ``Parton distributions from
  high-precision collider data'',} \textit{ Eur. Phys. J. C} \textbf{ 77}
  (2017)
  \href{http://dx.doi.org/10.1140/epjc/s10052-017-5199-5}{\doi{10.1140/epjc/s10052-017-5199-5}},
\href{http://www.arXiv.org/abs/1706.00428}{\texttt{arXiv:1706.00428}}.

\bibitem{Agostinelli:2002hh}
\hrefCMSnoop {}{{GEANT4} Collaboration, ``{\GEANTfour}---a simulation
  toolkit'',} \textit{ Nucl. Instrum. Meth. A} \textbf{ 506} (2003) 250,
\href{http://dx.doi.org/10.1016/S0168-9002(03)01368-8}{\doi{10.1016/S0168-9002(03)01368-8}}.

\bibitem{CMS:2020uim}
\hrefCMSnoop {}{{CMS Collaboration}, ``{Electron and photon reconstruction and
  identification with the CMS experiment at the CERN LHC}'',} \textit{ JINST}
  \textbf{ 16} (2021) P05014,
  \href{http://dx.doi.org/10.1088/1748-0221/16/05/P05014}{\doi{10.1088/1748-0221/16/05/P05014}},
  \href{http://www.arXiv.org/abs/2012.06888}{\texttt{arXiv:2012.06888}}.

\bibitem{Sirunyan:2018}
\hrefCMSnoop {}{{CMS Collaboration}, ``Performance of the {CMS} muon detector
  and muon reconstruction with proton-proton collisions at $\sqrt{s} =
  13${\TeV}'',} \textit{ JINST} \textbf{ 13} (2018) P06015,
  \href{http://dx.doi.org/10.1088/1748-0221/13/06/P06015}{\doi{10.1088/1748-0221/13/06/P06015}},
\href{http://www.arXiv.org/abs/1804.04528}{\texttt{arXiv:1804.04528}}.

\bibitem{Rehermann:2010vq}
\hrefCMSnoop {}{K.~Rehermann and B.~Tweedie, ``Efficient identification of
  boosted semileptonic top quarks at the {LHC}'',} \textit{ JHEP} \textbf{ 03}
  (2011) 059,
  \href{http://dx.doi.org/10.1007/JHEP03(2011)059}{\doi{10.1007/JHEP03(2011)059}},
\href{http://www.arXiv.org/abs/1007.2221}{\texttt{arXiv:1007.2221}}.

\bibitem{Dokshitzer:1997in}
\hrefCMSnoop {}{Y.~L. Dokshitzer, G.~D. Leder, S.~Moretti, and B.~R. Webber,
  ``Better jet clustering algorithms'',} \textit{ JHEP} \textbf{ 08} (1997)
  001,
  \href{http://dx.doi.org/10.1088/1126-6708/1997/08/001}{\doi{10.1088/1126-6708/1997/08/001}},
\href{http://www.arXiv.org/abs/hep-ph/9707323}{\texttt{arXiv:hep-ph/9707323}}.

\bibitem{Wobisch:1998wt}
\href {https://inspirehep.net/record/484872}{M.~Wobisch and T.~Wengler,
  ``Hadronization corrections to jet cross sections in deep inelastic
  scattering'',} in \textit{ {Proceedings of the Workshop on Monte Carlo
  Generators for HERA Physics, Hamburg, Germany}}, p.~270.
\newblock 1998.
\newblock
\href{http://www.arXiv.org/abs/hep-ph/9907280}{\texttt{arXiv:hep-ph/9907280}}.
\newblock

\bibitem{Dasgupta:2013ihk}
\hrefCMSnoop {}{M.~Dasgupta, A.~Fregoso, S.~Marzani, and G.~P. Salam, ``Towards
  an understanding of jet substructure'',} \textit{ JHEP} \textbf{ 09} (2013)
  029,
  \href{http://dx.doi.org/10.1007/JHEP09(2013)029}{\doi{10.1007/JHEP09(2013)029}},
\href{http://www.arXiv.org/abs/1307.0007}{\texttt{arXiv:1307.0007}}.

\bibitem{Butterworth:2008iy}
\hrefCMSnoop {}{J.~M. Butterworth, A.~R. Davison, M.~Rubin, and G.~P. Salam,
  ``Jet substructure as a new {Higgs}-search channel at the {LHC}'',} \textit{
  Phys. Rev. Lett.} \textbf{ 100} (2008) 242001,
  \href{http://dx.doi.org/10.1103/PhysRevLett.100.242001}{\doi{10.1103/PhysRevLett.100.242001}},
\href{http://www.arXiv.org/abs/0802.2470}{\texttt{arXiv:0802.2470}}.

\bibitem{Larkoski:2014wba}
\hrefCMSnoop {}{A.~J. Larkoski, S.~Marzani, G.~Soyez, and J.~Thaler, ``Soft
  drop'',} \textit{ JHEP} \textbf{ 05} (2014) 146,
  \href{http://dx.doi.org/10.1007/JHEP05(2014)146}{\doi{10.1007/JHEP05(2014)146}},
\href{http://www.arXiv.org/abs/1402.2657}{\texttt{arXiv:1402.2657}}.

\bibitem{Sirunyan:2018jin}
\hrefCMSnoop {}{{CMS Collaboration}, ``Identification of heavy, energetic,
  hadronically decaying particles using machine-learning techniques'',}
  \textit{ JINST} \textbf{ 15} (2020) P06005,
  \href{http://dx.doi.org/10.1088/1748-0221/15/06/P06005}{\doi{10.1088/1748-0221/15/06/P06005}}.

\bibitem{BTV-16-002}
\hrefCMSnoop {}{{CMS Collaboration}, ``{Identification of heavy-flavour jets
  with the CMS detector in pp collisions at 13~TeV}'',} \textit{ JINST}
  \textbf{ 13} (2018) P05011,
  \href{http://dx.doi.org/10.1088/1748-0221/13/05/P05011}{\doi{10.1088/1748-0221/13/05/P05011}},
\href{http://www.arXiv.org/abs/1712.07158}{\texttt{arXiv:1712.07158}}.

\bibitem{Bols:2020bkb}
E.~Bols\hrefCMSnoop {}{ {et~al.}, ``{Jet flavour classification using
  DeepJet}'',} \textit{ JINST} \textbf{ 15} (2020) P12012,
  \href{http://dx.doi.org/10.1088/1748-0221/15/12/P12012}{\doi{10.1088/1748-0221/15/12/P12012}},
  \href{http://www.arXiv.org/abs/2008.10519}{\texttt{arXiv:2008.10519}}.

\bibitem{CMS-DP-2018-058}
\href {http://cds.cern.ch/record/2646773}{{CMS Collaboration}, ``{Performance
  of the DeepJet b tagging algorithm using 41.9/fb of data from proton-proton
  collisions at 13\TeV with Phase 1 CMS detector}'',} CMS Detector Performance
  Note CMS-DP-2018-058, 2018.

\bibitem{Thaler:2010tr}
\hrefCMSnoop {}{J.~Thaler and K.~Van~Tilburg, ``Identifying boosted objects
  with {N}-subjettiness'',} \textit{ JHEP} \textbf{ 03} (2011) 015,
  \href{http://dx.doi.org/10.1007/JHEP03(2011)015}{\doi{10.1007/JHEP03(2011)015}},
\href{http://www.arXiv.org/abs/1011.2268}{\texttt{arXiv:1011.2268}}.

\bibitem{Sirunyan:2017mzl}
\hrefCMSnoop {}{{CMS Collaboration}, ``Measurement of normalized differential $
  \mathrm{t}\overline{\mathrm{t}} $ cross sections in the dilepton channel from
  pp collisions at $ \sqrt{s}=13 $ {TeV}'',} \textit{ JHEP} \textbf{ 04} (2018)
  060,
  \href{http://dx.doi.org/10.1007/JHEP04(2018)060}{\doi{10.1007/JHEP04(2018)060}},
\href{http://www.arXiv.org/abs/1708.07638}{\texttt{arXiv:1708.07638}}.

\bibitem{Khachatryan:2016mnb}
\hrefCMSnoop {}{{CMS Collaboration}, ``Measurement of differential cross
  sections for top quark pair production using the lepton+jets final state in
  proton-proton collisions at 13 {TeV}'',} \textit{ Phys. Rev. D} \textbf{ 95}
  (2017) 092001,
  \href{http://dx.doi.org/10.1103/PhysRevD.95.092001}{\doi{10.1103/PhysRevD.95.092001}},
\href{http://www.arXiv.org/abs/1610.04191}{\texttt{arXiv:1610.04191}}.

\bibitem{Sirunyan:2018iff}
\hrefCMSnoop {}{{CMS Collaboration}, ``Search for a heavy resonance decaying to
  a pair of vector bosons in the lepton plus merged jet final state at $
  \sqrt{s}=13 $ {TeV}'',} \textit{ JHEP} \textbf{ 05} (2018) 088,
  \href{http://dx.doi.org/10.1007/JHEP05(2018)088}{\doi{10.1007/JHEP05(2018)088}},
\href{http://www.arXiv.org/abs/1802.09407}{\texttt{arXiv:1802.09407}}.

\bibitem{Rosenblatt}
\hrefCMSnoop {}{M.~Rosenblatt, ``Remarks on some nonparametric estimates of a
  density function'',} \textit{ Ann. Math. Statist.} \textbf{ 27} (1956) 832,
  \href{http://dx.doi.org/10.1214/aoms/1177728190}{\doi{10.1214/aoms/1177728190}}.

\bibitem{Silverman}
B.~W. Silverman, ``Density estimation for statistics and data analysis''.
\newblock Chapman and Hall, 1986.
\newblock ISBN~0412246201.

\bibitem{Scott}
D.~W. Scott, ``Multivariate density estimation: theory, practice, and
  visualization''.
\newblock John Wiley and Sons, 1992.
\newblock ISBN~0471547700.

\bibitem{Oreglia:1980cs}
\href {http://www.slac.stanford.edu/cgi-wrap/getdoc/slac-r-236.pdf}{M.~J.
  Oreglia, ``A study of the reactions {$\psi^\prime \to \gamma \gamma
  \psi$}''}.
\newblock PhD thesis, Stanford University, 1980.
\newblock
{SLAC} Report {SLAC-R-236}.

\bibitem{Gaiser:1982yw}
\href
  {http://www-public.slac.stanford.edu/sciDoc/docMeta.aspx?slacPubNumber=slac-r-255.html}{J.~Gaiser,
  ``Charmonium spectroscopy from radiative decays of the {J$/\psi$ and
  $\psi^\prime$}''}.
\newblock PhD thesis, Stanford University, 1982.
\newblock
{SLAC} Report {SLAC-R-255}.

\bibitem{CMS:2021xjt}
\hrefCMSnoop {}{{CMS Collaboration}, ``{Precision luminosity measurement in
  proton-proton collisions at $\sqrt{s} =$ 13 TeV in 2015 and 2016 at CMS}'',}
  \textit{ Eur. Phys. J. C} \textbf{ 81} (2021) 800,
  \href{http://dx.doi.org/10.1140/epjc/s10052-021-09538-2}{\doi{10.1140/epjc/s10052-021-09538-2}},
  \href{http://www.arXiv.org/abs/2104.01927}{\texttt{arXiv:2104.01927}}.

\bibitem{CMS-PAS-LUM-17-004}
\href {https://cds.cern.ch/record/2621960}{{{CMS}} Collaboration, ``{CMS
  luminosity measurement for the 2017 data-taking period at $\sqrt{s} =
  13~\mathrm{TeV}$}'',} {CMS Physics Analysis Summary} CMS-PAS-LUM-17-004,
  2018.

\bibitem{CMS-PAS-LUM-18-002}
\href {https://cds.cern.ch/record/2676164}{{{CMS}} Collaboration, ``{CMS
  luminosity measurement for the 2018 data-taking period at $\sqrt{s} =
  13~\mathrm{TeV}$}'',} {CMS Physics Analysis Summary} CMS-PAS-LUM-18-002,
  2019.

\bibitem{Cacciari:2003fi}
M.~Cacciari\hrefCMSnoop {}{ {et~al.}, ``The $\ttbar$ cross-section at 1.8 and
  1.96{\TeV}: a study of the systematics due to parton densities and scale
  dependence'',} \textit{ JHEP} \textbf{ 04} (2004) 068,
  \href{http://dx.doi.org/10.1088/1126-6708/2004/04/068}{\doi{10.1088/1126-6708/2004/04/068}},
\href{http://www.arXiv.org/abs/hep-ph/0303085}{\texttt{arXiv:hep-ph/0303085}}.

\bibitem{Catani:2003zt}
\hrefCMSnoop {}{S.~Catani, D.~de~Florian, M.~Grazzini, and P.~Nason,
  ``Soft-gluon resummation for {Higgs} boson production at hadron colliders'',}
  \textit{ JHEP} \textbf{ 07} (2003) 028,
  \href{http://dx.doi.org/10.1088/1126-6708/2003/07/028}{\doi{10.1088/1126-6708/2003/07/028}},
\href{http://www.arXiv.org/abs/hep-ph/0306211}{\texttt{arXiv:hep-ph/0306211}}.

\bibitem{Baker:1983tu}
\hrefCMSnoop {}{S.~Baker and R.~D. Cousins, ``Clarification of the use of chi
  square and likelihood functions in fits to histograms'',} \textit{ Nucl.
  Instrum. Meth.} \textbf{ 221} (1984) 437,
\href{http://dx.doi.org/10.1016/0167-5087(84)90016-4}{\doi{10.1016/0167-5087(84)90016-4}}.

\bibitem{Cowan:2010js}
\hrefCMSnoop {}{G.~Cowan, K.~Cranmer, E.~Gross, and O.~Vitells, ``Asymptotic
  formulae for likelihood-based tests of new physics'',} \textit{ Eur. Phys. J.
  C} \textbf{ 71} (2011) 1554,
  \href{http://dx.doi.org/10.1140/epjc/s10052-011-1554-0}{\doi{10.1140/epjc/s10052-011-1554-0}},
  \href{http://www.arXiv.org/abs/1007.1727}{\texttt{arXiv:1007.1727}}.
[Erratum: \DOI{10.1140/epjc/s10052-013-2501-z}].

\bibitem{Junk:1999kv}
\hrefCMSnoop {}{T.~Junk, ``Confidence level computation for combining searches
  with small statistics'',} \textit{ Nucl. Instrum. Meth. A} \textbf{ 434}
  (1999) 435,
  \href{http://dx.doi.org/10.1016/S0168-9002(99)00498-2}{\doi{10.1016/S0168-9002(99)00498-2}},
\href{http://www.arXiv.org/abs/hep-ex/9902006}{\texttt{arXiv:hep-ex/9902006}}.

\bibitem{Read:2002hq}
\hrefCMSnoop {}{A.~L. Read, ``Presentation of search results: The {\CLs}
  technique'',} \textit{ J. Phys. G} \textbf{ 28} (2002) 2693,
\href{http://dx.doi.org/10.1088/0954-3899/28/10/313}{\doi{10.1088/0954-3899/28/10/313}}.

\bibitem{Oliveira:2014kla}
\hrefCMSnoop {}{A.~Carvalho, ``{Gravity particles from Warped Extra Dimensions,
  predictions for LHC}'',} 2014.
  \href{http://www.arXiv.org/abs/1404.0102}{\texttt{arXiv:1404.0102}}.

\bibitem{Gouzevitch:2013qca}
M.~Gouzevitch\hrefCMSnoop {}{ {et~al.}, ``Scale-invariant resonance tagging in
  multijet events and new physics in {Higgs} pair production'',} \textit{ JHEP}
  \textbf{ 07} (2013) 148,
  \href{http://dx.doi.org/10.1007/JHEP07(2013)148}{\doi{10.1007/JHEP07(2013)148}},
\href{http://www.arXiv.org/abs/1303.6636}{\texttt{arXiv:1303.6636}}.

\end{thebibliography}\endgroup
